\documentstyle[12pt,astron,psfig,huub]{article_huub}

\begin{document}

\centerline{\Large\bf The Gaseous Environments of Radio Galaxies in the }
\centerline{\Large\bf Early Universe: Kinematics of the Lyman $\alpha$ Emission} 
\centerline{\Large\bf and Spatially Resolved HI Absorption
\footnote{Based on observations collected at the European Southern Observatory,
La Silla, Chile, with the WHT on La Palma, Spain and with the AAT, Australia.}}

\begin{verse}
R. van Ojik$^1$, H.J.A. R\"ottgering$^{1,2,3}$, G.K. Miley$^1$, R.W. Hunstead$^4$
\end{verse}

{\small\it
\begin{quote}
\item $^1$Leiden Observatory, P.O. Box 9513, 2300 RA, Leiden, The Netherlands
\item $^2$Mullard Radio Astronomy Observatory, Cavendish Laboratory, Cambridge CB3 0HE, England
\item $^3$Institute of Astronomy, Madingley Road, Cambridge CB3 0HA, England
\item $^4$School of Physics, University of Sydney, NSW2006, Australia
\end{quote}
}
\vspace{1cm}

$^*$ Based on observations collected at the European Southern Observatory,
La Silla, Chile, with the WHT on La Palma, Spain and with the AAT, Australia.

\vspace{4cm}

\noindent R. van Ojik et al.: The gaseous environment of radio galaxies in 
the early Universe: Kinematics of the Lyman $\alpha$ emission and spatially
resolved HI absorption

\noindent Main Journal

\noindent 3. Extragalactic Astronomy

\noindent 11.01.2, 11.09.1, 11.19.3

\noindent H. J. A. R\"ottgering, 
Leiden Observatory, Niels Bohrweg 2, P.O. Box 9513, 2300 RA
Leiden, The Netherlands, tel. 071-275835

\newpage

\addtocounter{page}{-1}
\begin{abstract}
\noindent
We present intermediate resolution ($\sim3$ \AA ) spectra of the
Ly$\alpha$ emission from 15 high redshift radio galaxies ($z>2$).
Together with previously published spectra we analyze data for a sample
of 18 objects. 

In 11 of the 18 radio galaxies we find deep troughs in the Ly$\alpha$
emission profile, which we interpret as H\,{\sc i} absorption  with
column densities in the range $10^{18}$--$10^{19.5}$ cm$^{-2}$. Since in
most cases the Ly$\alpha$ emission is absorbed over its entire spatial
extent (up to 50 kpc), the absorbers must have a covering fraction close
to unity. Under plausible assumptions for the temperature and density of
the absorbing gas this implies that the absorbing material must consist
of $\sim 10^{12}$ clouds of typical size $\sim 0.03$ pc with a total mass
of $\sim 10^8$ M$_{\odot}$. 

Our observations show that strong H\,{\sc i} absorption occurs in
$>60$\% of the high redshift radio galaxies, while from the statistics of
quasar absorption lines there is only a 2\% probability of such a strong
H\,{\sc i} absorption line falling by chance in the small redshift
interval of the Ly$\alpha$ emission line. These absorbers are therefore
most likely to be physically associated with the galaxy hosting the radio
source or its direct environment. 

There are strong correlations between the properties of the Ly$\alpha$
emission of the galaxies and the size of the associated radio source: (i)
Of the smaller ($<50$ kpc) radio galaxies 9 out of 10 have strong
associated H\,{\sc i} absorption, whereas only 2 of the 8 larger ($>50$
kpc) radio galaxies show such strong absorption. (ii) Larger radio
sources tend to have larger Ly$\alpha$ emission regions and (iii) a
smaller Ly$\alpha$ velocity dispersion than smaller radio sources. The
sizes of the Ly$\alpha$ regions range from $\sim$15 to $\sim$130 kpc and
radio sizes from $\sim$20 to $\sim$180 kpc. The Ly$\alpha$ velocity
dispersions range from $\sim$700 km s$^{-1}$ (FWHM) in the largest radio
sources to $\sim$1600 km s$^{-1}$ (FWHM) in the smallest. In the smaller
radio sources Ly$\alpha$ is often observed to be more extended than the
radio emission. 

We have also defined several parameters to describe the spatial
distortion of the Ly$\alpha$ velocity field and of the radio structure.
We find a strong correlation between the amount of distortion present in
the Ly$\alpha$ spectra and the distortion of the radio structure.

These strong correlations show that the radio sources have a profound
influence on their environments. The correlations of radio size with the
gas velocity dispersion and the Ly$\alpha$ size, are evidence for
interaction of the radio jet with the ionized gas. 

Three different scenarios to explain these correlations are discussed:

\begin{enumerate}
\item The first is based on the properties of the environment. The smaller
radio sources are situated in denser (cluster) environments than the
larger sources. The relatively dense environment is responsible for the
strong extended H\,{\sc i} absorptions. Kinetic energy is transferred
from the radio plasma to the gaseous medium, resulting in larger line
widths and reduced propagation velocities for the radio lobes.  The
turbulence associated with the radio plasma may expose more gas from
dense clouds to the ionizing radiation, causing the Ly$\alpha$ emission
region to have an extent similar to the radio source. 

\item The second scenario linking radio size and Ly$\alpha$ is based on
evolution of the radio source. When the radio  source is small it 
interacts strongly with the dense central gas, while as it tunnels
through the medium to larger sizes the interaction and thus the gas
velocity dispersion decreases. The radio size -- Ly$\alpha$ size relation
has the same origin as in scenario 1.

\item The third scenario is based on orientation effects. If radio
galaxies are quasars whose nuclei and broad line regions are obscured
towards our line of sight, larger line widths may be expected if we
observe closer to the edge of the ionization cone. In this scenario the
radio size -- Ly$\alpha$ size relation would  be simply due to projection
effects. The largest radio sources have the largest Ly$\alpha$ sizes with
the smallest velocity dispersions because they are oriented closest to
the plane of the sky and the broad line regions are most strongly
obscured. This would not explain, however, why the largest sources show
almost no associated H\,{\sc i} absorption. 

\end{enumerate}

\end{abstract}


\markboth{Kinematics of Ly$\alpha$ and spatially resolved HI absorption}
{Kinematics of Ly$\alpha$ and spatially resolved HI absorption}

\section{Introduction} During the last few years we have carried out a
programme to enlarge the sample of high-redshift radio galaxies ($z>2$,
HZRGs hereafter), by selecting ultra-steep-spectrum (USS) radio sources
\cite{rot93,rot94}. This project has included radio imaging, optical
broad-band imaging and spectroscopy. Most of the optical observations
were carried out as a Key Programme at the European Southern Observatory
in Chile. Until now our USS radio source survey has resulted in the
discovery of 29 radio galaxies at redshifts $z>2$
\cite{rot93,rot95,rot96,oji94a,oji96}, i.e. about half of the total
number of galaxies known at such high redshifts. The Ly$\alpha$ emission
from HZRGs can be very bright, with fluxes $\sim10^{-15}$ erg s$^{-1}$
cm$^{-2}$ (luminosities $\sim 10^{44}$ erg s$^{-1}$), and can extend over
more than $10''$ ($\sim100$ kpc).  These Ly$\alpha$ halos are therefore
ideal targets for ground-based optical telescopes, providing a powerful
tool for studying the gas around the most massive galaxies known in the
early Universe. 

The morphologies and kinematics of the emission line gas in low-redshift
radio galaxies have been studied extensively (e.g. Baum {et al.}
1992\nocite{bau92}, and references therein). Only now are sufficiently
large numbers of HZRGs known to allow such detailed studies at high
redshifts so that the properties of the emission line halos at low and
high redshifts can be compared. Such a comparison may help to constrain
the evolution of the radio galaxies and their environments. 

Furthermore, these extended Ly$\alpha$ halos provide an opportunity to
observe H\,{\sc i} absorption clouds against a spatially resolved
background source. Since the late sixties, there have been extensive
studies of absorption lines against the strong continuum emission from
quasar nuclei. Because quasars are spatially unresolved, such studies
yield no direct information on the spatial scale of the
absorbers.\footnote{Limited information on the lateral extent of
absorbing clouds has come from quasar pairs and gravitationally-lensed
quasars with multiple images.}  The detection of such absorption clouds
against the Ly$\alpha$ emission of HZRGs can, however, provide us with
spatial information about the absorbing clouds. An additional advantage
of studying absorption against the strong Ly$\alpha$ of HZRGs is that no
strong ionizing continuum is (directly) observed from HZRGs that could
influence the properties of the absorber. 

We have previously presented observations of the Ly$\alpha$ kinematics of
two radio galaxies, with strong spatially extended absorption being seen
in 0943$-$242 at $z=2.9$ \cite{rot95a}, and complex kinematics in
1243+036 at $z=3.6$ (van Ojik et al. 1996)\nocite{oji96}. 
Earlier, a high resolution spectrum of the Ly$\alpha$ emission from 4C41.17 has
been presented by Chambers et al. (1990)\nocite{cha90}. We report
here on the results of high spectral resolution observations of a further
15 radio galaxies from our sample. 

In Sect. 2 we describe the selection of objects and the observations.
In Sect. 3 we present the two-dimensional and one-dimensional
Ly$\alpha$ spectra. The spectra are described in detail and parametrized
to obtain the properties of the line-emitting gas and H\,{\sc i}
absorption systems; correlations between the Ly$\alpha$ and radio
properties are investigated. In Sect. 4 we discuss the physical
properties of the Ly$\alpha$-emitting gas (including its kinematics and
dynamics) and the H\,{\sc i} absorption systems, and their relationship
with the radio galaxy. We go on to discuss the implications of our work
for the dust content of HZRGs and, finally, we put forward three
scenarios which may account for the interaction between the radio source
and the Ly$\alpha$ emission. We summarize our results and conclusions in
Sect. 5.

Throughout this paper we assume a Hubble constant of $H_0=50$ km
s$^{-1}$ Mpc$^{-1}$ and a deceleration parameter of $q_0=0.5$.

\section{Source Selection and Observations}

The main criterion for the selection of objects from our sample of high
redshift radio galaxies was that spectra with good signal-to-noise (S/N)
ratio be obtained. We therefore concentrated on the objects with bright
Ly$\alpha$ emission and redshifts which placed Ly$\alpha$ in the most
sensitive range of the spectrographs and detectors. The S/N ratio at very
blue wavelengths is also limited by the transmission of the atmosphere.
We therefore selected objects with redshifts $z>2.1$. As a secondary
consideration we gave preference to objects whose spatial extent was
large on the low resolution spectra.

High resolution optical spectroscopy was carried out during several
different observing sessions. Most of the observations were made with the
New Technology Telescope (NTT) of the European Southern Observatory in
Chile. One session was with the Anglo-Australian Telescope (AAT) in
Australia and one session with the William Herschel Telescope (WHT) on La
Palma, Spain (see Table 1). The spectral
resolutions ranged from 1.5 to 3 \AA\ FWHM, corresponding to 100--200 km
s$^{-1}$, about a factor of ten better than in the low resolution
discovery spectra \cite{rot96} where the Ly$\alpha$ emission lines are
just resolved, with velocity dispersions $\sim$1000--2000 km s$^{-1}$
(FWHM). A resolution of 2--3 \AA\ is therefore well suited to studying
the overall kinematical structure of the emission-line gas and the
possible presence of strong H\,{\sc i} absorption in the Ly$\alpha$
emission profile.

The ESO NTT observations used the ESO Multi Mode Instrument (EMMI), with
the light path split into two arms optimized for observations in the blue
and red respectively. Our observations were made in the blue arm where
the detector was a Tektronix CCD having $1024 \times 1024$ pixels. The
scale along the slit was $0.37''$ per pixel; with a slit width of $2.5''$
and ESO grating 3 we achieved a spectral resolution of 2.8 \AA.
Observations at the WHT were made with the blue arm of the ISIS
spectrograph. A $1024 \times 1024$ Tektronix CCD was used, giving a scale
along the slit of $0.49''$ per pixel.  With a slit width of $2''$ and
grating R1200B, the spectral resolution was 1.7 \AA. At the AAT, the RGO
spectrograph was used with a $1024 \times 1024$ Tektronix CCD.  The
spatial scale along the slit was $0.74''$ per pixel, and a $1.6''$ slit
and 1200B grating gave a spectral resolution of 1.5 \AA\ (see also
R{\"o}ttgering {et al.} 1995a\nocite{rot95a}). Important parameters
for the different observing sessions are summarized in Table 1, 
where we give the dates of the different
sessions, telescopes and instruments, achieved resolutions and an
indication of the observing conditions. 

The objects were positioned in the slit using blind offsetting from
nearby ``bright'' stars. Due to inaccuracies in the blind offsetting ---
offsets were sometimes $>1'$ --- the measured line fluxes may be slightly
lower than those measured from our low resolution spectroscopy
\cite{rot96}.\footnote{The low resolution spectra were obtained by
first taking short images to position the object exactly in the slit.} 

The objects that we observed are listed in Table 2,
with the relevant total exposure times, slit position angles and 
cross-reference to the sessions in Table 1.
The spectra were flat-fielded, sky-subtracted and wavelength- and
flux-calibrated using the longslit package in the {\sc iraf} reduction
software of the U.S. National Optical Astronomy Observatory.

\section{Results and Analysis}

\subsection{Description and parameterization of the Ly$\alpha$ spectra}

In the following subsections we shall derive various properties and
parameters of the Ly$\alpha$ spectra. The two-dimensional spectra are
shown in Fig. 1. They have been smoothed slightly with a
Gaussian function of $1'' \times 2$ \AA\ FWHM to enhance the most
extended Ly$\alpha$ emission. One-dimensional spectra were extracted
using apertures which included all extended Ly$\alpha$ emission visible
in Fig. 1. These are shown in Fig. 4.
Spatial profiles of the Ly$\alpha$ emission, summed over the full
spectral range where Ly$\alpha$ emission was detected, are in  Fig. 5. 
The profiles have been smoothed with a Gaussian function
of $1''$ FWHM.

\subsubsection{The Ly$\alpha$ profiles}

From the spectra it is immediately clear that many of the radio galaxies
have a double- or multi-peaked Ly$\alpha$ velocity profile. The troughs
between the peaks have widths of several hundred kilometres per second
and in most of them no emission is detected from the bottom of the
troughs. However, in several cases the depth of the troughs change with
spatial location; a good example of this is 0200+015, where one end of
the trough is shallower than the other (see also below).

Because of the steepness of the deepest troughs and the relatively smooth
overall velocity profile of Ly$\alpha$ emission, we conclude that the
troughs are unlikely to be caused by genuine velocity structure of the
Ly$\alpha$ emitting gas but are caused instead by the absorption of
Ly$\alpha$ photons by neutral hydrogen very close in redshift. The
troughs appear deeper on the one-dimensional spectra of Fig. 4  
than on the smoothed two-dimensional ones (Fig. 1), 
consistent with the absorption features being
relatively narrow. The profiles of the deep troughs in the Ly$\alpha$
profiles appear similar to strong absorption lines observed in the
spectra of distant QSOs. Most of the QSO absorption lines at such
redshifts are interpreted as arising in neutral hydrogen clouds between
us and the QSO. It therefore seems likely that the troughs in the
Ly$\alpha$ emission profiles of our radio galaxies are also due to
H\,{\sc i} absorption systems. Although
dust mixed through the emission
line gas can also effectively absorb Ly$\alpha$ emission due to the large
pathlength of the resonant scattering Ly$\alpha$ photons, 
dust would not be expected 
to produce such a narrow absorption feature but 
the entire Ly$\alpha$ emission profile would be depressed and have a
more chaotic structure (e.g. 0211$-$122, see below and van Ojik {et
al.} 1994 \nocite{oji94a}). We therefore conclude that the deep troughs
in the Ly$\alpha$ emission profiles are caused by H\,{\sc i} absorption. 

In some cases where the S/N is relatively low, we cannot exclude the
possibility that the double peaked profile is caused by two separate
velocity components of the ionized gas. However, since these profiles are
very similar to the profiles with good S/N in which a sharp H\,{\sc i}
absorption trough is observed, we conclude that the double peak in
Ly$\alpha$ profiles observed with lower S/N are also probably due to
H\,{\sc i} absorption.

We will model the Ly$\alpha$ profile assuming that the underlying
Ly$\alpha$ emission line has a Gaussian velocity profile and that the
troughs are due to H\,{\sc i} absorption characterised by Voigt profiles.
We used an iterative scheme that minimized the sum of the squares of the
differences between the model and the observed spectrum (see also
R\"ottgering {et al.} 1995a\nocite{rot95a}). Input parameters for
this scheme are initial estimates for the redshift, strength and width of
the unabsorbed Gaussian emission line and estimates of the redshifts,
Doppler parameters and column densities ($z$, $b$ and N(H\,{\sc i})) of
the absorption systems. Most objects show only one obvious absorption
feature, but in a few cases there are several (0200+015, 0828+193,
2202+128). The iterative scheme solved for the parameters of the model
and we thereby derived the best fit shape of the Gaussian emission line
and the redshifts, column densities and Doppler parameters of the H\,{\sc
I} absorbers. 

The modelling of the troughs was not always satisfactory. If an
absorption feature appears too complex to be one single absorber, but
more likely consists of several absorbers close to each other in
redshift, the best fit from the model fitting procedure is not unique and
the derived parameters should be treated with caution. Features in the
spectra that are not sharp troughs but are more like ``shoulders'' in the
profile are observed in several objects with the best S/N spectra. These
``shoulders''  may not be H\,{\sc i} absorption but could be due to true
velocity structure in the emitting gas. In spite of these possible small
deviations from a Gaussian emission profile, the Gaussian approximation
appears to be a reasonable description of the overall underlying emission
profiles. Modelling these ``shoulders'' in the Ly$\alpha$ profile with
H\,{\sc i} absorption requires large Doppler parameters ($>200$ km
s$^{-1}$) and relatively low H\,{\sc i} column densities (N(H\,{\sc i})
$\sim 10^{15}$ cm$^{-2}$). Large Doppler parameters imply macroscopic
motion of the absorbing gas, as a thermal origin would require a gas
temperatures of $>10^5$ K for $b>50$ km s$^{-1}$ and $>10^6$ K for
$b>200$ km s$^{-1}$ (see also Section 4). To ensure that the iterative
scheme did not diverge for those objects with a ``shoulder'' in the
Ly$\alpha$ profile and a strong narrow (N(H\,{\sc i}) $>10^{17}$
cm$^{-2}$) absorption feature, such broad ``shoulders'' were also
modelled by H\,{\sc i} absorption. In the section on individual objects
we illustrate these difficulties for a few cases. 

The one-dimensional Ly$\alpha$ profiles and the adopted model-fits of the
objects are displayed in Fig. 4. The derived parameters
($z$, $b$ and N(H\,{\sc i})) of the absorption systems are listed in
Table 3. In this table $z_{Ly\alpha}$ is the redshift of the peak of
the original emission line profile (assumed Gaussian) that was modelled
with the iterative fitting of emission and absorptions. The absorbers for
each object are numbered in order of the H\,{\sc i} column density, $z_1$
being the redshift of the strongest H\,{\sc i} absorption system in an
object. The redshifts of the emission and absorbers were determined after
converting the wavelengths to vacuum values and have an estimated
accuracy of $\sim 0.0002$. The resultant width of the original
(unabsorbed) Ly$\alpha$ emission line is listed in Table 4, which lists
all determined parameters of the Ly$\alpha$ emission.

\subsubsection{Extent and mass of the Ly$\alpha$ gas}
From the two-dimensional spectra the spatial extent of the 
Ly$\alpha$
emission regions varies from $2''$--$17''$
between the sources. We can define
a total extent by simply measuring the maximum extent to which Ly$\alpha$ 
emission is seen at $> 2 \sigma$ level in Fig. 1. We 
denote this extent by D$_{Ly\alpha}^{tot}$. However, this measure is not a 
good indicator of the scale size of the Ly$\alpha$ extent
because the spectra have different integration times and sensitivities.
Furthermore, the flux level of the most extended Ly$\alpha$ emission may depend 
on the central Ly$\alpha$ flux of an object and thus could be below the 
detection limit of our spectra for the objects with low Ly$\alpha$ luminosity. 
To obtain a more robust measure of the Ly$\alpha$ extent, we have 
extracted spatial profiles of the Ly$\alpha$ by summing the columns over the
wavelength range where Ly$\alpha$ was detected in Fig. 1. 
From these spatial profiles we define a scale size of the Ly$\alpha$ emission,
D$_{Ly\alpha}^{20\%}$, as 
the extent between the spatial points where the Ly$\alpha$ emission flux is 
20\% of the peak level. This 20\% 
level is chosen because it can still be measured reliably in the spectra with 
the lowest S/N ratio in our sample.
Both D$_{Ly\alpha}^{20\%}$ and D$_{Ly\alpha}^{tot}$ are listed in Table 4.

We can estimate the mass of the Ly$\alpha$
emitting gas from the integrated Ly$\alpha$ luminosity as measured 
from our spectra.
After obtaining the density of the ionized
gas assuming case B recombination through 
$L=4 \times 10^{-24}n_e^2 f_v V$ erg s$^{-1}$ \cite{mcc90a}, 
where $V$ is the total volume occupied
by the emission line gas and $f_v$ the volume filling factor of the gas, 
the mass in ionized gas is M$\approx n_e m_{proton} f_v V$ \cite{mcc90a}.
The volume filling factor of the ionized gas must be estimated because 
we have no means of determining it directly for our HZRGs.
From direct measurements, using sulphur lines,
of the density of line emitting gas in low redshift
radio galaxies, filling factors in the range 10$^{-4}$--10$^{-6}$ and
typically of order $\sim10^{-5}$ have been deduced \cite{bre85d,hec82}. We 
will assume this value for the Ly$\alpha$ emission line regions of our HZRGs.
For a more detailed discussion of the ionization of the Ly$\alpha$ emission
line region in HZRGs see e.g. McCarthy {et al.} (1990), Chambers {et al.}
(1990), McCarthy (1993), van Ojik {et al.} 
(1996)\nocite{mcc90a,cha90,mcc93a,oji96}.
Calculating the volume of the measured Ly$\alpha$ emission from the 
defined D$^{20\%}_{Ly\alpha}$ and the width of the slit and taking the depth
of the Ly$\alpha$ region the same as the transverse size (i.e. the slit width)
we find that the masses of Ly$\alpha$ emitting gas in the radio galaxies 
of our sample are are typically a few times 10$^8$ M$_{\odot}$ (see Table 4).

\subsubsection{Spatial structure in the Ly$\alpha$ spectra}
When examining the two-dimensional spectra it is apparent that
in many cases the spatial position of the maximum intensity of the 
Ly$\alpha$ emission varies as a function of wavelength.
In some cases the peak position changes smoothly with wavelength
(e.g. 0748+134) while in others it varies in a more irregular manner (e.g.
0828+193, 2202+128). We have measured this spatial position 
of the maximum Ly$\alpha$ intensity
per wavelength bin (S$_{\rm M}$($\lambda$)), 
where the extent of a wavelength bin varied 
from 2--4 pixels
to obtain a reasonable S/N per bin so that
the position could be determined with an accuracy of no worse than a few
tenths of an arcsecond. Note that these measurements of the spatial
position of the maximum intensity of Ly$\alpha$ at different wavelengths 
(S$_{\rm M}$($\lambda$))
is not the same as used to measure velocity structure in spectra, where at 
every spatial position the wavelength of the maximum intensity 
of the emission line profile is determined (thus $\lambda _{\rm M}$(S)).
For the objects with
brightest Ly$\alpha$ the position variations were measured on the unsmoothed 
frames, while the fainter were measured from the smoothed frames of Fig. 1. 
The errors
on the positions depend on the S/N level of the Ly$\alpha$ emission in
each wavelength bin.
The regions of the spectra where a strong H\,{\sc i} absorption
is present were excluded.
The galaxy 1707+105 was excluded from these 
measurements because it has at least two spatial maximums at
the position of two galaxies situated along the radio axis.
We also give positional changes from the two-dimensional
Ly$\alpha$ spectrum of radio galaxy 1243+036 (van Ojik et al. 1996).
\nocite{oji96}
The results are shown in  Fig. 6.

From Fig. 6 we see that indeed most objects show significant
changes in the spatial positions as a function of wavelength (or velocity).
The amplitude of these changes vary from
a few tenths of an arcsecond to several arcseconds.
In some spectra the spatial position changes gradually while some others 
have several spatial ``wiggles''. We will parameterize the positional changes
in two ways. The first is the amplitude, 
$\Delta$S = S$_{\rm M}^{\rm max} - $S$_{\rm M}^{\rm min}$, 
which 
is simply the distance between the most extreme
spatial positions of maximum Ly$\alpha$ intensity measured 
(S$_{\rm M}^{\rm max}$ and S$_{\rm M}^{\rm min}$),
whose accuracy is still better than $\sim0.5''$.
This amplitude does not indicate whether there is
a smooth gradient or if there are several discrete ``wiggles''.
Therefore, we introduce
another parameter that indicates the amount of ``wiggles'' in the positional 
changes relative to the velocity range over which the positions are measured.
This ``wiggling index'', $w_S = n_S / \Delta v$, 
is (an estimate of) the number of different positional gradients along the
velocity profile, $n_S$ (or the number of times that the positional gradient
changes, plus one), divided by the velocity range ($\Delta v$, in thousands km
s$^{-1}$) of the interval measured.

In Table 4 we list all measured Ly$\alpha$ properties.
In this table we have also included the additional objects 1243+036
(van Ojik et al. 1996)\nocite{oji96}, 
0943-242 \cite{rot95a} and 4C\,41.17 \cite{cha90} from which
similar high resolution spectra have been published.
In some cases we found that the
Ly$\alpha$ fluxes at high resolution were smaller than those measured in the 
low resolution spectra \cite{rot96,oji94a,oji96}
from which the redshifts were determined. Various factors can contribute
to this discrepancy, e.g. differing seeing or photometric conditions, or
small errors in the telescopes blind offsetting for the high 
resolution spectroscopy.
Therefore the objects for which we have determined the 
Ly$\alpha$ fluxes only from the high resolution spectroscopy are marked
in Table 4. Due to the large
sizes of the Ly$\alpha$ emission regions, it is likely that the limited 
width of the slit ($2.5''$) of the spectrograph did not cover the entire
transverse extent of the Ly$\alpha$ emission (see e.g. 1243+036). 
We estimate that the true
Ly$\alpha$ fluxes of the galaxies may well be a factor 2 higher than those
determined from our high resolution spectroscopy.
The columns give the following parameters:
\newline
(1) the source name \newline
(2) the velocity width (FWHM) of the Gaussian fitted to the 
    Ly$\alpha$ emission profile (section 3.1.1) \newline
(3) indication whether strong H\,{\sc i} absorption ($>10^{17}$ cm$^{-2}$) is present: 
    1: yes; 0: no \newline
(4) 20\% projected linear size of the Ly$\alpha$ emission
    region (20\% of the peak level in Fig. 5) \newline
(5) total observed projected linear 
    size of the Ly$\alpha$ region ($>2 \sigma$ in Fig. 1) \newline
(6) logarithm of the Ly$\alpha$ luminosity \newline
(7) mass of the Ly$\alpha$ emitting gas \newline
(8) the total H\,{\sc i} mass derived from the column density and size of the absorber (Section 4) \newline
(9) the distance between the most extreme spatial positions of maximum 
    Ly$\alpha$ intensity in the spectrum \newline
(10) The number of different positional gradients of the maximum Ly$\alpha$ 
     intensity along the velocity
     profile as seen in Fig. 6 \newline
(11) The ``wiggling index'', $w_S = n_S / \Delta v$, the number of positional
     gradients of Ly$\alpha$ (column 10) divided by the velocity interval 
     over which the Ly$\alpha$ positions were measured

\subsection{Remarks on individual objects}

In this section we discuss in detail the Ly$\alpha$ spectra of several
of the most interesting objects from the sample. This will illustrate
that in at least some of these systems there is clear observational
evidence that dust, broad emission lines or accompanying galaxies play
an important role. It will also illustrate some of the limitations of
the fitting procedure that we applied to obtain parameters of the
absorption systems.  For the full list off remarks on objects we refer
to an \underline{Electronic Appendix}.

\noindent {\bf 0211$-$122}: This radio galaxy has a peculiar optical spectrum
in which the Ly$\alpha$ emission is anomalously weak compared to the higher 
ionization lines. This anomaly 
has been interpreted as being produced by dust mixed through the emission 
line gas which partly absorbs the Ly$\alpha$ emission \cite{oji94a}.
The two-dimensional high resolution 
spectrum of the Ly$\alpha$ shows a clearly different structure than the rest 
of the galaxies in our sample. 
There is one small region with strong Ly$\alpha$ 
emission that is relatively narrow (300 km s$^{-1}$ FWHM) and is responsible
for more than one third of the flux. Furthermore, there are several weaker 
patches of Ly$\alpha$ emission distributed around the bright peak.
The bright narrow peak is spatially offset by $\sim 1''$ from the 
peak of the continuum emission.
This somewhat peculiar two-dimensional Ly$\alpha$ profile is consistent with the
interpretation of van Ojik et al. (1994)\nocite{oji94a} 
that dust is mixed through the
Ly$\alpha$ emitting gas of 0211$-$122.
The bright narrow peak may be a region where the dust
content is sufficiently low so that the Ly$\alpha$ photons can escape. 
In spite of the different appearance of the velocity field, the
integrated line profile does show a double peaked shape.
Although this double peaked appearance may be produced by dust within 
the Ly$\alpha$ region, 
it is also possible that an H\,{\sc i} absorption system plays a role.

If we model the observed
profile as being due to H\,{\sc i} absorption of an originally Gaussian profile, we 
obtain a fit as displayed in Fig. 2 with a very narrow 
($\sim$700
km s$^{-1}$ FWHM) Ly$\alpha$ having a strong 
H\,{\sc i} absorption of column density N(H\,{\sc i})=10$^{19.5}$ cm$^{-2}$. 
However, the centre of this absorption is not as opaque
as it should be with such a strong absorption (see also the two-dimensional 
spectrum in Fig. 1). There is more Ly$\alpha$ emission from
the bottom of the trough than in any of the other galaxies with strong HI
absorption. Thus it seems unlikely that this simple model
is correct. An alternative fit is given by a less luminous
original emission profile, thus
ignoring most of the emission from the bright narrow peak and giving an HI
column density of $\sim10^{18}$ cm$^{-2}$ (displayed in Fig. 4). 
However, this fit is also not well matched to the observed profile. 
It may be that the profile is purely due to dust 
mixed with the ionized gas, extinguishing the resonant scattering 
Ly$\alpha$ photons, although we cannot exclude that
H\,{\sc i} absorptions contribute to the shape of the Ly$\alpha$ profile
of 0211$-$122.

\noindent {\bf 0828+193}: 
The Ly$\alpha$ profile has a
spectacular shape. The emission is very strong but drops steeply
on the blue side of the peak, while slightly further bluewards some Ly$\alpha$
emission is visible again. Thus, it appears that almost the entire
blue wing of the Ly$\alpha$ emission profile has been absorbed. Also the most 
extended and fainter emission shows the same sudden drop at 4343 \AA.
This is a remarkable radio galaxy that 
has a close ($\sim3''$) companion along its radio axis \cite{rot95}.
The presence of a close companion, from which no Ly$\alpha$ emission is 
detected, suggests that a neutral gaseous halo of this galaxy might be
responsible for the absorption of the blue wing of the Ly$\alpha$ emission from
0828+193. However, it is not certain that the companion is at the same redshift,
because no emission lines are detected.
Also in the red wing of the Ly$\alpha$ profile, a broad shoulder is
observed that may be due to multiple H\,{\sc i} absorption systems or may be caused by 
intrinsic velocity structure of the Ly$\alpha$ emitting gas.

The steepness of the absorption trough next to the Ly$\alpha$
peak requires the absorber to have 
an H\,{\sc i} column density of $\sim10^{18.3}$ cm$^{-2}$. But the 
broadness of the absorption, extinguishing nearly all emission in the blue 
wing of Ly$\alpha$, requires the Doppler parameter for a single absorber to 
be $\sim162$ km s$^{-1}$. This fit is displayed in Fig. 2.
This absorption fit has removed too much of the original Ly$\alpha$
profile, as there is clearly emission observed just blueward of the sharp drop 
to zero at $\sim4340$ \AA. Thus, the absorption may be 
due to the combination of several absorption systems at slightly different 
velocities with respect to the Ly$\alpha$ peak.
The few small emission peaks that are left of the
Ly$\alpha$ blue wing, are significant and can be well modelled by assuming
three distinct absorbers (see Fig. 4).
The main absorber has a column density of $\sim10^{18.1}$ cm$^{-2}$
and the other two are $\sim10^{16.3}$ cm$^{-2}$ with more reasonable 
Doppler parameters of 16 to 80 km s$^{-1}$. Although the absorption may
be even more complex, this model is the simplest one that gives a satisfactory 
fit so we adopted these values. The broad shoulder in the red wing 
is also modelled by HI
absorption with a large gas velocity dispersion, but we cannot exclude that 
it is caused by true velocity structure in the Ly$\alpha$ emitting gas.

\noindent {\bf 1357+007}: This galaxy which is associated 
with a small ($3''$) radio source
has a deep trough in the Ly$\alpha$ profile.
The Ly$\alpha$ profile has a relatively low S/N.
The one-dimensional spectrum of this object 
is a good example for demonstrating the difference 
between modelling the troughs as being due to H\,{\sc i} absorption and as being due 
to true velocity structure.
The limitations of the models are illustrated by comparing the best fits
obtained by a two velocity component emission model (Fig. 3)
with those from an absorption model (Fig. 4).
The fit of the Ly$\alpha$ emission profile by a combination of two
Gaussian emission profiles at different velocities
is less satisfactory than the H\,{\sc i} absorption model. This is
because the trough between the two peaks on the profile is steeper than the
wings on the outside of the emission peaks. Although we cannot exclude the 
possibility of a non-Gaussian but symmetric double peaked emission profile,
this difference in steepness of
the trough and the outer wings of the emission profile is accounted for by a 
Voigt absorption profile which is also the best fit for the troughs in the 
profiles with the best S/N.
The actual situation may be more complicated than implied by the idealized 
assumptions of the models, but it seems that also in 1357+007
H\,{\sc i} absorption is the more plausible
interpretation in spite of the lower S/N ratio than in our best
spectra.

\noindent {\bf 1436+157}: This is a quasar--galaxy pair oriented along the 
radio axis (like the galaxy-galaxy pair 0828+193). 
There is no direct evidence that the galaxy is at the same 
redshift as the quasar; only
the quasar shows strong Ly$\alpha$ emission. 
Apart from very broad Ly$\alpha$
emission and strong continuum, as is common for quasars, the Ly$\alpha$ has
a spatially extended narrow component. This is why we include it in our sample.
A strong H\,{\sc i} absorption feature is present in the narrow line component 
over the entire Ly$\alpha$ extent. 
The extended Ly$\alpha$ emission is
larger than the radio source. Part of the emission
may be due to the companion galaxy or a tidal interaction if the companion 
galaxy is indeed associated with the quasar. 
 
The strong H\,{\sc i} absorption feature in the narrow line Ly$\alpha$
component requires a very large Doppler parameter of $\sim$200 km s$^{-1}$
when modelled by one single absorber. Although we cannot exclude the 
possibility of an absorber with such a large
intrinsic velocity dispersion, a better fit to the data is obtained 
by a model of two or more absorbers. 
In Fig. 3 the fit is shown that 
we obtain for two adjacent absorbers, 
each with an H\,{\sc i} column density of about 10$^{19.3}$ cm$^{-2}$ and 
Doppler parameters of 12--75 km s$^{-1}$.
We have also extracted a spectrum of the off-nuclear Ly$\alpha$ emission
by only summing the spectra beyond a distance of $2''$ from the peak of the 
continuum (shown in Fig. 4).
Although there is still some
contamination from the quasar continuum in this spectrum, the contribution
of broad Ly$\alpha$ emission is negligible.
To fit the off-nuclear profile, either a single absorber with large Doppler 
parameter ($\sim$250 km s$^{-2}$) is needed or more than one 
absorber. As a most plausible fit we have
adopted a two absorbers model, whose parameters are in Table 3.

\subsection{Correlations between the physical parameters}
The data presented here for the first time allow the properties of the
Ly$\alpha$ to be compared with other properties of high redshift 
radio galaxies to search for statistical correlations.
In Table 5 we list
some of the main radio properties of the objects
from R\"ottgering {et al.} (1994)\nocite{rot94}, Carilli {et
al.} (1996)\nocite{car95} and Chambers {et al.} (1990)\nocite{cha90}.

The radio galaxies for which high resolution radio maps are available
\cite{car95} show large variation in the complexity of the radio morphologies,
from a simple straight double structure with or without a nucleus to bent 
morphologies with double hot spots.
To indicate the amount of distortion of the radio morphology we define
a ``distortion index'', $\xi_{radio}$, as the number of 
lines with significantly different position 
angles that connect the different components in a radio source. This index not
only indicates the presence of 
bending on the scale of the total extent of the radio source
but also takes into account the presence ``wiggles'' caused by 
multiple hot spots. Although 
$\xi_{radio}$ is somewhat subjective, dependent on the size of a radio source
and the resolution on the radio maps, 
it does indicate the relative differences in 
distortion between the radio sources.
The columns of Table 5 contain the following: \newline
(1) the source name \newline
(2) largest projected angular size of the radio emission from 1.5 GHz VLA maps,
 which are available for all sources. The sizes measured at 8 GHz (for the 
sources that were also observed at this frequency) do not differ
significantly from those at 1.5 GHz. \newline
(3) transverse radio size; projected size perpendicular to the main radio axis (from 2),
 measured from 1.5 or 8 GHz maps \newline
(4) the radio ``distortion index'', $\xi_{radio}$ \newline
(5) radio spectral index between 178 or 365 or 408 MHz and 2.7 or 5 GHz 
(see R\"ottgering {et al.} 1994\nocite{rot94}) \newline
(6) radio core fraction (ratio of core flux to total radio flux) at a 
rest frequency of 20 GHz \newline
(7) ratio $Q$ of the distances between the radio core at 4.7 GHz, or optical 
    identification (superposed on the 4.7 GHz maps or 1.5 GHz) and the brightest
    feature in each radio lobe \newline
(8) ratio of the integrated flux densities of the radio lobes (maximum/minimum)
    \newline
(9) logarithm of the monochromatic radio power at a rest frequency 1.5 GHz,
    estimated from the flux at 1.5 GHz and the spectral index.

For more details on the definitions of the radio properties see Carilli 
{et al.} 1996; R\"ottgering {et al.} 1994 and McCarthy 
{et al.} 1991\nocite{car95,rot94,mcc91a}.

\subsubsection{Ly$\alpha$ size and velocity dispersion}
We have investigated if there are correlations between the various parameters
as listed in Tables 4 and 5.
A strong correlation were found between the radio size
and Ly$\alpha$ size, D$_{Ly\alpha}^{20\%}$, and a strong anti-correlation
between the radio size and the
velocity width (FWHM) of the Ly$\alpha$ emission.
These two correlations
are shown in Fig. 7.
We chose to show only the
correlation with D$_{Ly\alpha}^{20\%}$  as this is the most objective measure 
of the scale of
Ly$\alpha$ extent of the sources, although the total observed Ly$\alpha$ size
(D$_{Ly\alpha}^{tot}$) also correlates strongly.
A Spearman rank correlation analysis shows that the anti-correlation of radio
size with the Ly$\alpha$ velocity width has significance level of 99.8\% and 
the correlation between the Ly$\alpha$ size and the radio size has a 
significance higher than 99.9\%.

There is also a somewhat weaker anti-correlation between the Ly$\alpha$ size 
and the Ly$\alpha$ velocity width (see Fig. 8), 
that has a significance of 99.6\% in a Spearman rank 
correlation analysis. A partial correlation analysis of the Ly$\alpha$ size
and the Ly$\alpha$ velocity width, controlling for the radio size, gives a 
low significance of only 71.4\% for this anti-correlation, 
an order of magnitude worse than for the two 
strong correlations when controlling for the third property.
This suggests that the apparent anti-correlation between the Ly$\alpha$ size
and the Ly$\alpha$ velocity width just reflects the first 
two correlations.
Thus, we conclude that the size and velocity dispersion of the Ly$\alpha$
emitting gas in our sample of
high redshift radio galaxies has a strong
dependence on the size of the associated radio source.
The smaller radio sources have smaller Ly$\alpha$ emission regions but
with a higher Ly$\alpha$ velocity dispersion.

We further note that in 11 of the 18 of the objects the total measured Ly$\alpha$
extent (D$_{Ly\alpha}^{tot}$) is larger than the radio source.

\subsubsection{HI absorption}
Strong H\,{\sc i} absorption (N(H\,{\sc i})$>10^{18}$ cm$^{-2}$) mainly occurs
in the objects with the smaller associated radio sources and Ly$\alpha$ 
emission region (see Fig. 7). 
In 9 out of the 10 radio galaxies smaller than 50 kpc
strong H\,{\sc i} absorption is found, while only in 2 out of 8 with radio sources 
larger than 50 kpc. Note that the latter include the peculiar galaxy 
0211$-$122 where the absorption of Ly$\alpha$ may not be entirely due to 
H\,{\sc i}.

The Ly$\alpha$
luminosity of the objects and the Ly$\alpha$ velocity width show no significant
correlation (Fig. 8). 
A notable feature from this plot is that 
the average Ly$\alpha$ luminosity of the radio galaxies with strong 
associated H\,{\sc i} absorption
is about a factor $2$ lower than those without a strong associated 
H\,{\sc i} absorber. 
Although this suggests that the H\,{\sc i} systems might be responsible for absorbing 
$\sim50$\% of the Ly$\alpha$ flux,
the large dispersion in the luminosities (the errors on
the mean luminosities of both samples are more than a factor of 2), this 
difference is not statistically significant.

In Fig. 9 we show a histogram of the relative velocity of the 
strong absorption systems with respect to the peak of 
the original (Gaussian) Ly$\alpha$ emission profile as obtained from the
modelling. The median velocity is at a blueshift of $\sim 100$ km s$^{-1}$.
It is noteworthy that in almost all  cases the strong H\,{\sc i} absorption
is within $\sim250$ km s$^{-1}$ of the Ly$\alpha$ peak. Only in
one case (0828+193) the absorption is at a substantially larger 
velocity of $\sim 540$ km s$^{-1}$.
We have tested whether this might be a selection effect, in that we would
not be able to distinguish the absorption systems if they had larger 
relative velocities. After restoring the original Ly$\alpha$ emission spectra
for those objects with strong H\,{\sc i} absorption,
by filling in the absorption troughs using the absorption line profiles
obtained from the Ly$\alpha$ profile fitting, and simulated the similar
absorptions
at larger relative velocities. We find that we would have clearly distinguished
the H\,{\sc i} absorption systems in the Ly$\alpha$ profiles out to velocities of
almost half the FWHM of the Ly$\alpha$ emission,
i.e. 450--700 km s$^{-1}$. In the spectra with the highest
S/N, absorption systems would have been distinguished out to
velocities even further than half the FWHM.
Thus we conclude that the velocity distribution with 90\% of the absorption
systems within $\sim250$ km s$^{-1}$, is real.

\subsubsection{Ly$\alpha$ position shifts}
We have found that the spatial position of the 
maximum intensity of the Ly$\alpha$ emission changes as a function of
wavelength and sometimes shows several ``wiggles'' over the wavelength
range of the emission line (Fig. 6).
Neither the maximum amplitude of the position shifts, 
$\Delta$S, nor the ``wiggling index'', $w_S$,
correlate with any of the other Ly$\alpha$ parameters, such as size and 
velocity dispersion.
But, although $\Delta$S  does not show an apparent correlation with the largest
linear size (i.e. measured along the radio axis) of the radio source, 
there is a (weak) correlation (96.5\% from a Spearman rank analysis) with
the transverse size (projected size perpendicular to the radio axis) 
of the radio source (Fig. 10).
It appears that the radio sources with larger transverse sizes tend to have 
larger Ly$\alpha$ shifts.
Note that sources for which there is only an upper limit of the 
transverse radio size have relatively small Ly$\alpha$ position shifts.
This implies that the actual relation would probably be stronger 
if the true transverse sizes of these sources were known.

A strong correlation ($>99.9$\%) exists  between 
the distortion index of the radio ($\xi_{radio}$),
and the Ly$\alpha$ ``wiggling index'' ($w_S$) 
(see Fig. 10). Thus, the
sources with the most distorted radio morphologies also show the most
``wiggles'' in the Ly$\alpha$ peak position.

Distortion of radio sources, i.e. bending and multiple hot spots, increases
the transverse size of a radio source, producing
the correlation between $\Delta$S and the transverse radio size.

\section{Discussion}

\subsection{The Ly$\alpha$ emission}

\subsubsection{Physical properties}
From previous low resolution studies it is known that HZRGs usually
have extended (up to 100 kpc), luminous ($\sim 10^{44}$ erg s$^{-1}$)
Ly$\alpha$ halos with velocity dispersions of $>1000$ km s$^{-1}$ FWHM
(see McCarthy 1993 and references therein)\nocite{mcc93a}.
Our high spectral resolution study confirms these general properties.
The total detected spatial extent of the Ly$\alpha$ halos ranges from $\sim15$ 
kpc to 135 kpc, while the velocity
dispersions range from 700 km s$^{-1}$ to 1600 km s$^{-1}$ (FWHM).

The detected Ly$\alpha$ emission often extends further than the radio lobes.
In the cases where this very extended emission is detected with good S/N,
it appears to have a lower velocity dispersion than the inner 
Ly$\alpha$ halo. This is most clearly seen in the objects 1243+036 
\cite{oji96}, 0200+015, 0828+193 and 4C41.17 \cite{cha90}.
In these cases the Ly$\alpha$ emission can be divided into two components:
an ``inner halo'' which has dimensions smaller than the radio source and
which has a large velocity dispersion (700--1600 km s$^{-1}$ FWHM), and
an ``outer halo'', outside the radio structure, which is relatively 
quiescent ($\sim300$ km s$^{-1}$ FWHM).
This relatively quiescent outer halo combined with more violent
kinematics inside the radio structure is most clearly apparent in the 
spectacular radio galaxy 1243+036 at $z=3.6$ (van Ojik et al. 1996).
\nocite{oji96}

McCarthy {et al.} (1991)\nocite{mcc91a} found that the extended emission
line gas of 3C radio galaxies (median redshift 0.4) is generally asymmetric and
brightest on the side of the radio lobe closest to galaxy nucleus, which also
tends to be the brightest radio lobe (46 out of 70). In our sample of high
redshift ($z>2$) radio galaxies, 7 of the 17 galaxies have clearly asymmetric
Ly$\alpha$ emission regions, i.e. the surface brightness of the extended
Ly$\alpha$ emission is larger on one side of the galaxy than on the other.
Only 3 of these 7 have the brightest extended Ly$\alpha$ emission on the side
of the closest radio lobe and thus do not show the correlation with 
lobe arm length asymmetry that McCarthy {et al.}
(1991)\nocite{mcc91a} found for the 3C galaxies (37 out of 39).
However, 6 of these 7 have the brightest extended Ly$\alpha$ emission on the
side of the brightest radio lobe, while McCarthy {et al.} found
this correlation only weakly present in his 3C sample. 
We note that the statistics of our sample
are poor with only 7 measured asymmetric emission line regions against 39
galaxies of McCarthy {et al.} (1991). One of the reasons why we do not
find the same asymmetry correlation may be because the extended emission line
regions of HZRGs are usually much larger and smoother, sometimes larger than
the radio sources, than those of the intermediate redshift 3C sample of
McCarthy {et al.} (1991), who also noted this difficulty in measuring the
asymmetry of Ly$\alpha$ in HZRGs.
Another possibility is that the inconsistency is due to the fact that we measure
the Ly$\alpha$ emission, while McCarthy {et al.} (1991) measured the extent of
[OII]$\lambda$3727. Different mechanisms may determine the spatial distribution
of Ly$\alpha$ and [OII] emitting gas.

It has been suggested that the emission line gas in radio galaxies may be 
responsible for the depolarization of the radio emission (e.g. van Breugel
{et al.} 1985a and 1985b\nocite{bre85d,bre85c} 
and references therein). The depolarization of the radio emission 
is also often observed to be asymmetric \cite{car95}.
A strong anti-correlation between asymmetric depolarization of the 
radio emission from quasars with the side of the quasar on which 
the radio jet is observed \cite{gar88,lai88} and
large rotation measures of the polarized radio emission in HZRGs 
\cite{car94,car95} have been interpreted as evidence that a hot ($10^7$ K)
magnetoionic halo surrounding the objects causes the depolarization and that 
the asymmetry is caused by the orientation of the radio source in the hot halo.
If the depolarization of the radio emission is caused by the emission line gas
or an associated hotter gas component,
one would expect that asymmetry of the extended emission line regions would 
correspond to asymmetric depolarization of the radio emission. 
For most of the galaxies in our sample
high resolution radio polarization measurements have been carried out by
Carilli {et al.} (1996)\nocite{car95}. Most of the radio galaxies have 
asymmetric polarization, where one radio lobe is much more depolarized than 
the other. Polarization information is available for only
5 of the 7 sources with asymmetric Ly$\alpha$ emission
in our sample. In only 2 of these 5,
the fractional polarization is lowest on the side of the most extended
Ly$\alpha$ emission, indicating that line asymmetry -- polarization asymmetry
at least does not have a strong correlation.
However, this result is statistically too poor to indicate whether
the Ly$\alpha$ emitting gas plays an important role in the depolarization of 
radio emission in our sample of HZRGs or not.

\subsubsection{Ionization of the Ly$\alpha$ emission line region}

Several authors have argued that the large equivalent width
of the Ly$\alpha$ emission as well as the line ratios of the emission
lines associated with radio galaxies in general are not well
reproduced by shock ionization nor by hot stars and that anisotropic
photoionization by an obscured nucleus appears to be the dominant
mechanism of ionization
\cite{bau89b,bau89c,bau92,mcc93a,fer86,fer87,mcc93a,chr93,ant85,ant93,mcc90a,cha90,mcc93a,oji96}.
In our sample of HZRGs we have found a strong relation between the
radio size and Ly$\alpha$ size, which suggests that the radio plasma
interacts with the ambient gas. Shock ionization associated with the
impact of the radio jet and with the expansion of the radio lobes
might explain this direct size relation.  However, the spatial
emission profiles show that Ly$\alpha$ emission is strongest near the
center of the galaxies and decreases rapidly with distance in the
halo, instead of being brightest at the extremities where the radio
lobe has just passed.  Furthermore, the presence of Ly$\alpha$
emission from the region outside the radio structure in more than half
our objects cannot be produced by shocks associated with the radio
sources, but argues strongly for photoionization of the emission line
gas by a central continuum source.  The relation between the sizes of
Ly$\alpha$ and the radio source must be produced by a different
mechanism (see Section 4.3). 
We conclude that the Ly$\alpha$ emission line region is most likely 
produced by (anisotropic) photoionization by a central continuum source.
   
\subsubsection{Kinematics and dynamics}
The velocity dispersion of the Ly$\alpha$ emitting gas in our
sample of HZRGs varies from $\sim$700--1600 km s$^{-1}$. Such large
velocity dispersions were already known from low resolution studies
(see McCarthy 1993\nocite{mcc93a} and references therein) and are similar to 
the velocity
dispersion of extended Ly$\alpha$ emission seen in some radio-loud quasars
\cite{hec91b}. Interaction between the radio source with the emission line 
gas has been suggested as a cause for high velocity dispersions in radio 
galaxies the \cite{you81,bre85d,you86,mcc90a,cha90}.
In the high redshift quasars, Heckman {et al.} (1991)\nocite{hec91b}
were unable to establish a definite connection between
the velocities of the extended emission 
line gas in quasars and the radio source.
They note that such large velocity dispersions may also have a gravitational 
origin if gas is falling in from large radii in the process of galaxy 
formation (see also Fall \& Rees 1985\nocite{fal85}).

Apart from the velocity dispersions we have observed detailed velocity 
structure in the Ly$\alpha$ emission line region.
We have found that in several of the individual objects the velocity profile
shows small deviations from a Gaussian profile (``shoulders'' on
the wings of the profiles) and occasionally there are signs of
rotation. Furthermore, the spatial position of the maximum intensity of 
Ly$\alpha$ often varies with wavelength (``wiggles''). 
These features in the velocity structure indicate that the Ly$\alpha$ emission 
line region is not one system with purely chaotic gas motions,
but also contains some large scale motions.

We have found several strong correlations between 
the velocity structure of the Ly$\alpha$ emitting gas and the properties 
of the radio source: 
\newline
(i) an anti-correlation between the size of the radio source and the
velocity dispersion of the Ly$\alpha$ emitting gas. The smaller radio
sources have larger Ly$\alpha$ velocity dispersion.
\newline
(iii) The number of ``wiggles'' in the Ly$\alpha$ position with wavelength
correlates with the amount of distortion of the radio structure.
\newline
(ii) the component of the Ly$\alpha$ emitting gas that is extended beyond 
the radio source has a lower velocity dispersion than the gas inside the
radio structure. 

As noted by  Heckman {et al.} (1991)\nocite{hec91b} and 
Fall \& Rees (1985)\nocite{fal85}, the high velocity dispersions typically 
found in HZRGs can be gravitational in origin. If the radio galaxies are at the
bottom of a deep potential well, i.e. a dense cluster, gas falling in from
large radii can reach velocities of more than 1000 km s$^{-1}$.
Individual infalling gas regions, ionized by the active nucleus as they
come near the center of the radio galaxy, could also produce spatial
``wiggles'' in the Ly$\alpha$ velocity structure.
However, such a gravitational origin of the Ly$\alpha$ velocity structure
does not produce any of the observed correlations with the radio source
properties. Furthermore, the relatively constant velocity dispersion inside 
the radio structure and the sudden decrease outside the radio source,
is inconsistent with a gravitational infall scenario.

The three correlations indicate that the kinematics of the Ly$\alpha$ are
connected with the radio source.
We have earlier presented convincing evidence that the radio
source accelerates and disturbs the emission line gas in the radio galaxy
1243+036 (van Ojik et al. 1996)\nocite{oji96}. 
In this galaxy the radio jet is bent at the position of
an ``accelerated'' region of emission line gas. Further, 
1243+036 has a quiescent (250 km s$^{-1}$ FWHM) outer Ly$\alpha$ halo, 
while inside the radio source the Ly$\alpha$ emitting 
gas has a large velocity dispersion (1575 km s$^{-1}$ FWHM). Our current 
observations have provided two more objects (0200+015, 0828+193) where the 
velocity dispersion of the Ly$\alpha$ emitting gas outside is lower 
than inside the radio structure. Similar quiescent Ly$\alpha$ gas is also
present outside the radio structure in 4C41.17 \cite{cha90}.
We conclude that the interaction of the radio 
plasma with the gas in our HZRGs is responsible for the large velocity 
dispersions inside the radio structure and the spatial ``wiggles'' in the
Ly$\alpha$ velocity structure.

On the basis of spectroscopy of low redshift radio galaxies, 
Baum {et al.} (1992)\nocite{bau92} classified these radio galaxies 
in three kinematical classes: (i) rotators, associated with recent 
mergers, (ii) calm non-rotators that are identified with cooling flows and 
(iii) violent
non-rotators, that have optical and radio morphologies similar to the 
rotators, but whose kinematics are 
dominated by large chaotic gas motions overshadowing any signs of rotation.
Because of the  similarities in their optical and radio properties,
Baum {et al.} (1992)\nocite{bau92} suggested that the rotators and
violent non-rotators may be kinematically linked, but that original rotation
in the violent non-rotators
has been ``washed out'' by the interaction with the powerful radio source.
The HZRGs that we have investigated here show relatively large velocity 
dispersions ($>1000$ km s$^{-1}$ FWHM) of the
ionized gas when compared to the low redshift radio galaxies that have 
velocity dispersions a few hundred km s$^{-1}$ (FWHM) of Baum 
{et al.} (1992)\nocite{bau92}. Only in a few cases 
(0748+134, 1410$-$001 and 1558$-$003) there is a hint of a velocity
shear over the extent of the Ly$\alpha$ emission region, that could be 
attributed to a relic of rotation beneath the large turbulent velocities 
of the ionized gas.
In 1243+036 (van Ojik et al. 1996)\nocite{oji96} 
we found ``calm'' rotational kinematics of the 
Ly$\alpha$ gas outside the radio structure, while inside the radio plasma the
ionized gas has a large (1500 km s$^{-1}$ FWHM) velocity dispersion and the
kinematics are dominated by chaotic motions.
In the rest of our sample of HZRGs there is often Ly$\alpha$ emission extended
beyond the radio structure with apparently a lower velocity dispersion, but
these outer halos do not show signs of rotation, except in 0200+015. 
In this galaxy the Ly$\alpha$ emitting gas extending beyond the radio source 
appears to have a velocity shear at least on one side of the galaxy where the 
H\,{\sc i} absorption also remains present over the same extent.
Using the criteria of Baum {et al.} (1992)\nocite{bau92} none of these
galaxies would classify as rotators, because the kinematics are dominated by
large turbulent motions and evidence for rotation
is only weakly present in a few cases.
The interpretation of 1243+036 (van Ojik et al. 1996)\nocite{oji96} 
that turbulence induced by the
propagating radio source ``washes out'' the signs of rotation, and the few
hints of rotation in several galaxies with lower than average gas velocity
dispersions, supports the idea of Baum {et al.} (1992)\nocite{bau92} for
the low redshift radio galaxies, that the rotators and violent non-rotators
are linked. Hence, all our powerful high redshift radio galaxies may well have
originally had rotating gas halos, due to a galaxy interaction or from the
accretion of gas from the primeval environment. However, any clear rotation
signature has been washed out by large turbulent motions due to interaction
with the propagating radio plasma.


\subsection{HI absorption systems}

\subsubsection{Physical properties and kinematics}
Our observations show that high redshift radio galaxies often have
very strong and extended H\,{\sc i} absorption systems seen against their Ly$\alpha$
emission. Evidence of such absorption was given by the observations of
Hippelein \& Meisenheimer (1993)\nocite{hip93}, who imaged the Ly$\alpha$
emission of 4C41.17 using a Fabry--P\'erot system.
They found 
a relatively weak H\,{\sc i} absorption, of $\sim 10^{15}$
cm$^{-2}$ column density, with a projected size of $\sim 10 \times 10$ kpc$^2$,
covering a part of the Ly$\alpha$ emission region of
4C41.17 at $z=3.8$. They interpret this absorption 
as due to a fortuitously intervening Ly$\alpha$ forest absorption. 

A small hint of this absorption is visible in the
high resolution spectrum of Chambers {et al.} (1990)\nocite{cha90}.
The strong absorptions that we find in our sample of radio galaxies
appear to be different from the weak absorption in 4C41.17 in that they
are more than 3 orders of magnitude stronger and up to 5 times
more spatially extended.
Our observations show that 
in the cases where strong absorption (N(H\,{\sc i}) $> 10^{18}$ cm$^{-2}$)
is seen against the Ly$\alpha$ emission of the radio galaxy, the absorber 
extends over the entire size of the emission line region with observed 
projected sizes up to $\sim 50$ kpc.
However, the strength of the absorption
is not always the same over all spatial scales. The largest Ly$\alpha$ emitting 
regions with H\,{\sc i} absorption show more Ly$\alpha$ emission from the bottom of 
the absorption trough (e.g. 0200+015)
than the objects with small Ly$\alpha$ emission regions (e.g. 2202+128).
The depth of the absorptions over such different scales
indicates that the covering factor of absorbing material must be close
to unity over scales of $\sim$20 kpc, and still of the order of 0.9 on 
$\sim$50 kpc scales.

From the sizes and derived column densities of the absorbers and assuming 
that the
size of the absorbers along the line of sight is the same as the observed
transverse size, we find that the mass in neutral hydrogen of an absorption 
system is M(H\,{\sc i})=10$^8$ $R_{35}^{2} N_{19}$ M$_{\odot}$, where
$R_{35}=35$ kpc is the average size the absorption systems in our sample
and $N_{19}=10^{19}$ cm$^{-2}$ the characteristic H\,{\sc i} column density.
The H\,{\sc i} masses deduced for the individual objects is listed in Table 4 
and
range from $\sim10^7$--$10^8$ M$_{\odot}$.  
The total hydrogen mass of the absorption systems depends on the ionization
state of the gas. Photoionization models of absorption systems with column 
densities in the range of those observed in our radio galaxies 
(N(H\,{\sc i})$=10^{18}$--$10^{20}$ cm$^{-2}$) by 
Bergeron (1988) and Bergeron \& Stasi\'nska (1986)\nocite{ber88,ber86},
indicate that the
ratio of H\,{\sc II}/H\,{\sc i} averaged over the whole absorption cloud is in the range 4--250.
Although the conditions in the extended absorption systems in our HZRGs may be
different from those in the line of sight to quasars, they are shielded from
the direct ionizing radiation from the radio galaxy. The range of ionization 
parameters found for the gas in H\,{\sc i} absorption systems by Bergeron \& 
Stasi\'nska (1986)\nocite{ber86} that are assumed to be ionized by the 
intergalactic UV radiation field,
is the same as that estimated in the narrow emission line 
regions of quasars and radio galaxies (see McCarthy 1993)\nocite{mcc93a}, 
although the ionization source is different.
Therefore, the H\,{\sc iI}/H\,{\sc i} ratio in the absorption systems in our HZRGs
may not be very different from the absorption systems
modelled by Bergeron (1988) and Bergeron \& Stasi\'nska 
(1986)\nocite{ber88,ber86}. Thus, the total mass in hydrogen in 
the observed absorption systems is probably in the range of 0.4--250 x $10^8$
M$_{\odot}$. 
This total mass in the absorption systems covering the extent
of the Ly$\alpha$ emission is similar 
to the masses we estimated for the Ly$\alpha$ emitting gas itself in 
these HZRGs.

Assuming that the clouds of the absorption systems are
confined by external pressure, we can try to derive some properties of these
clouds (see also R\"ottgering {et al.} 1995a\nocite{rot95a}). 
There is strong evidence from X-ray observations and depolarization 
measurements that radio galaxies at low and high redshifts
are surrounded by hot ($\sim10^7$ K) gas (e.g. Arnaud {et al.} 
1984\nocite{arn84}; Henry and Henriksen 1986\nocite{hen86}; 
Crawford and Fabian 1993\nocite{cra93}; Carilli {et al.}
1994\nocite{car94}).
From the balance of the minimum
pressures of the radio plasma of HZRGs and the external hot gas, we find 
a typical external pressure $nT\sim 10^6$ cm$^{-3}$ K (e.g. Chambers 
{et al.} 1990; McCarthy 1993; Carilli {et al.} 1994; van Ojik 
{et al.} 1996).\nocite{cha90,mcc93a,car94,oji96}
As mentioned above, 
absorption systems of H\,{\sc i} column densities $10^{18}$--$10^{20}$ cm$^{-2}$ 
are predominantly ionized, and their
temperatures are maximally a few times 10$^4$ K \cite{ber88,ber86}. 
Observations of  quasar 
absorption systems with low column densities (10$^{12}$--10$^{14}$ cm$^{-2}$)
indicate that temperatures in those absorption systems
may be lower than $\sim$5000--10\,000 K
\cite{pet90a}.

Because the various observations appear to indicate temperatures around
10$^4$ K, we shall assume a temperature of 10$^4$ K for the absorption
systems in our HZRGs as a working assumption in the following. 
For pressure balance between the absorbing gas and the external medium 
($nT \sim 10^6$ cm$^{-3}$ K),
the gas density would then be $\sim100$ cm$^{-3}$.
The mass, density and size of the absorption systems indicate that the
gas only occupies a small fraction of the volume of the absorption system.
The volume filling factor is given  by $f_{v} = N / Rn$, where $N$ is the HI
column density, $R$ is the size of the absorption system (assuming the depth
of the system is the same as the observed size along the slit) 
and $n$ is the density of the gas.
The volume filling factor for the extended absorption systems in our sample 
of radio galaxies then range from a few times 10$^{-7}$ to a few time 10$^{-6}$.
With an average size of an absorption system of $R_{35}=35$ kpc,
gas density $n_{100}=100$ cm$^{-3}$ and H\,{\sc i} column density $N_{19}=10^{19}$ 
cm$^{-2}$, the volume filling factor of the absorption systems is:

\[ f_{v}\sim 10^{-6} N_{19} R_{35}^{-1} n_{100}^{-1} \]
 
However,
the absorption systems may contain more ionized than neutral 
hydrogen, which would imply a higher volume filling factor.
From the typical mass in H\,{\sc i} of the absorption systems, $M_8 = 10^8$ M$_{\odot}$,
and the ratio of total hydrogen mass to H\,{\sc i} mass defined as $r=$ 
M(H)/M(H\,{\sc i})
the volume filling factor is:

\[ f_v\sim10^{-6} r M_8 R_{35}^{-3} n_{100}^{-1} \]
 
Although this is only a crude estimate of the filling factor, depending very
much on the external pressure, the temperature of the absorbing gas and
on the assumed geometry of the absorption systems, 
we adopt
the value of $f_v=10^{-6}$ to estimate the sizes of the individual clouds.
The size of an individual cloud is estimated by $d=Rf_v/f_c$. 
This size can be derived since we know the covering
fraction of the clouds in the absorption system $f_c$ to be close to unity.
For a typical
absorption system size of 35 kpc, the individual clouds have sizes of
$\sim0.035$ pc. To cover the projected surface of the absorption system, it
must contain $\sim10^{12}$ such clouds (corresponding to $\sim$4x10$^8$ 
M$_{\odot}$, consistent with an H\,{\sc iI}/H\,{\sc i} ratio of a few).
If the individual clouds are not spherical but filamentary, as may be 
likely, the number of such clouds would change, but the derived size would be
the average size of individual clouds along the line of sight.

The large Doppler parameters that are often required for an optimum fit to the
absorption profiles are too large to be only due to thermal motions of the gas,
because they would imply gas temperatures of $\sim 10^6$ K.
The width of the absorption lines may be due to a superposition of a large
number of absorbing clouds covering the velocity range or be due to macroscopic
gas motions in the absorbing clouds. We can estimate the
velocity dispersion of the gas from the Doppler parameter and the temperature of
the gas, where $b=(2kT/m + 2\sigma^2)^{1/2}$ where $m$ is the mass of hydrogen
and $\sigma$ is the gas velocity dispersion (see Cowie and Songaila 
1986\nocite{cow86}). Assuming a maximum temperature of the absorbing gas of
10$^4$ K, for Doppler parameters 50--200 km s$^{-1}$ the gas velocity dispersion
$\sigma$ is 35--140 km s$^{-1}$.

\subsubsection{Relation to the radio galaxy and Ly$\alpha$ emission}
From the statistics of quasar absorption lines, the chance of a random
intervening absorber with column density larger than 10$^{18}$ cm$^{-2}$ in
the small redshift interval of the Ly$\alpha$ emission line ($\Delta z \sim 
0.03$) is only $\sim$2--3\%
(e.g. Petitjean {et al.} 1993\nocite{pet93}). Our sample of 
high redshift radio galaxies has a 
high incidence of $\sim$60\% of such strong associated absorption, and 
even $\sim$90\% of the radio sources with radio size smaller than 50 kpc.
Thus these absorption systems are much more common than in quasars and cannot 
be random intervening absorbers, but must
be intrinsic to the radio galaxies or their direct environment.

One possibility might be that a strong absorption is caused by the gaseous
halo of a neighbouring galaxy if the radio sources reside in clusters.
There is evidence from some damped Ly$\alpha$ absorption systems (Briggs
{et al.} 1989\nocite{bri89b})
and from the statististics of Ly$\alpha$ absorption in
the line of sight of gravitationally lensed QSOs that high redshift galaxies
may have halos of more than 10 kiloparsec
(e.g. Smette {et al.} 1992 and references therein).\nocite{sme92}
Assuming that cluster galaxies have a hydrogen halo size 
of 35 kpc (the average size
of the observed extent of the H\,{\sc i} absorbers and consistent with the size 
estimates derived from the statistics of Ly$\alpha$ absorptions in QSOs)
and a typical cluster size of 1 Mpc, this would imply $\sim 200$
such galaxies to cover all lines of sights to the radio galaxy. 
Such a density of galaxies with large hydrogen halos 
in a cluster around the radio source seems unlikely.
Furthermore, the typical velocity dispersion in low and intermediate redshift
clusters is $\sim 1000$ km s$^{-1}$. 
However, the H\,{\sc i} absorption systems 
in our HZRGs are almost always within 250 km s$^{-1}$ of the peak of the 
Ly$\alpha$ emission profile (see Section 3.3, Fig. 9).

Alternatively, the absorption may be caused by a large tidal remnant
of an interaction with a companion galaxy (R\"ottgering et al
1996b). However, 
it is implausible
that, even if all our radio galaxies have had an interaction, such
tidal remnants would in 60--90\% of the cases be exactly located in
our line of sight to the Ly$\alpha$ emission and covering its entire
extent.  Furthermore, if quasars are in similar cluster
environments as radio galaxies, one would expect a similar high
incidence of associated absorptions in quasars if they are caused by
cluster galaxies or tidal remnants, which is not observed.

In orientation-based  
unification models radio galaxies and radio loud quasars
are regarded as intrinsically similar objects, viewed from different angles
(see Antonucci 1993; Barthel 1989\nocite{ant93,bar89} and references therein).
Due to an obscuring torus of dust and gas surrounding the active nucleus, the
radiation from the nucleus can only be observed within a cone which is 
oriented along the radio axis. This anisotropic radiation 
ionizes the gas clouds in its path resulting in an extended emission line
region. In quasars the radio/cone axis is oriented close to the line of sight
and we therefore observe
the bright ionizing continuum and broad-line emission region close to the
nucleus, while in radio galaxies the nucleus is obscured so that we cannot 
observe the ionizing radiation directly.

In this scenario, 
both quasars and radio galaxies must have a large envelope of gas 
surrounding the entire systems. 
In radio galaxies the ionizing cone is close to the plane of the sky and
we see the emission line region from the side. Our line of
sight to the Ly$\alpha$ emission region would pass through the region of 
gas outside the cone of ionizing radiation (see Fig. 11). 
This gas is not affected by the propagating radio plasma, that is presumed
to have 
stirred up the emission line gas, causing high velocity dispersions.
This much more quiescent gas outside the ionizing cone is the most likely
cause of the strong absorption in the Ly$\alpha$ emission line profiles.
If indeed both the Ly$\alpha$ emitting gas and the absorbing gas are components
of the extended gaseous envelope of the galaxy, 
it would also explain
why in all observed cases the absorption systems extend over the entire Ly$\alpha$ 
emission region.
Suggestive evidence for this interpretation is also that the estimated masses
for the amount of emission line gas are roughly similar to the estimated 
total masses (neutral and ionized gas) of the absorption systems (typically a 
few times 10$^8$ M$_{\odot}$). The presence of H\,{\sc i}
outside the ionization cone has been directly observed in the low redshift 
($z=0.022$) Seyfert 2 galaxy NGC\,5252 \cite{alm93}.

We note that although this interpretation 
may work as a general picture for the HZRGs, there are several individual
cases where the situation must be more complicated (e.g. 1243+036, 0828+193,
1436+157).
The presence and strength of H\,{\sc i} absorption in an individual case
may also depend on
e.g. the total amount of gas present in a radio galaxy, the power of the
nuclear ionizing continuum, differences in cone opening angle
and the presence of an extra (isotropic) ionization source (starburst) and
interaction with companion galaxies.

If the gas of the absorption systems is indeed from the same gaseous envelope
as the Ly$\alpha$ emitting gas, the high covering factor (0.9--1) of the
absorbing gas does not imply that the gas located within the ionization cone
has the same high covering factor in clouds that are {\it optically thick} to
the nuclear ionizing continuum radiation. 
If the degree of ionization increases by a factor of
10 inside the ionization cone, the covering fraction of optically thick clouds
would decrease to $\sim0.1$, the value that is estimated for extended
Ly$\alpha$ emission observed in high redshift radio loud quasars \cite{hec91a}.
Note that this same mechanism probably causes the ``inverse'' or ``proximity''
effect in quasars, where there is a strong decrease of the number of 
Ly$\alpha$ absorption
systems close to the redshift of the quasars (e.g. Murdoch {et al.} 1986;
Lu {et al.} 1991\nocite{mur86,lu91}).

It is interesting that, although there does not seem to be a large excess of
associated absorption in QSOs in general, there is a tendency for associated
absorption to occur preferentially in steep
spectrum radio loud quasars \cite{fol88,and87,bar90b,hec91b}.
The incidence rate of associated absorptions in radio loud quasars
is about 16\% within 2000 km s$^{-1}$ of the quasar redshift.
Radio loud quasars
also have more extended line emission than radio quiet QSOs \cite{hec91a}. These
properties suggest that both radio loud galaxies and radio loud quasars 
at high redshifts reside in denser environments than radio quiet and flat
spectrum objects.
Furthermore, 
their larger relative velocities and lower
occurrence rate compared to the
associated absorption in our sample of HZRGs suggests that they may not 
have the same origin 
and are more consistent with being due to
neighbouring cluster galaxies rather than being intrinsic to the quasars 
themselves.

We note that the associated absorption systems in these radio galaxies are 
quite different from the broad (several thousands km s$^{-1}$)
absorption line systems that are observed in
some quasars.

\subsubsection{Implications for the dust content of HZRGs}
A comparison of optical with infrared measurements show that
high redshift radio galaxies 
have a Ly$\alpha$/H$\alpha$ ratio lower than was expected
for case B recombination by a factor of 2--10 \cite{mcc92b,eal93b,mcc93a}.
This has been interpreted as being most likely due to dust
mixed with the emission line gas, selectively destroying the resonant
scattering Ly$\alpha$ photons. Other evidence for the presence of dust comes
from optical polarization measurements, that indicate that much of the extended
continuum light may be scattered continuum radiation from an obscured AGN
\cite{tad92,cim93,cim94,ser94}.

Our observations of the Ly$\alpha$ profiles have shown that neutral hydrogen 
is present in many HZRGs and can absorb $\sim$50\% of
the original Ly$\alpha$ emission. Thus, the low observed Ly$\alpha$/H$\alpha$
ratios are at least in part due to associated H\,{\sc i} absorption systems (see
also R\"ottgering {et al.} 1995a\nocite{rot95a}).

We note that to effectively destroy Ly$\alpha$ emission by dust, the dust
must be distributed homogeneously through the ionized gas. Because of the
low volume filling factor of the emission line gas, a Ly$\alpha$ emission
region with low dust content or non-homogeneous dust distribution will
produce strong Ly$\alpha$ emission from the galaxy.
The radio galaxy 0211$-$122 appears to be an exceptional case where
the dust distribution meets this stringent geometrical requirement well
enough to effectively extinguish most of the Ly$\alpha$ emission (see
van Ojik et al. 1994)\nocite{oji94a}).
While very little dust distributed homogeneously can extinguish all Ly$\alpha$
emission, for different geometries galaxies could contain large amounts of
dust without destroying the Ly$\alpha$ emission. Differences between the
distribution of dust and Ly$\alpha$ emitting gas are not unlikely because dust
is probably produced by star formation in the galaxy itself and transported
outwards into the gaseous halo while the Ly$\alpha$ halo gas may well
originate from accretion from the primeval environment.

The geometrical requirement of H\,{\sc i} absorption of Ly$\alpha$ is much less
stringent than for absorption by dust. The H\,{\sc i} clouds must be non-uniformly
distributed around the emission line gas and located in our line of sight to
reduce the observed Ly$\alpha$ emission flux. Only if neutral gas clouds were
exactly uniformly distributed around the radio galaxies, the observed flux in
our line of sight would not be reduced, in spite of the random scattering of
the Ly$\alpha$ photons in the H\,{\sc i} clouds.

Thus, because of the presence of 
associated H\,{\sc i} absorption in HZRGs, which reduces the observed Ly$\alpha$ flux,
the dust content of HZRGs derived from the H$\alpha$/Ly$\alpha$ ratio may 
have been overestimated.
In the near future, more accurate polarization measurements with 8-metre
class telescopes and continuum measurements with new mm-wavelength 
instrumentation and the ISO satellite along with spectroscopy of more IR 
emission lines, 
will be crucial to determine the true dust content of these primeval galaxies.

\subsection{The Ly$\alpha$ -- radio source connection: scenarios}
Perhaps the most remarkable results from our study relate the Ly$\alpha$ 
emission and absorption to the radio properties of high redshift radio
galaxies. 
Firstly, the small radio sources 
almost always have strong associated H\,{\sc i} absorption. 
The large radio sources show very little associated H\,{\sc i} absorptions
and have narrower Ly$\alpha$ velocity widths.
Secondly, there is a  strong anti-correlation
between the radio size and velocity width of Ly$\alpha$. 
Thirdly, there is a strong correlation between the radio size and the
size of the Ly$\alpha$ emission region. The bigger radio sources have larger 
Ly$\alpha$ emission regions associated with them.
Fourthly, variations in the spatial position of the maximum intensity of
Ly$\alpha$ as a function
of wavelength (``wiggles'') correlate with the amount of distortion of the
radio morphology.

These relations between the Ly$\alpha$ and radio properties are 
strong evidence for a kinematic interaction between the radio jet and the 
ambient gas. 
In the case of 1243+036 there is direct and convincing morphological
evidence for vigorous interaction between the radio source and the Ly$\alpha$
gas (van Ojik et al. 1996).\nocite{oji96}
We note that a similar relation between radio power and 
[OIII] width was noted by Heckman {et al.} (1981)\nocite{hec81} for
low redshift radio sources and Seyfert galaxies 
and was taken as evidence for a radio--gas
interaction.

As we have argued in Section 4.1, shock-ionization associated with the
expansion of the radio lobes is probably not the cause of the Ly$\alpha$ size
-- radio size correlation, because there are several indications that the
Ly$\alpha$ is mainly ionized by a central continuum source. The radio plasma
may have another effect on the ambient gas, namely to increase the amount of
Ly$\alpha$ emission from the region inside the radio lobes by the mechanism 
suggested by Bremer {et al.} (1996)\nocite{brem96}: Because 
the luminosity depends on the
volume filling factor of the gas, interaction of the radio jet and turbulence
in the radio lobes may rip apart neutral cores of emission line clouds
therefore increasing the filling factor, exposing more gas to the ionizing
continuum and increasing the Ly$\alpha$ emission. Outside the radio structure,
gas (with lower filling factor) can also be photoionized but Ly$\alpha$
emission is at a lower flux level. This mechanism has the advantage that it
can explain the radio size -- Ly$\alpha$ size relation, without the 
inconsistencies of a shock-ionization interpretation (see Section 4.1).

The ``wiggles'' in the Ly$\alpha$ position as a function of wavelength 
indicate that
there are large regions of gas having velocities different from the mean
velocity at different positions with respect to the centre of the galaxy.
The ``wiggles'' in the Ly$\alpha$ position usually occur on scales of less than an
arcsecond, while the distortions of the radio structure tend to be
large scale phenomena. In several objects the radio
source has double hot spots at its extremities, at $\sim20$ kpc from the
nucleus while the Ly$\alpha$ ``wiggles'' are within 5 kpc of the centre of the
galaxy. This indicates that, although the Ly$\alpha$ ``wiggles'' correlate
with the degree of distortion of the radio sources, they do not
necessarily coincide spatially. Possibly, interaction of the radio jet with
the ionized gas near the centre of the galaxy at an earlier epoch has resulted
in the current radio distortions on larger scales, while the current
Ly$\alpha$--radio interaction induces a future bent morphology and multiple
hot spots of the radio source.

\subsubsection*{Scenarios}
Apart from the possible nature of the ``wiggles'' in the Ly$\alpha$ position
as a function of wavelength
and the relation between Ly$\alpha$ size and radio size which were discussed
above, we need to explain the  correlations of the radio size 
with the velocity dispersion of the Ly$\alpha$ emitting gas and 
the presence of associated H\,{\sc i} 
absorption.
We propose three different scenarios, one environmental, one
evolutionary and one based on orientation:

{\bf 1.} {\it Environmental scenario}
The environment of a massive galaxy in which a radio source resides, 
will influence
the size evolution and morphology of the radio source and the observed
Ly$\alpha$ properties. In a dense (proto-cluster) environment a radio
source will vigorously interact with the ambient gaseous medium. Kinetic
energy will be transferred from the radio source to the gas, increasing the 
velocity dispersion of the gas and reducing the propagation speed of the
radio source.
Therefore the radio source stays 
relatively small. The Ly$\alpha$ emission from the gas that is ionized by 
photoionization and possibly shocks shows strong H\,{\sc i} absorptions from the dense
gas in the direct environment.
In a low density environment, the propagating radio source will encounter
much less resistance and reach to a large size. There will be relatively little
interaction with the environment, a smaller velocity dispersion
and no clouds large and dense enough to produce
strong H\,{\sc i} absorption of the Ly$\alpha$ emission.

This would be consistent with the work of  
Baum {et al.} (1989)\nocite{bau89c} who at low redshifts
finds that the smaller powerful 3C sources are
generally more depolarized, have more complex emission line morphologies
and more distorted radio structures. Fifty percent of these objects are known to be at
the centers of rich (cooling flow) clusters.
The larger 3C radio galaxies have more normal polarization and less distortion.
These galaxies are isolated or reside in small groups of galaxies, where 
there may be less material to interact with.
Thus the coincidence of larger Ly$\alpha$ velocity dispersion
and H\,{\sc i} absorption with
the smaller and sometimes more distorted radio structures in our $z>2$ sample,
may be evidence that these galaxies are at the centers of dense 
(proto-)clusters.
The larger radio sources have simpler Ly$\alpha$ kinematics 
and no strong absorptions and they are probably more isolated galaxies or
reside in less dense clusters.

In a denser environment one might expect the Ly$\alpha$ luminosity to be higher
because this scales strongly with the density of the emission line gas
(the Ly$\alpha$ luminosity scales as $L \sim n_e^2 f_v V$ erg s$^{-1}$, where
$ n_e$ is the density of the gas, $f_v$ the volume filling factor and $V$ the
volume occupied by the emission line gas, see McCarthy {et al.} 
1990\nocite{mcc90a}). We do 
not find any evidence for such a correlation in our observations. However, 
not only the gas density but also the total volume of an emission line region
determines the Ly$\alpha$ luminosity.
While the environment of the larger radio sources has a lower density, 
the sizes and 
therefore the volume occupied by the Ly$\alpha$ emission line region is much
larger than for the smaller sources.
Therefore the total Ly$\alpha$ luminosity of the larger sources may not be 
very different from the smaller sources.
Furthermore, the presence of strong H\,{\sc i} absorption
also reduces the observed Ly$\alpha$ luminosity of the smaller sources. 
Finally, also the large variation of the
Ly$\alpha$ luminosities (by a factor of 30 between the individual objects) 
makes it difficult to any possible evidence of a correlation between the 
density of the environment and the Ly$\alpha$ luminosity.

{\bf 2.} {\it Evolutionary scenario}
In this scenario the small radio sources are at different
stages in their evolution than the larger radio sources. 
As the radio source propagate outwards it first
encounters a dense gaseous medium whose velocity dispersion is increased via
transfer of kinetic energy from the radio jet and 
we observe a small radio source with Ly$\alpha$ emission of relatively large
velocity width. The dense parts of the gaseous medium that have not (yet) been
affected by the interaction with the radio source would produce strong
absorption in the Ly$\alpha$ emission. As the radio source propagates further
outwards there is less interaction as the ambient density decreases with radius
and the radio jet has cleared a path through the denser inner parts of the
galaxy's halo. The velocity dispersion of the gas then reduces due to
dissipation and less direct interaction with the radio source.

The main problem of this evolutionary scenario is to explain the decrease
of extended HI
absorption clouds as the radio source propagates outwards. In the case of shock
ionization the clouds may become ionized as the boundary of the radio lobe
passes through the neutral gas. However, as described above, shock ionization is
unlikely to be dominant enough to ionize the H\,{\sc i} absorption systems.
Furthermore, the 
possible disruption of the neutral clouds by the passage of the radio lobes
will not prevent them from producing H\,{\sc i} absorption features, because
the gas will still be shielded from the central ionizing continuum.

{\bf 3.} {\it Orientation scenario} A third way of producing the
observed correlation between the projected radio and Ly$\alpha$ size
and the anti-correlation with the velocity dispersion of the gas might
be in the context of the orientation-based unification schemes of
quasars and radio galaxies (as explained earlier).  Close to the
active nucleus the velocity dispersion of the gas is high (many
thousands of kilometres per second), while further out the gas
velocities decrease.  Smaller radio sources have their radio axes
oriented preferentially closer to our line of sight.  In that case we
would also see a smaller emission line region and, as we observe
closer into the nuclear region, we may see more gas with a larger
velocity dispersion.  This scenario however does not explain why smaller radio
sources (oriented towards our direction) are more likely to have
extended H\,{\sc i} absorption than larger ones (those closer to the
plane of the sky). We should then also observe associated H\,{\sc i}
absorption in quasars as frequently as in radio galaxies, which is not
the case.  In fact the opposite would be more likely in an orientation
scenario, if the absorbing gas is outside the cone of ionizing
radiation from the AGN.

West (1993)\nocite{wes94} has argued that the radio axis of radio galaxies
and quasars may actually be oriented along the direction of mass distribution 
in (proto) clusters and that the 
formation of clusters and their brightest galaxy would
occur through infall along this direction. If this is the case there would
be more mass present 
along the direction of the radio axis so for objects whose axes are oriented
closer to our line of sight there would be more (neutral) gas in the
absorbing path.
However, as mentioned earlier, the velocity dispersion in a cluster is
of the order of 1000 km s$^{-1}$ while the strong H\,{\sc i} absorptions are
much narrower and almost always within $\sim250$ km s$^{-1}$ of the
Ly$\alpha$ emission peak (Fig. 9).

A possible indication of the angle between the radio axis and our line of sight
would be the radio core fraction (e.g. Browne \& Perley 1986; 
Browne \& Murphy 1987\nocite{bro86,bro87}).
For radio sources oriented closer to our line of sight it is expected that
the radio flux from the core is a larger fraction of the total radio flux than
for radio sources oriented closer to the plane of the sky. We see no evidence 
for a such relation between the Ly$\alpha$ velocity dispersion and the radio 
core fraction (Tables 4 and 5; see also Carilli {et al.} 
1996\nocite{car95}). This suggests that orientation effects do not play an 
important role in the observed correlations.

Because of the problems that the evolutionary and orientation scenario have
with explaining 
the presence of strong H\,{\sc i} absorption preferentially in the smaller radio sources,
we favour the environmental explanation for the observed velocity--size 
correlation. Additional evidence for the importance of environmental effects
comes from observations of 1243+036 at $z=3.6$ where the morphological 
correspondence between the radio and the Ly$\alpha$ emission indicates that
there is clear vigorous interaction between the radio source and the 
surrounding gas (van Ojik et al. 1996).\nocite{oji96}
Thus we find that it is likely that many steep spectrum HZRGs reside in
a dense environment, possibly a (proto) cluster.

\section{Conclusions}
Our high resolution spectroscopic study of Ly$\alpha$ emission in high
redshift radio galaxies has shown the following:
\newline
1) The Ly$\alpha$ velocity profiles
show considerable detailed structure much of which is indicative of 
H\,{\sc i} absorption.
\newline
2) Strong H\,{\sc i} absorption 
with column densities 10$^{18}$--10$^{19.5}$
is found against the Ly$\alpha$ emission in 11 of 18 objects. 
Almost all our galaxies (9 out of 10)
with (projected) radio sizes smaller than 50 kpc, have strong H\,{\sc i} absorption.
For the galaxies with larger radio sources, there is only occasional evidence 
for associated H\,{\sc i} absorption
\newline
3) The H\,{\sc i} absorbers are spatially extended over the entire Ly$\alpha$
emission region, indicating that they have a covering fraction near unity
over a region of $\sim$50 kpc.
\newline
4) The overall Ly$\alpha$ kinematics are dominated by turbulent motions. Only
in a few cases there is a hint of large scale organized dynamics, possibly
rotation. On smaller scales the Ly$\alpha$ spectra there is also evidence
that the kinematics are not purely chaotic,
such as ``shoulders''in the velocity profile and 
``wiggles'' in the spatial position of the maximum intensity of
Ly$\alpha$ as a function of velocity.
\newline
5) The Ly$\alpha$ emission is usually sharply peaked near the centre of the
radio galaxy and the Ly$\alpha$ emission often extends beyond the size of
the radio source. The detected extent of the Ly$\alpha$ emission regions 
range from 15 to 135 kiloparsec.
\newline
6) There is a strong correlations between the radio size and the Ly$\alpha$
size and a strong anti-correlation between the radio size and the velocity
dispersion of the emission line gas. The smaller radio sources have larger gas
velocity dispersions and smaller emission line gas regions than the larger
radio sources.
\newline
7) The amount of ``wiggling'' of the Ly$\alpha$ position as a function of
velocity correlates 
strongly with the amount of distortion of the radio source morphology and the 
amplitude of the wiggling correlates with the transverse radio size.

These observations have the following implications:
\newline
a) The frequent occurrence of strong H\,{\sc i} absorption systems is much too large 
to be caused by random intervening absorbers. The absorbers must be directly 
associated with the radio galaxy or its environment.
\newline
b) On the basis of orientation-unification scenarios of radio galaxies and 
quasars, the H\,{\sc i} absorption systems may well be due to 
gas from the gaseous halo of the galaxy, situated outside the
ionization cone.
Halos of neighbouring cluster galaxies 
or tidal tails from galaxy interactions seem very unlikely causes for the 
H\,{\sc i} absorptions.
\newline
c) The large velocity dispersions of the ionized gas and the
wiggles in the Ly$\alpha$ peak spatial position as a function of velocity
are best explained as due to interaction of the radio plasma with the gas.
\newline
d) The correlation between Ly$\alpha$ properties and radio 
properties and the frequent presence of associated H\,{\sc i} absorption indicate that
HZRGs are located in dense environments.
The smaller radio sources most likely reside in the densest (cluster) 
environments, 
where large amounts of neutral gas are present and most interaction with the 
radio source occurs. Transfer of kinetic energy
from the radio jet to the gas increases the gas velocity dispersion and
reduces the propagation speed of the radio source.

There is considerable scope for following up on these observations.
First, it is important to make similar measurements in a range of
position angles and not just confined to the radio position angle.
Secondly, more detailed imaging of the Ly$\alpha$ halos will help
delineate the geometry of the individual Ly$\alpha$ halos. Thirdly,
HST imaging and spectroscopic observations will allow the details 
of the gas and its kinematics to be studied in the important nuclear
regions. Fourthly, small high redshift radio sources with large gas velocity 
dispersions and strong associated H\,{\sc i} absorptions are excellent
targets in searching for clusters of galaxies in the early Universe.

One intriguing question that remains is what role the gas clouds
play in galaxy evolution. The typical sizes of 40 lightdays deduced
for individual clouds are comparable with that of the solar system.
It is tempting to speculate that these clouds are intimately 
associated with the early formation stages of individual stars and that
they delineate a fundamental phase in galaxy evolution.

\section*{Acknowledgements}

We would like to thank Malcom Bremer useful discussions.  We
acknowledge support from an EU twinning project, funding from the
high-z programme subsidy granted by the Netherlands Organization for
Scientific Research (NWO) and a NATO research grant. RWH acknowledges
financial assistance for this programme from the Australian Research
Council.

\newpage

\section{Figures}

\begin{figure}[p]
\hbox{
\psfig{figure=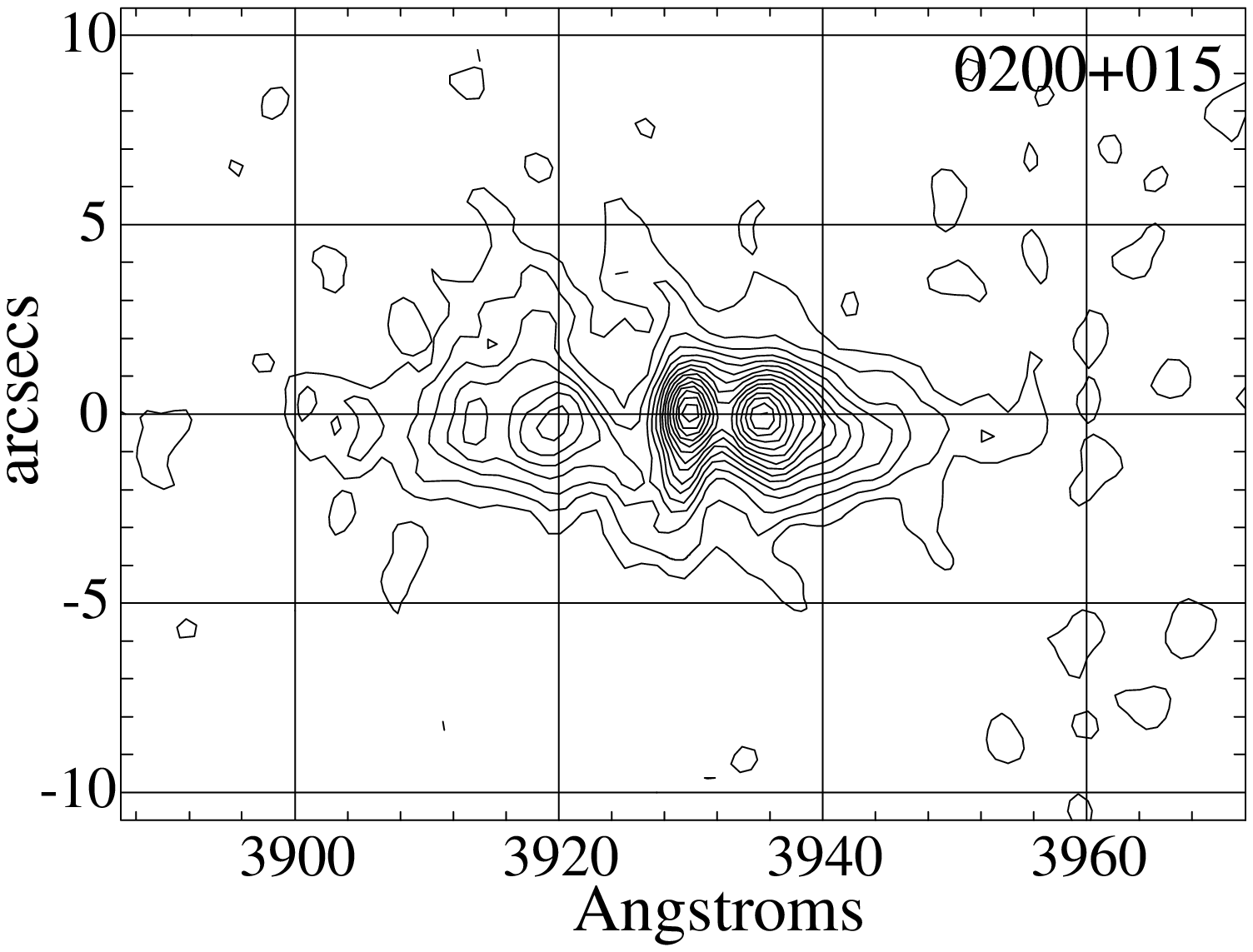,angle=90,width=8cm,height=6cm}
\psfig{figure=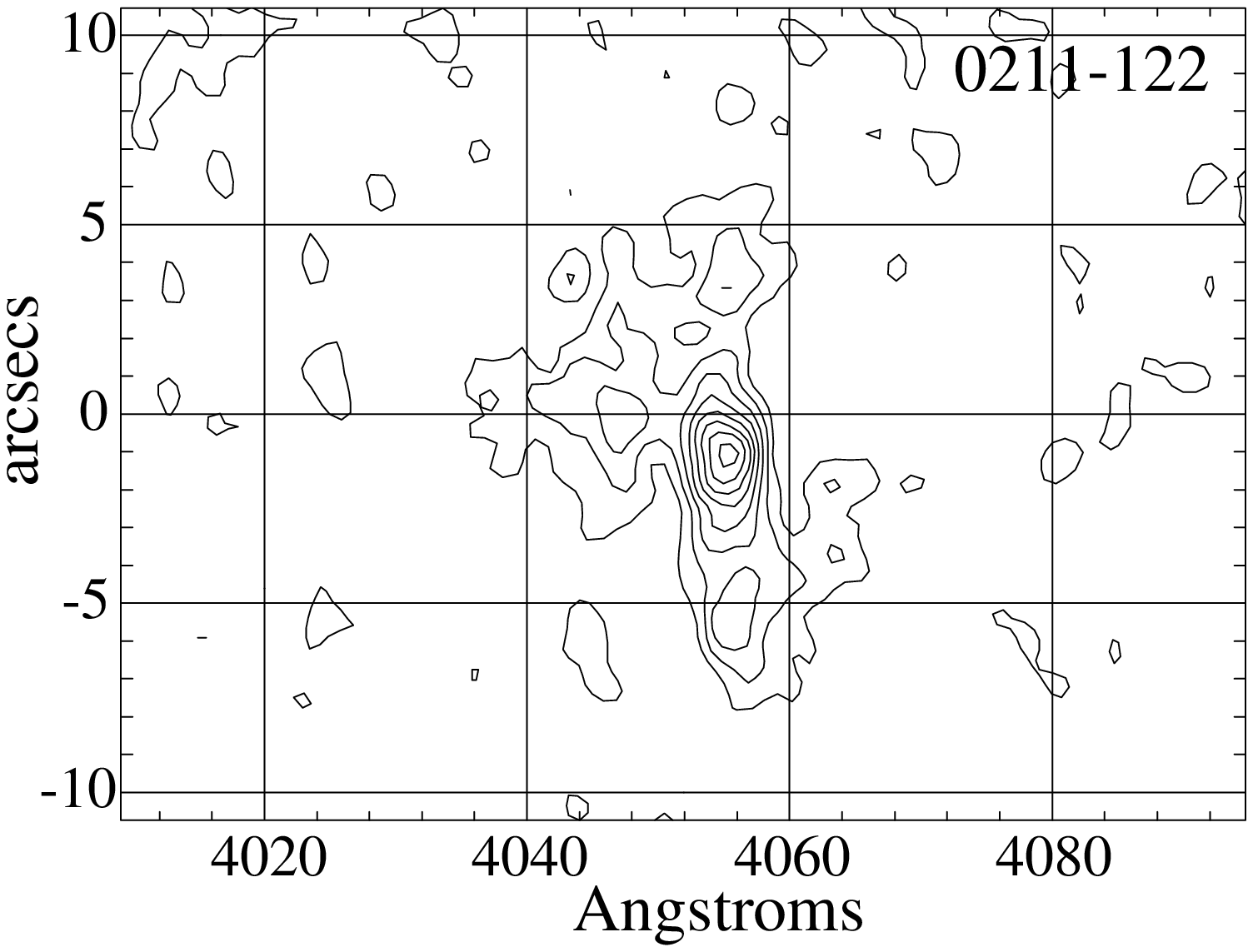,angle=90,width=8cm,height=6cm}
}
\hbox{
\psfig{figure=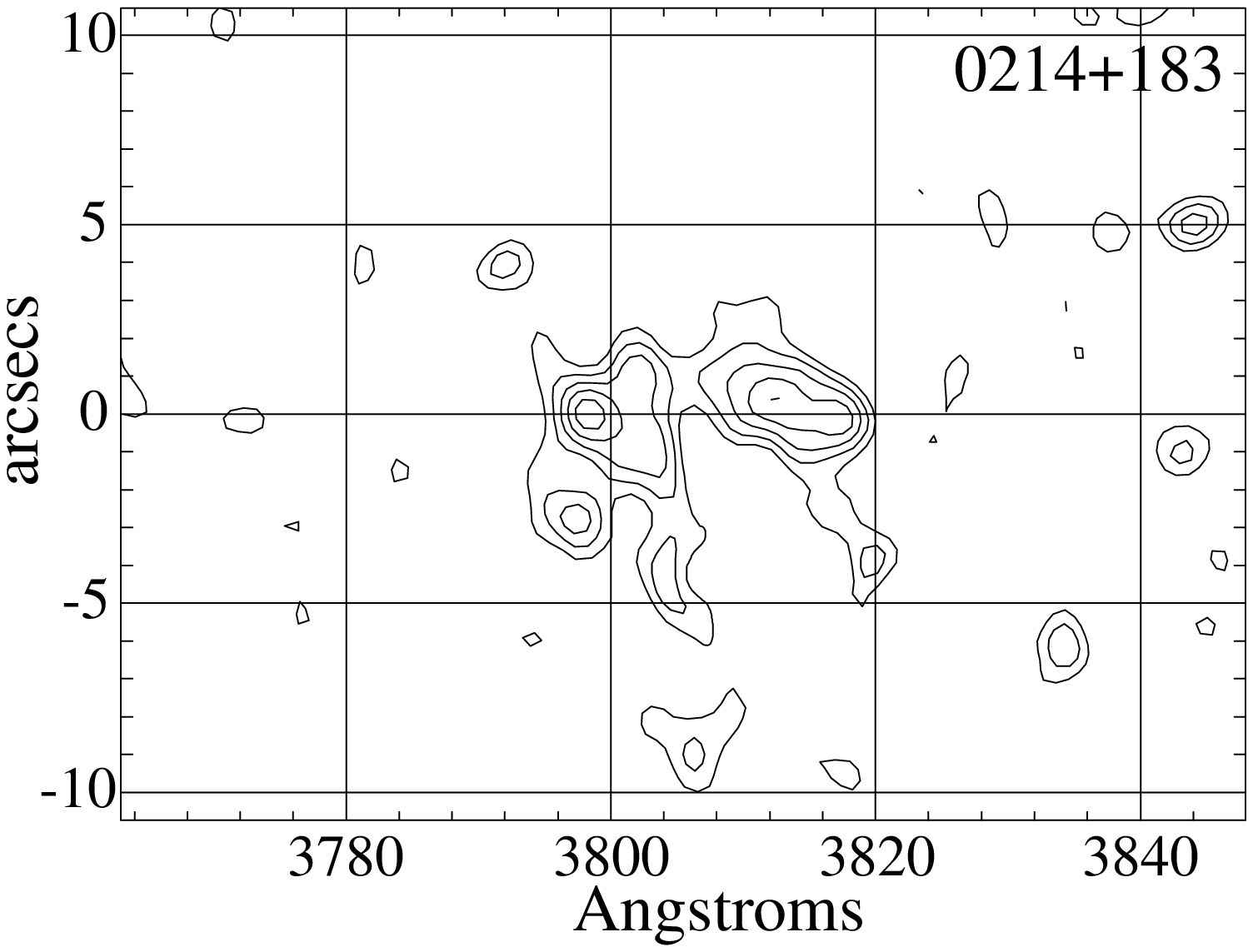,angle=90,width=8cm,height=6cm}
\psfig{figure=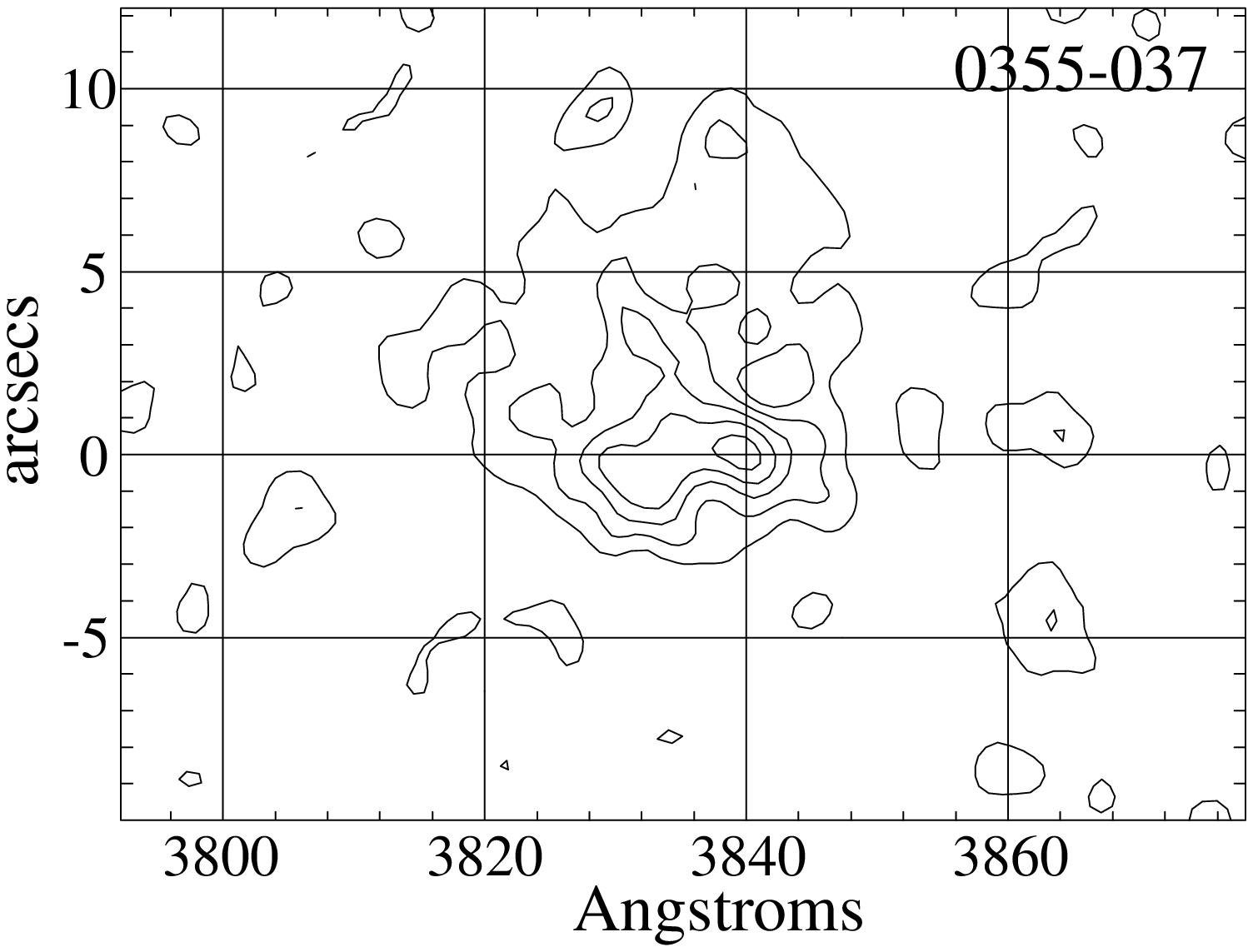,angle=90,width=8cm,height=6cm}
}
\hbox{
\psfig{figure=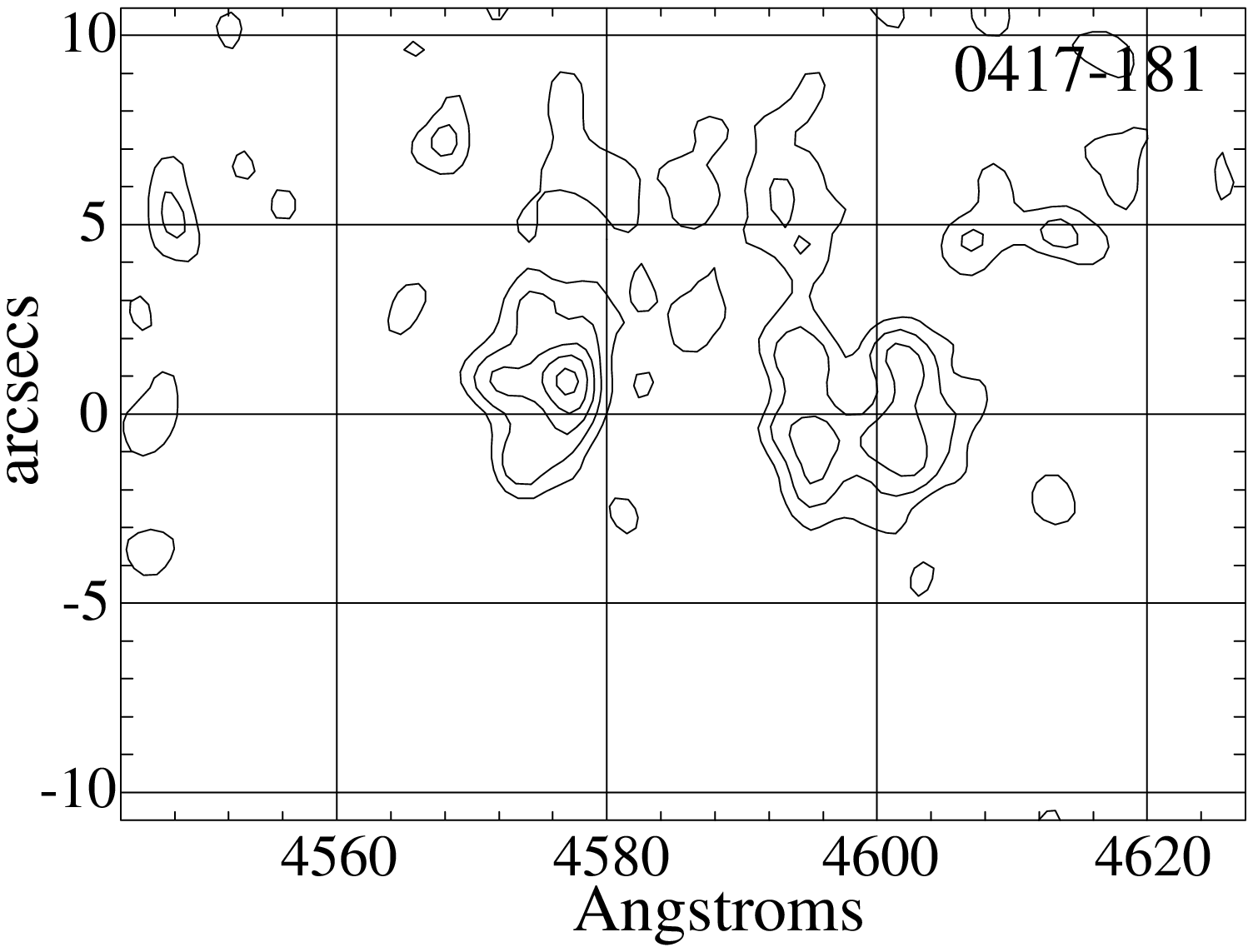,angle=90,width=8cm,height=6cm}
\psfig{figure=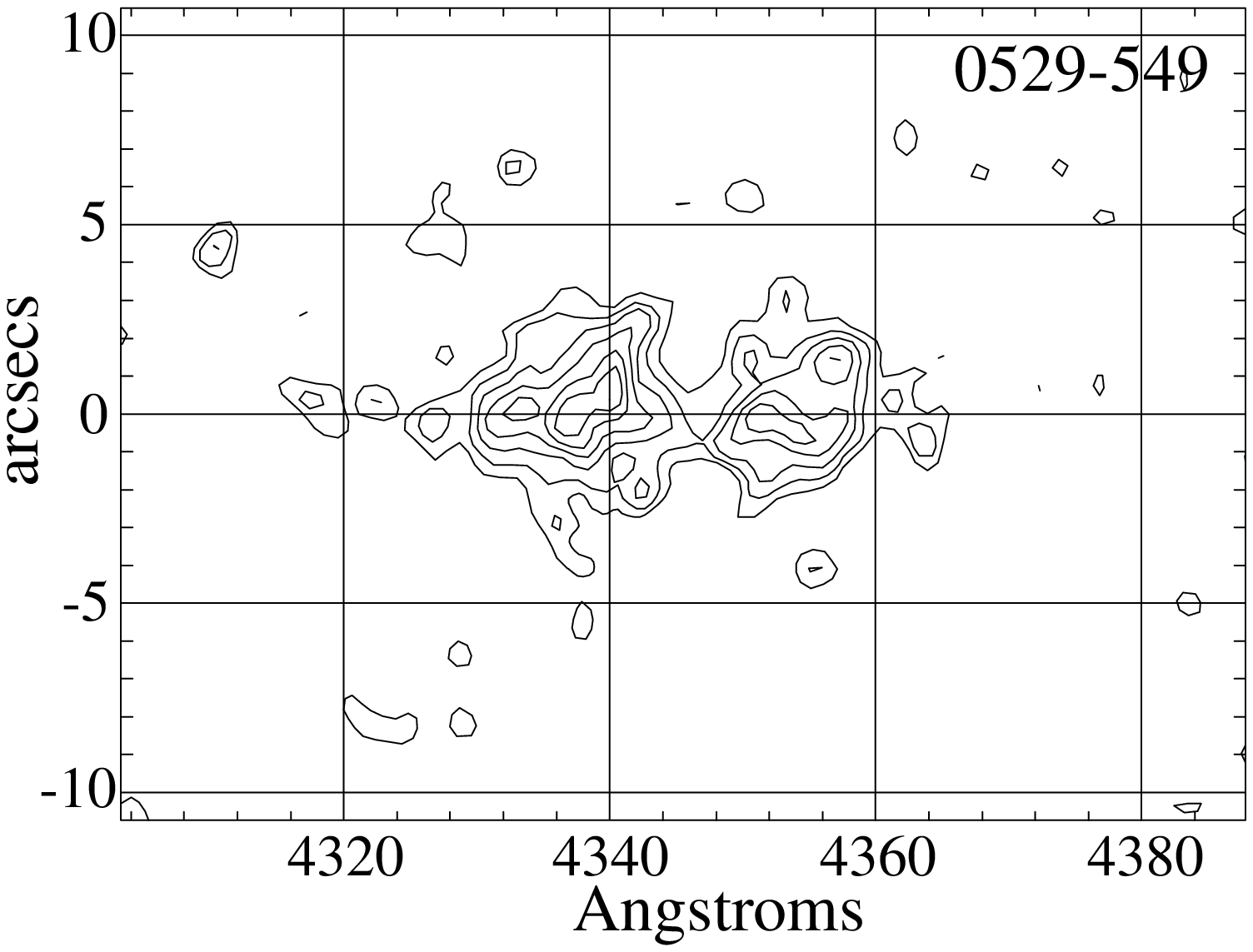,angle=90,width=8cm,height=6cm}
}
\noindent {\bf Fig. 1.} Two-dimensional high resolution spectra of the
Ly$\alpha$ emission regions. The two-dimensional spectra have been 
smoothed with Gaussian of $1''\times 2$ \AA\ (FWHM) to enhance the extended 
Ly$\alpha$ emission. The final resolution in these representations of
the two-dimensional spectra is $\sim 1.2'' \times 3.5$ \AA, except for those 
objects observed in $>1''$ seeing (see Table 1 and 2).
Because of the narrowness of the absorption features,
the smoothing also results in the absorption troughs appearing less deep 
than in the unsmoothed one-dimensional spectra.  
In the spectra with the strongest
Ly$\alpha$ lines, contours are linearly spaced at
2$\sigma$, 4$\sigma$, 6$\sigma$, etc., where $\sigma$ is the RMS 
background noise.
For the objects with fainter Ly$\alpha$ emission,
0214+183, 0417$-$181, 0529$-$549, 0748+134, 1357+007 and 
2202+128, contours are linearly spaced at 2$\sigma$, 3$\sigma$, 4$\sigma$, 
etc
\end{figure}

\clearpage \newpage 
\begin{figure}[p]
\hbox{
\psfig{figure=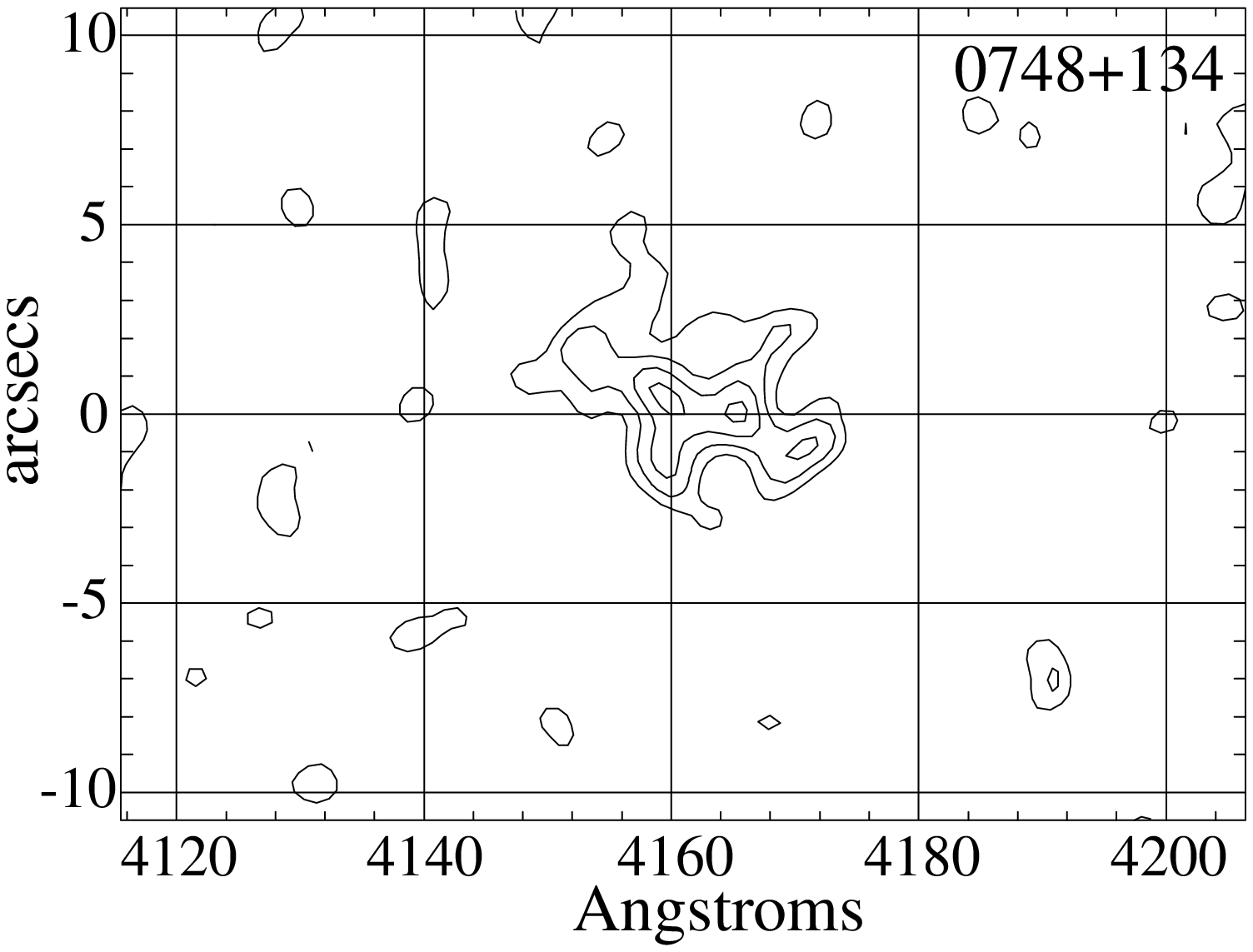,angle=90,width=8cm,height=6cm}
\psfig{figure=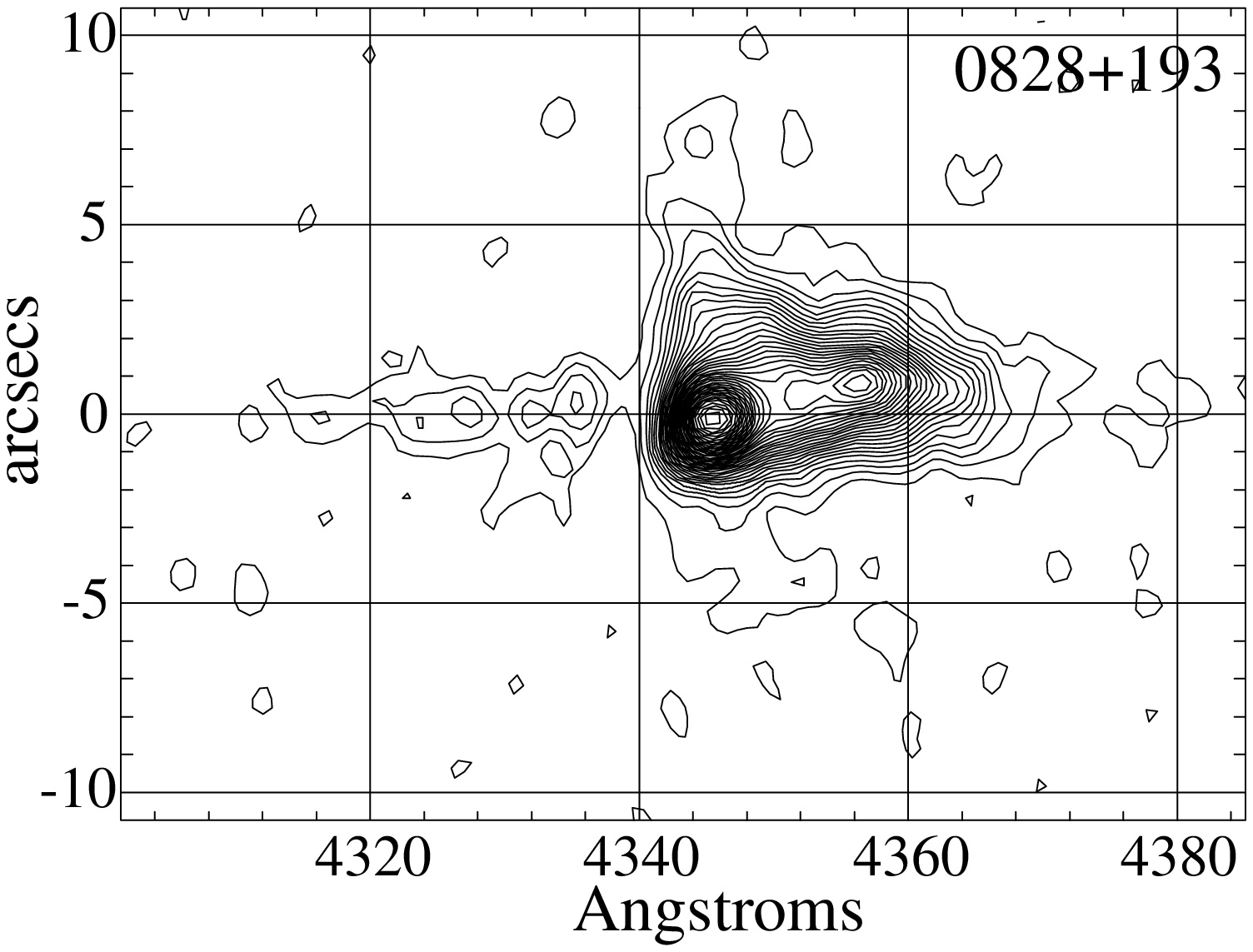,angle=90,width=8cm,height=6cm}
}
\vspace{0.5cm}
\hbox{
\psfig{figure=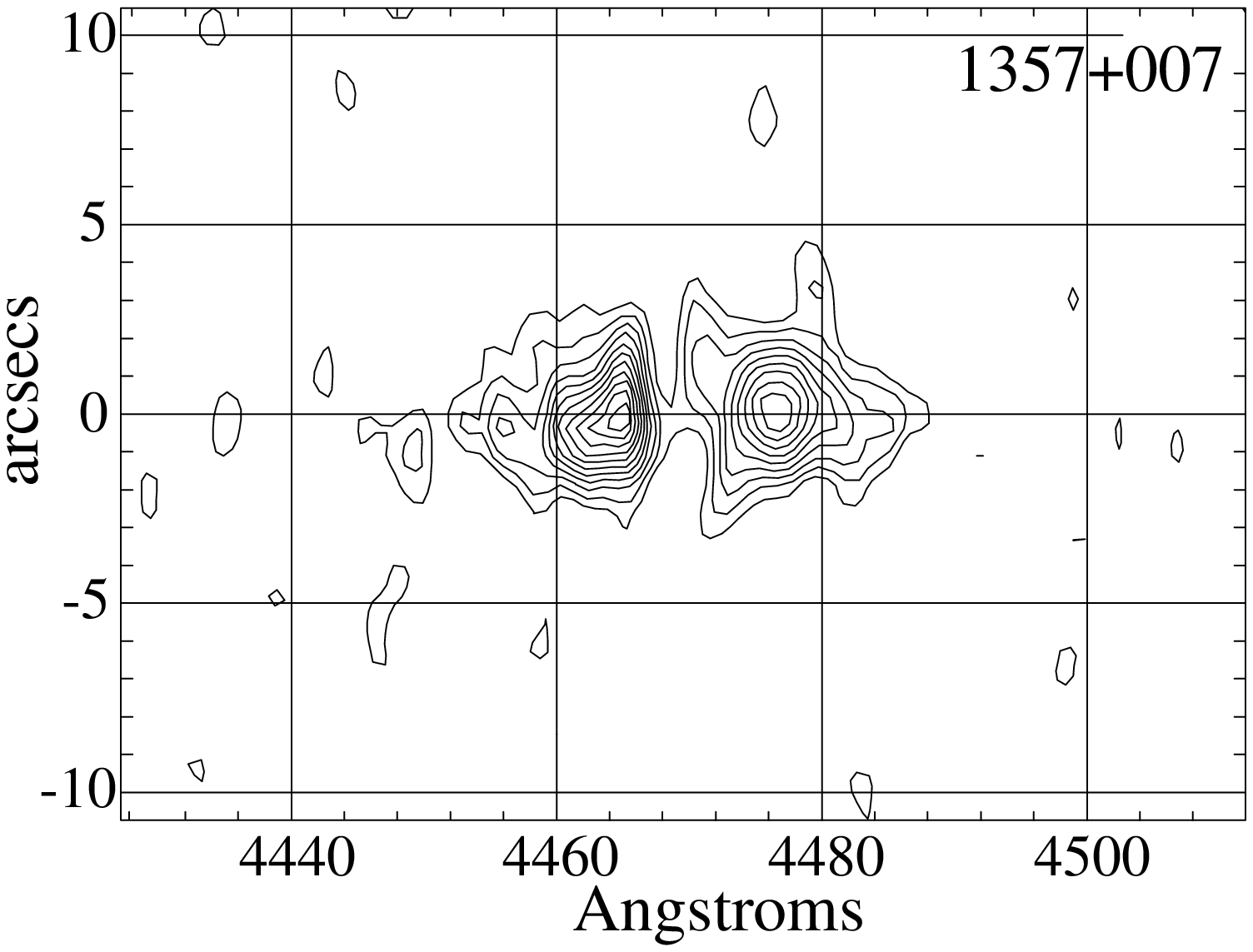,angle=90,width=8cm,height=6cm}
\psfig{figure=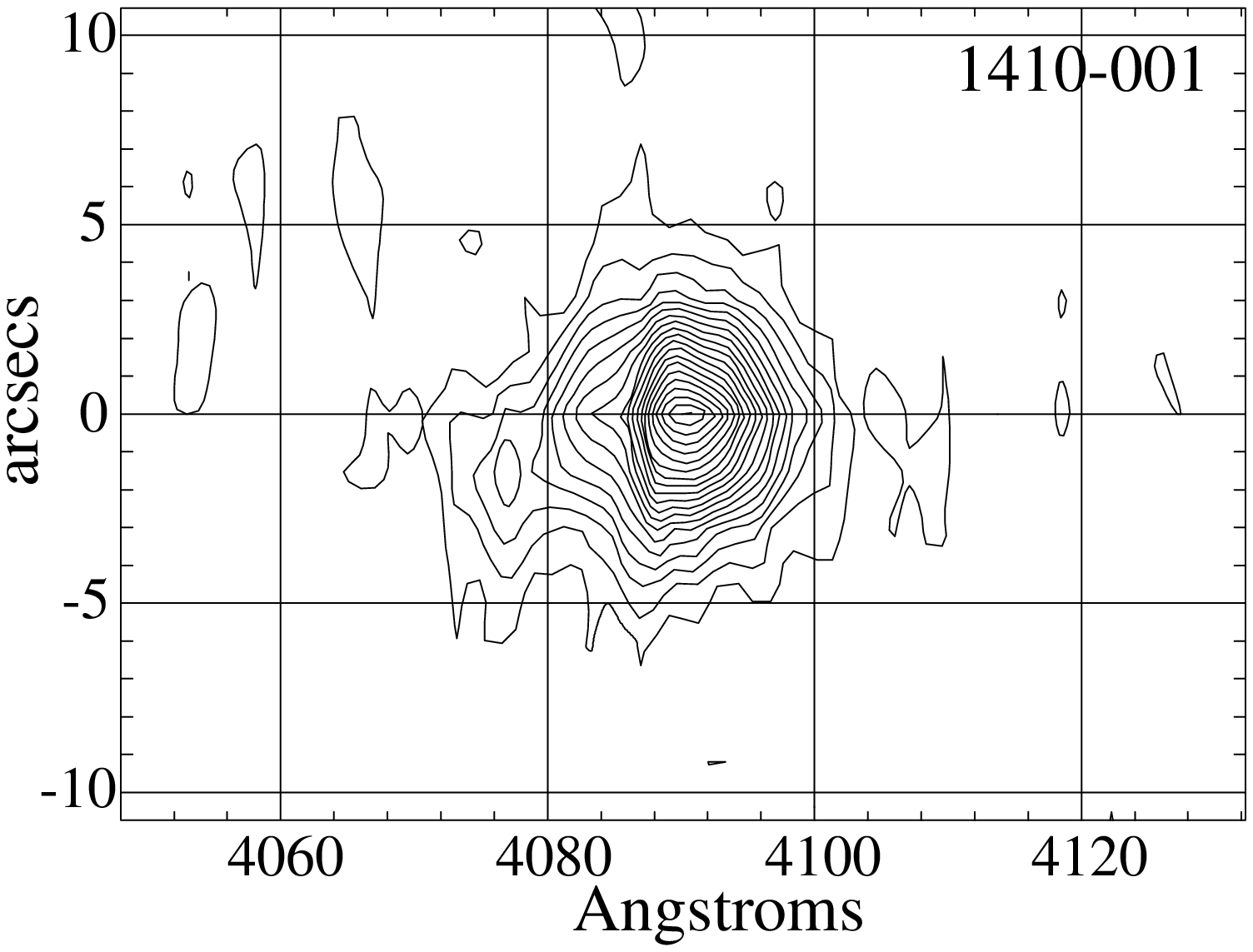,angle=90,width=8cm,height=6cm}
}
\vspace{0.5cm}
\hbox{
\psfig{figure=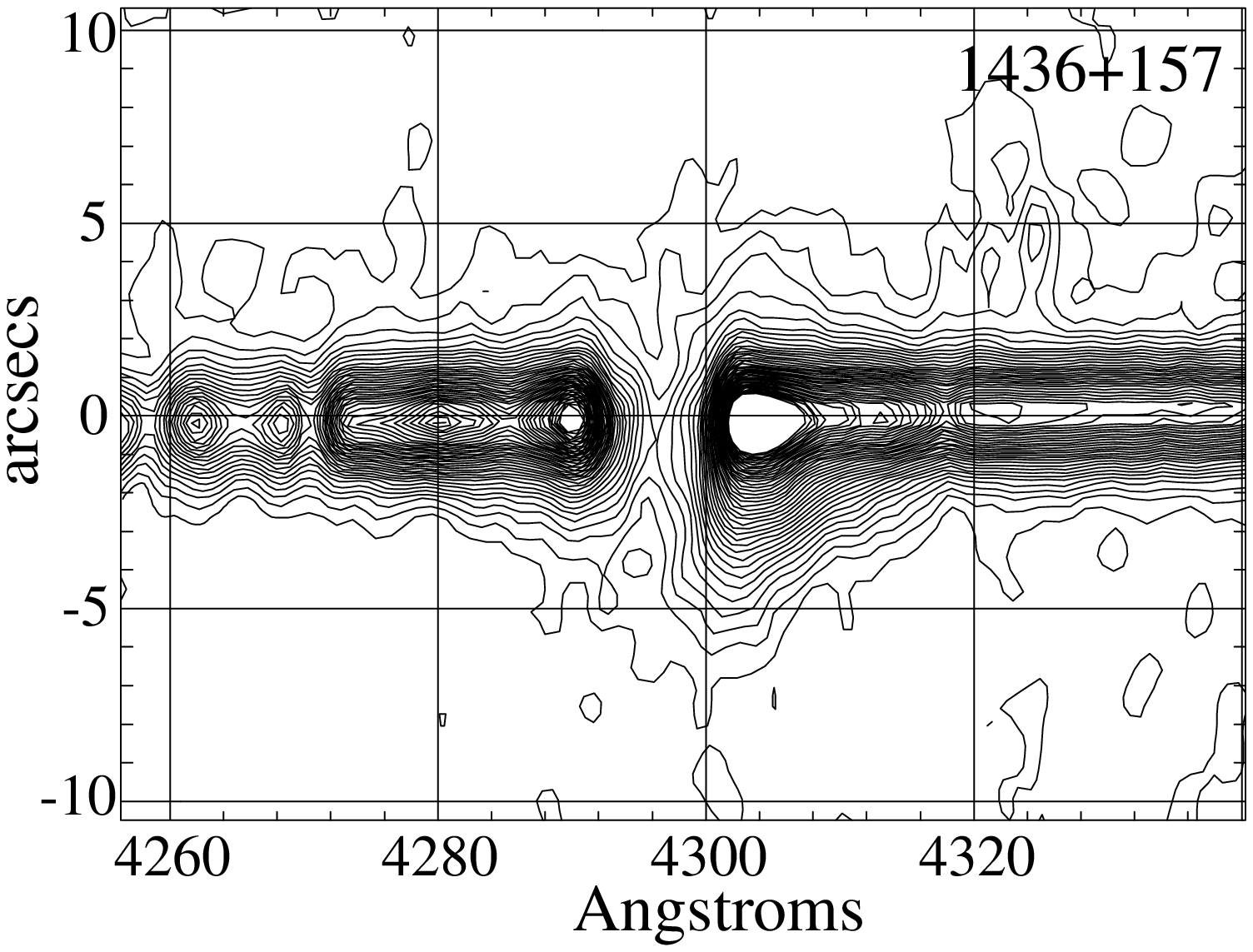,angle=90,width=8cm,height=6cm}
\psfig{figure=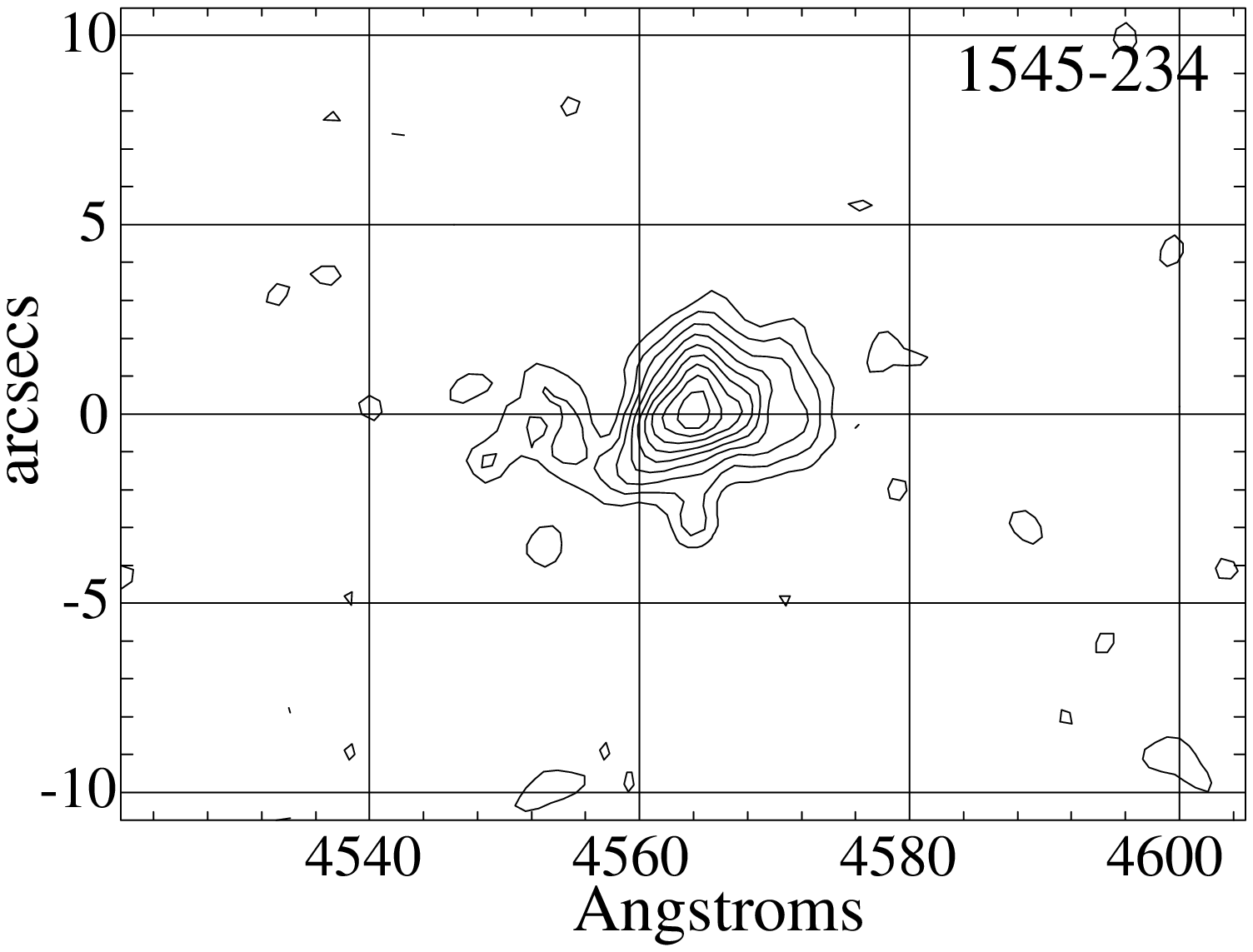,angle=90,width=8cm,height=6cm}
}
\noindent {\bf Fig. 1.} -- continued --.
\end{figure}

\clearpage \newpage

\begin{figure}[t]
\vspace{0.5cm}
\hbox{
\psfig{figure=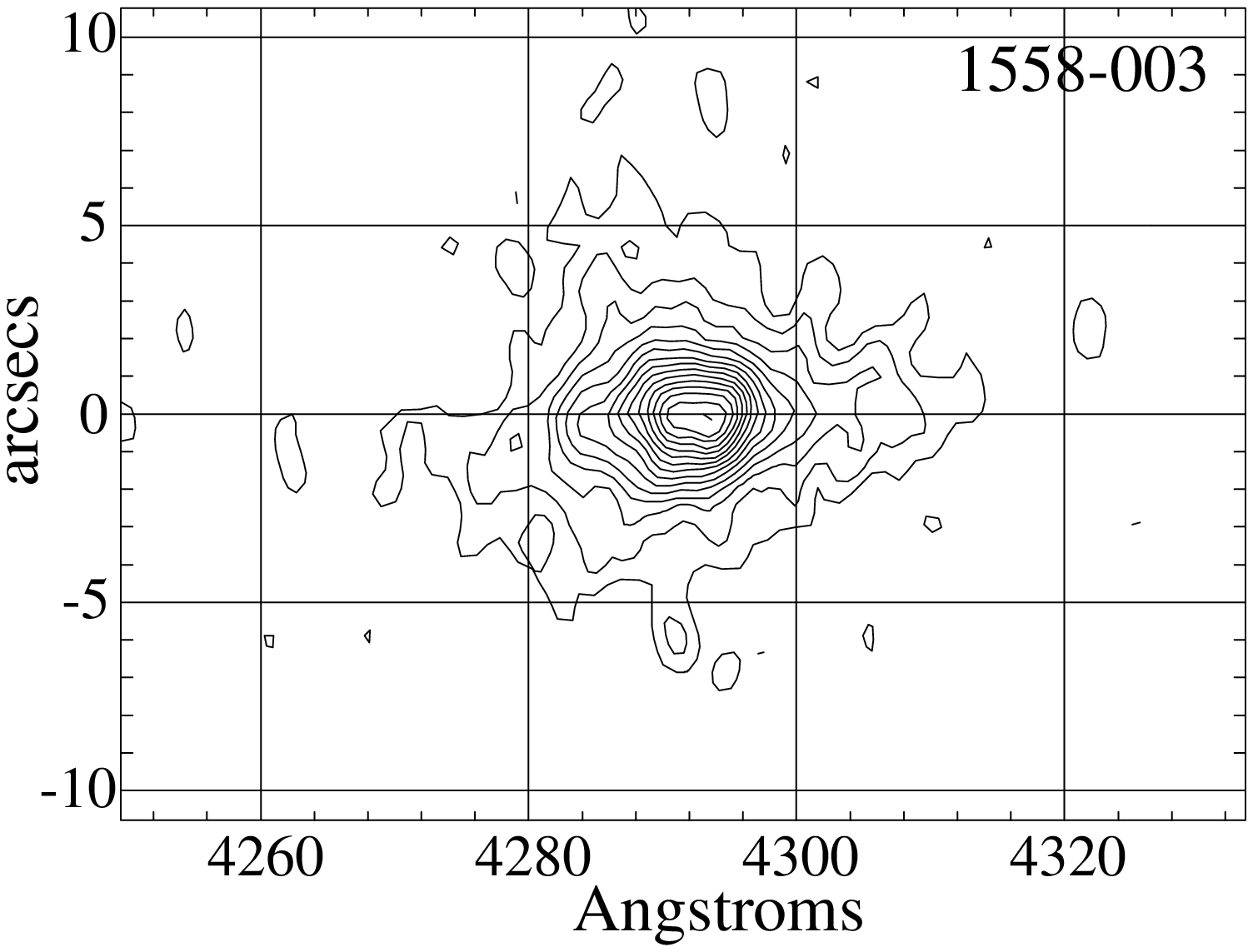,angle=90,width=8cm,height=6cm}
\psfig{figure=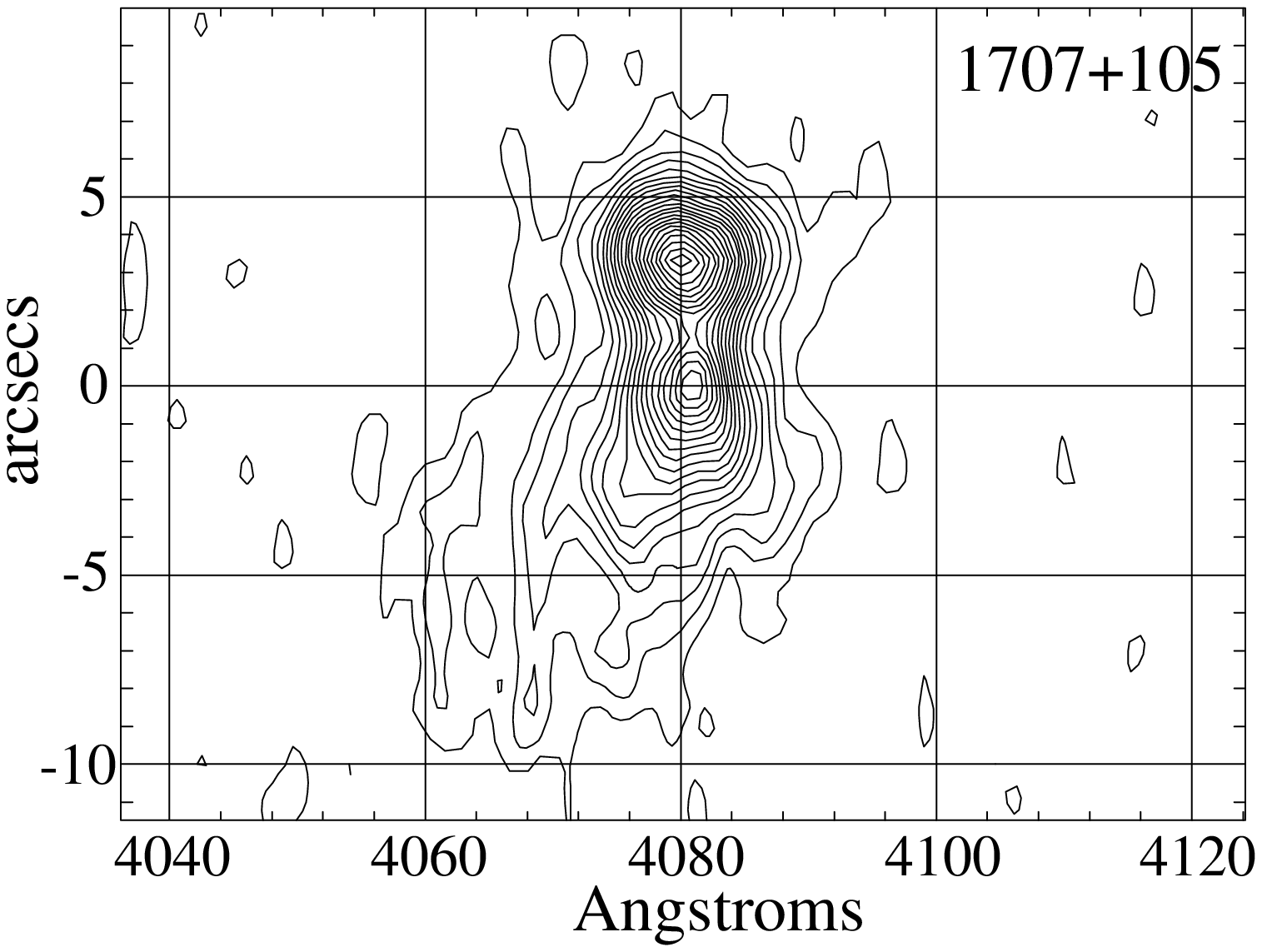,angle=90,width=8cm,height=6cm}
}
\hbox{
\psfig{figure=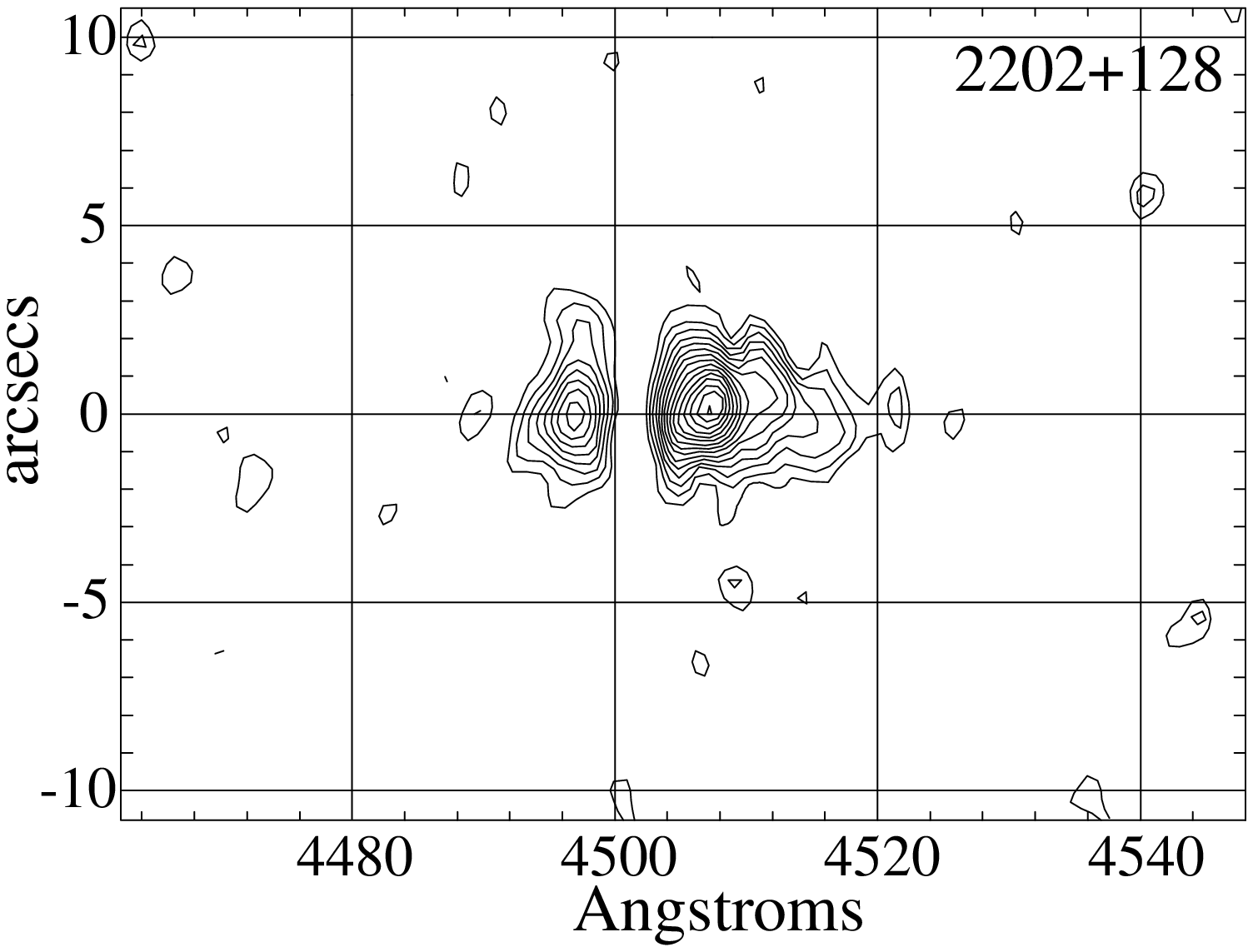,angle=90,width=8cm,height=6cm}
}
\noindent {\bf Fig. 1.} -- continued --.
\end{figure}

\clearpage \newpage

\begin{figure}
\hbox{
\psfig{figure=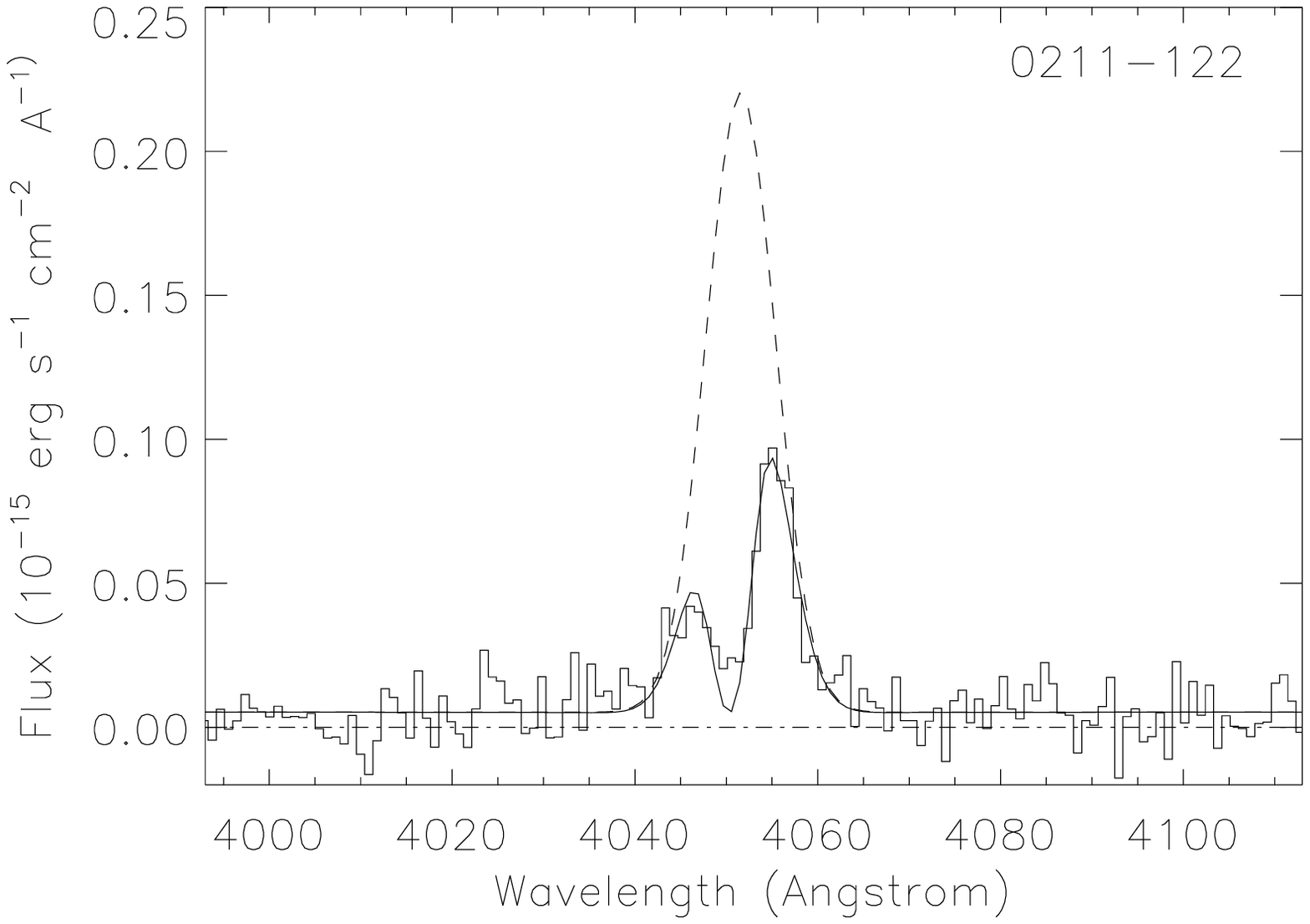,width=8cm}
\psfig{figure=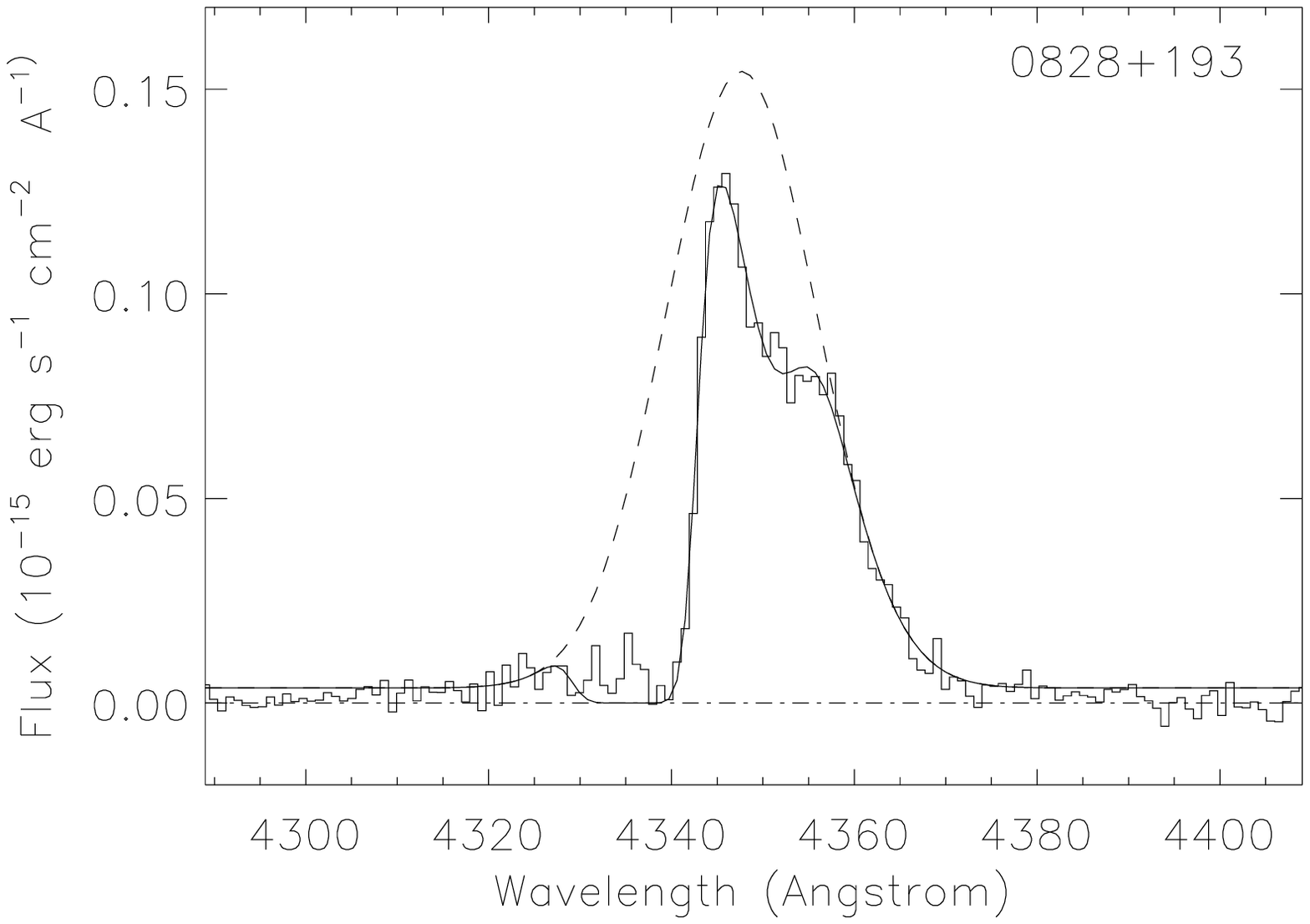,width=8cm}
}
\noindent {\bf Fig. 2.} {\it Left}: The profile of the ``dusty'' Ly$\alpha$ 
profile of 0211$-$122 fitted by an HI absorption with 
N(HI)=10$^{19.5}$ cm$^{-2}$. On the basis of our simple modelling 
assumptions, such strong HI absorption 
would be deeper than is observed.
The solid line is the resultant profile from the model fitting,
and the dashed line is the original Gaussian emission line profile calculated
from the model fitting.
{\it Right}: The absorption in the blue wing of Ly$\alpha$
in 0828+193 is modelled by one broad absorber. This model fits the steep drop
from the Ly$\alpha$ peak well, but absorbs too much of the 
Ly$\alpha$ emission from the blue wing

\end{figure}

\clearpage \newpage

\begin{figure}
\hbox{
\psfig{figure=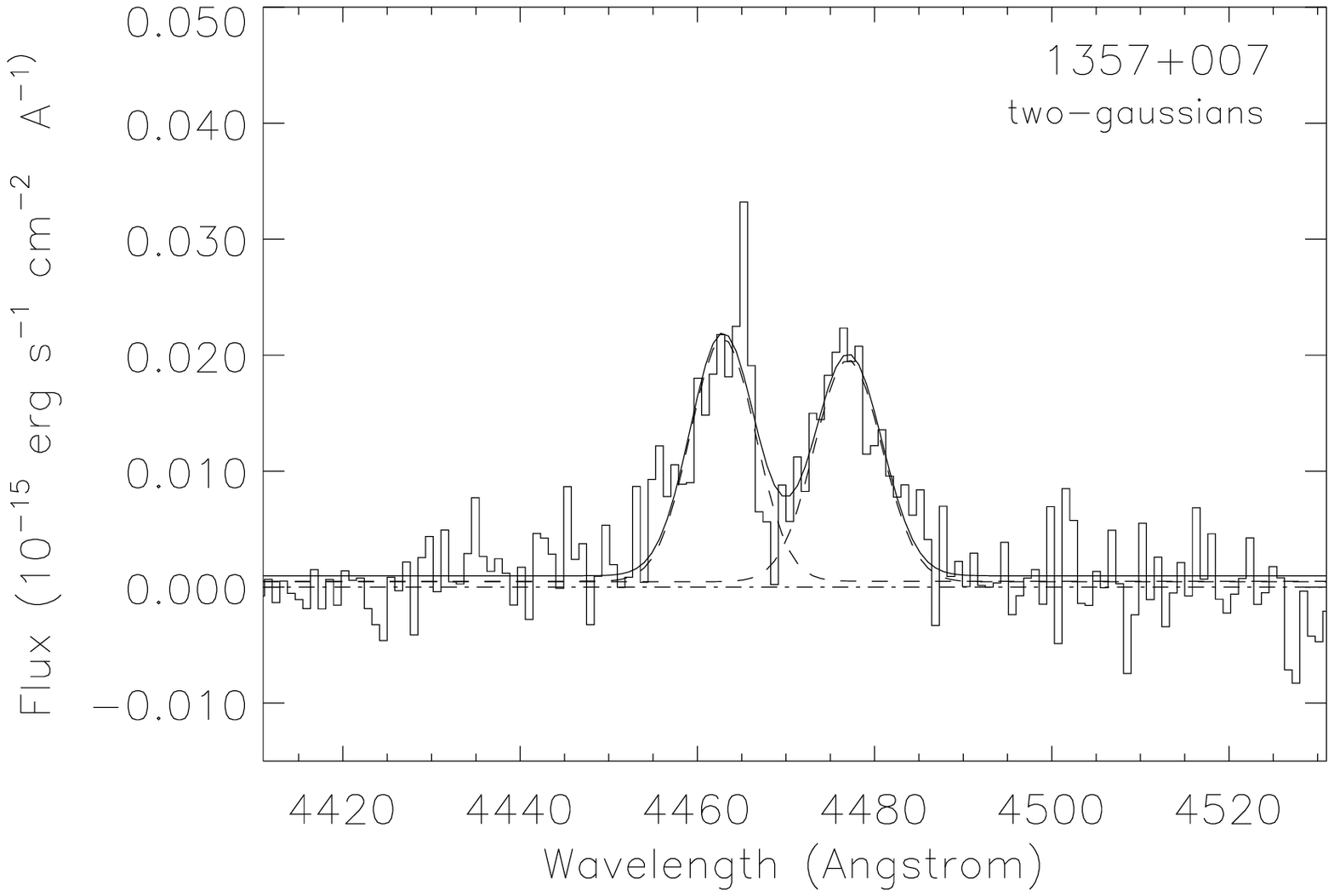,width=8cm}
\psfig{figure=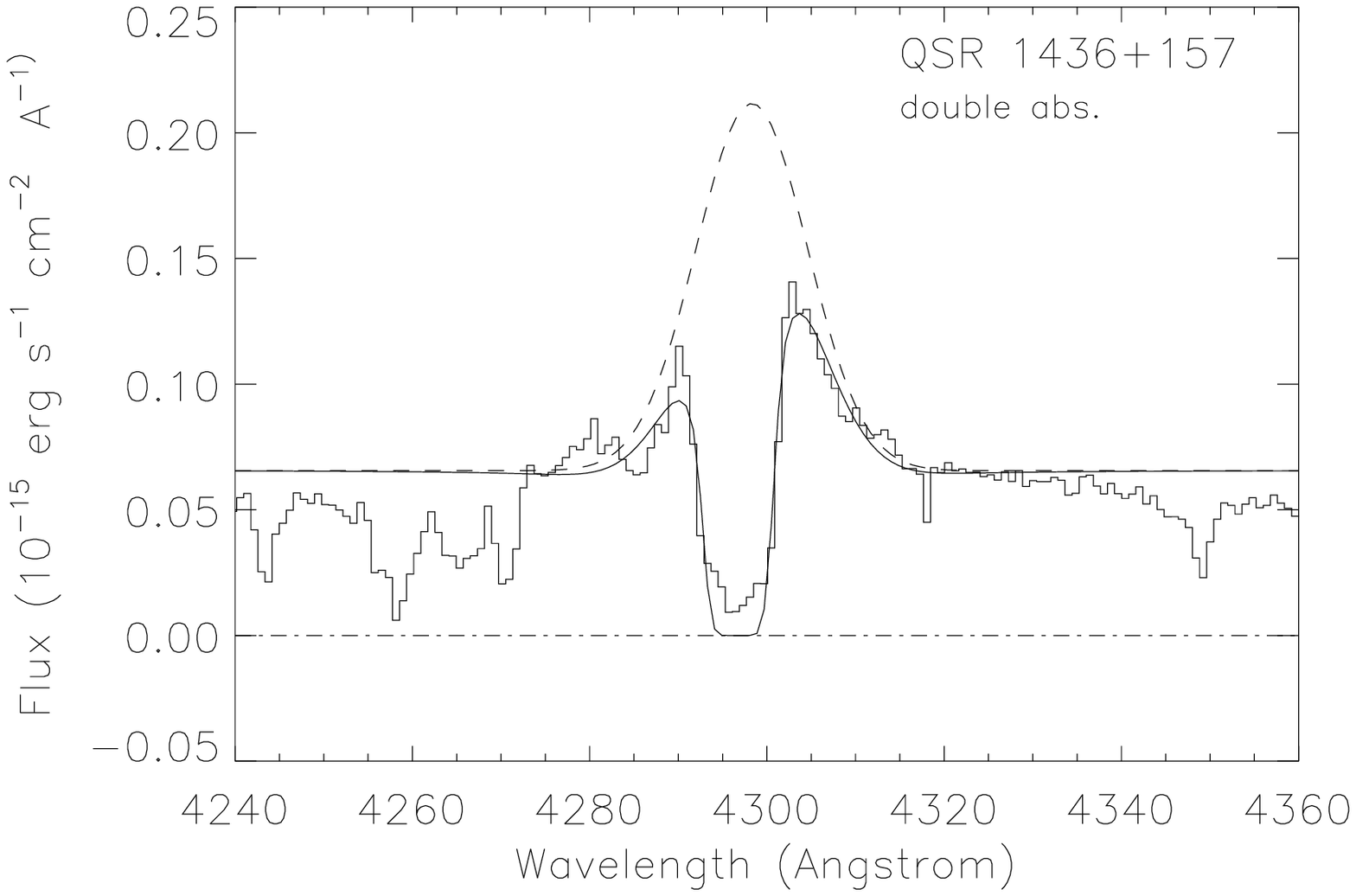,width=8cm}
}
\noindent {\bf Fig. 3.} {\it Left}: The double peaked Ly$\alpha$ profile 
of 1357+007 modelled by two 
separate Gaussian velocity components. 
The solid line is the resultant profile from the model fitting,
and the dashed lines are the original Gaussian emission line profiles calculated
from the model fitting. 
Note the inability of this model to closely fit
the steep trough between the Ly$\alpha$ peaks, while the wings on either side
are well fit by a Gaussian. 
{\it Right}: The strong absorption in the narrow Ly$\alpha$ component of the 
quasar
1436+157 modelled with two N(HI)$\sim$10$^{19.3}$ cm$^{-2}$ absorbers. A single
absorber would require a very large Doppler parameter and gives a worse fit

\end{figure}

\clearpage \newpage

\begin{figure}[p]
\vspace{-0.5cm}
\hbox{
\psfig{figure=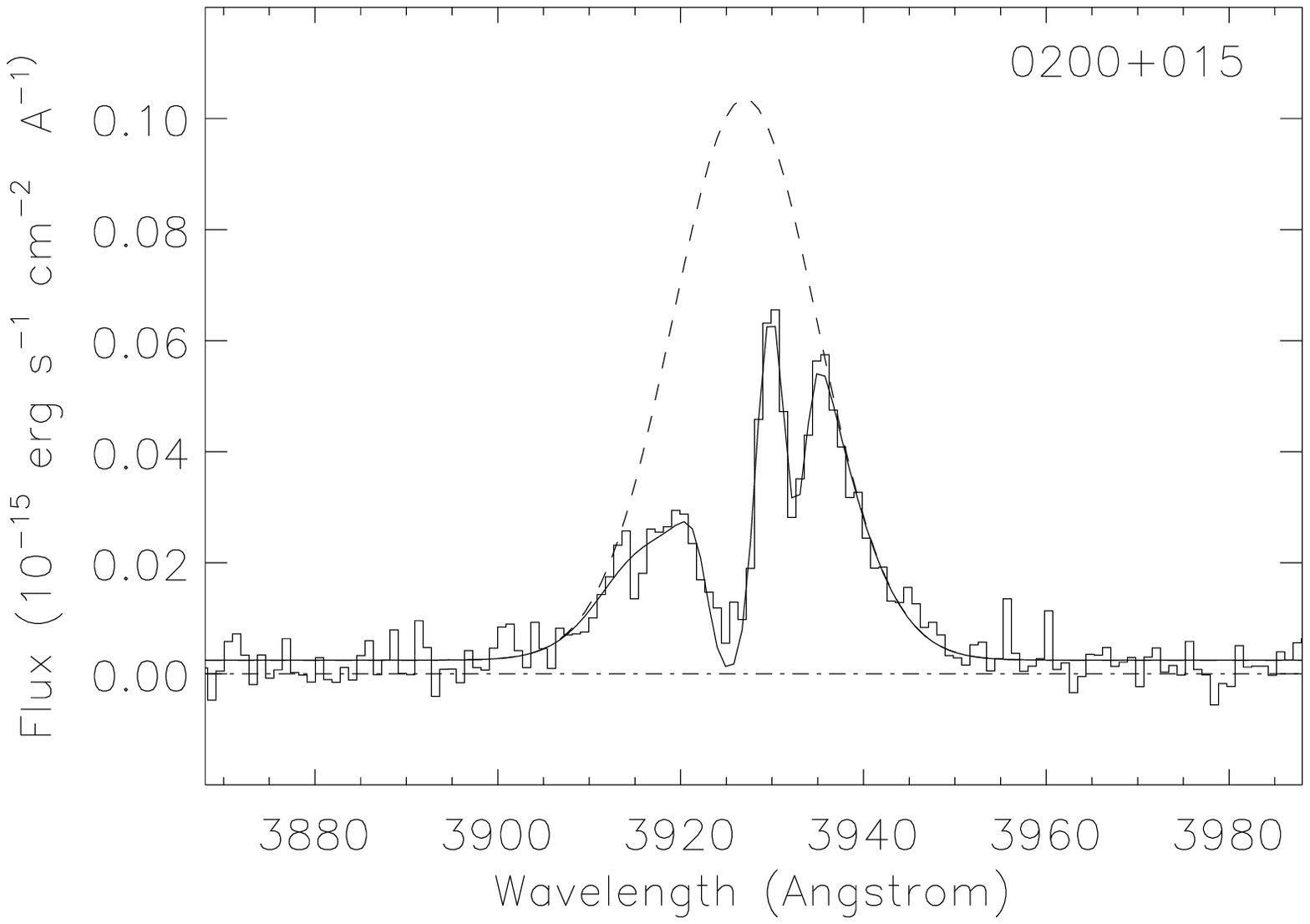,width=8cm,height=7cm}
\psfig{figure=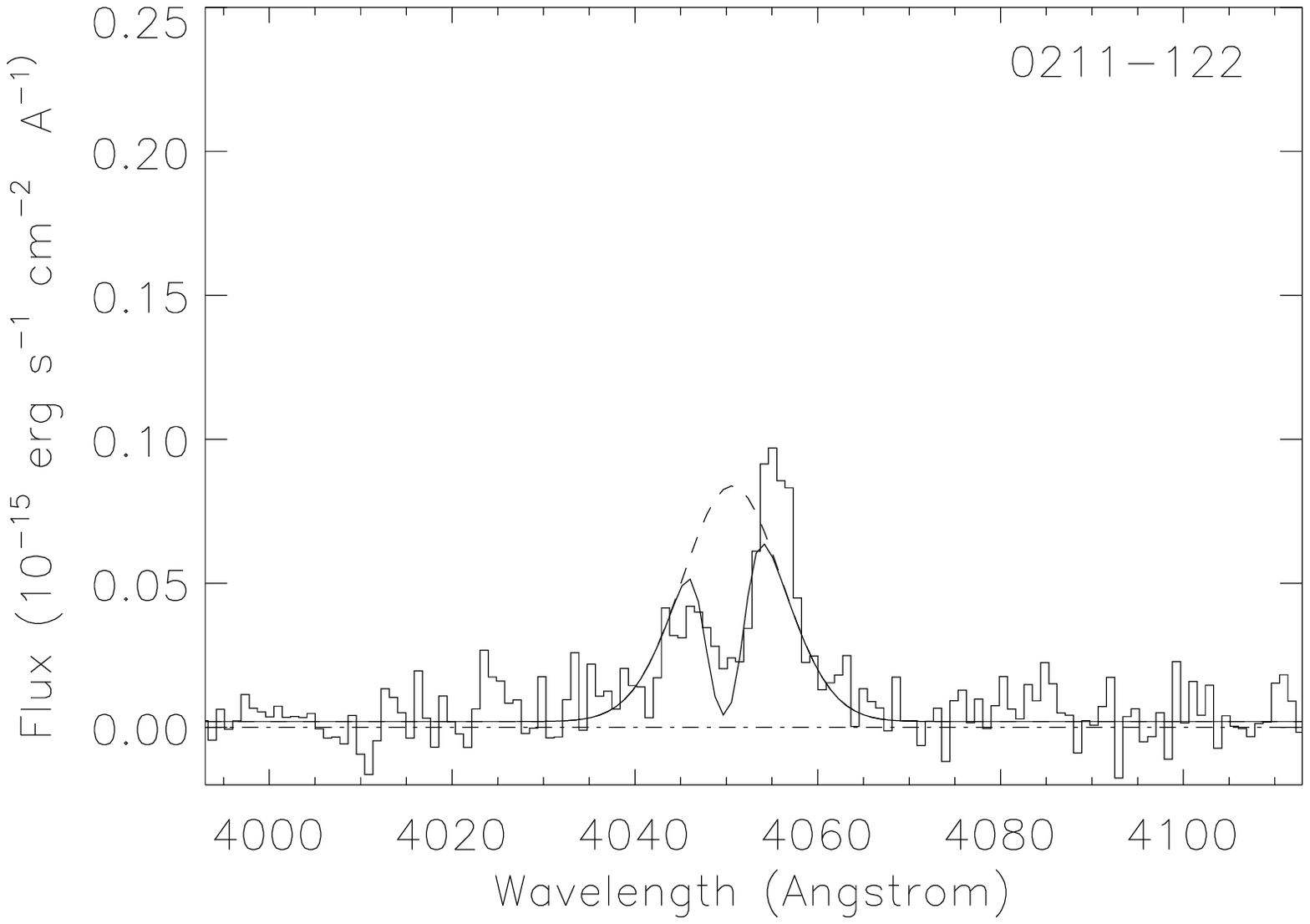,width=8cm,height=7cm}
}
\hbox{
\psfig{figure=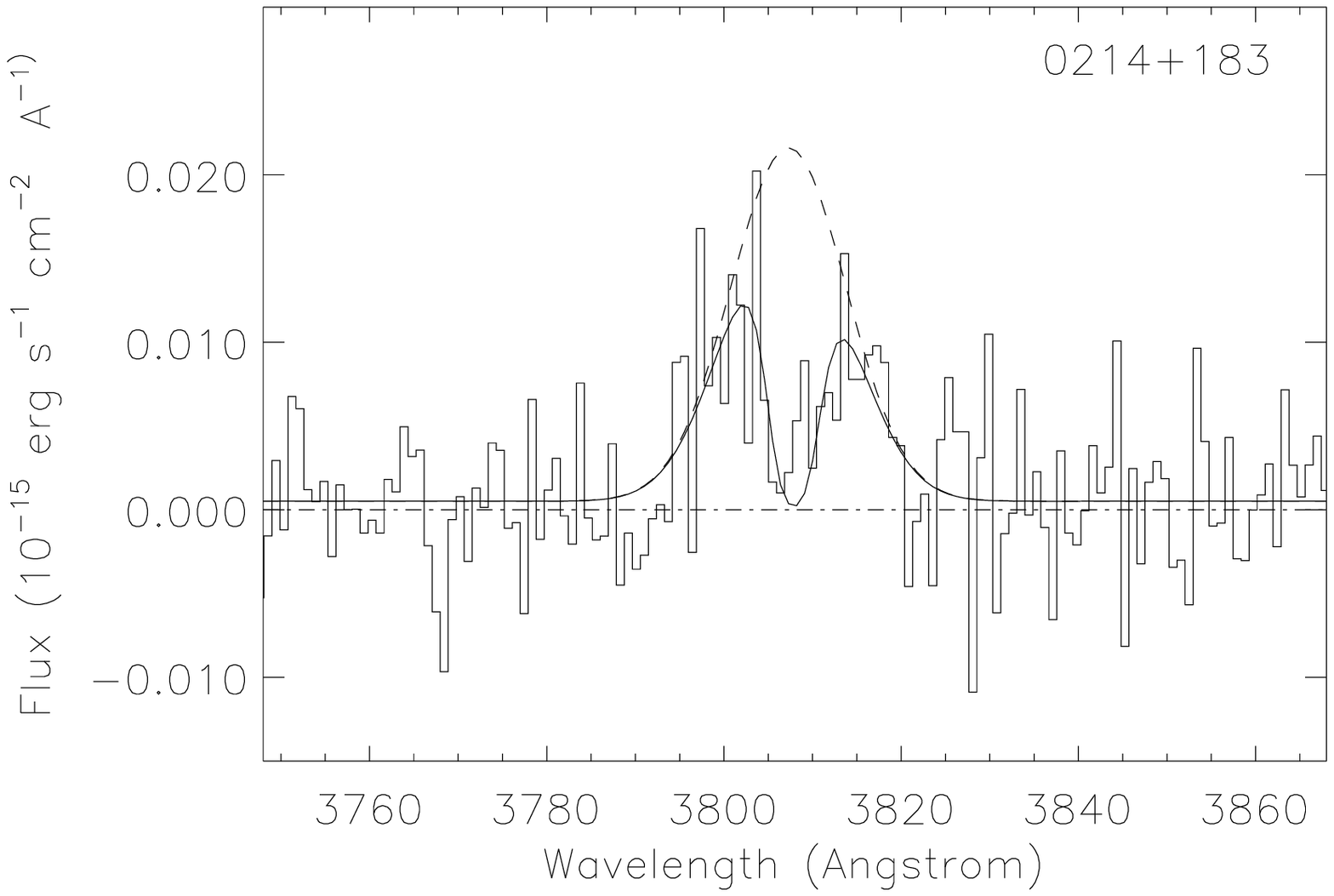,width=8cm,height=7cm}
\psfig{figure=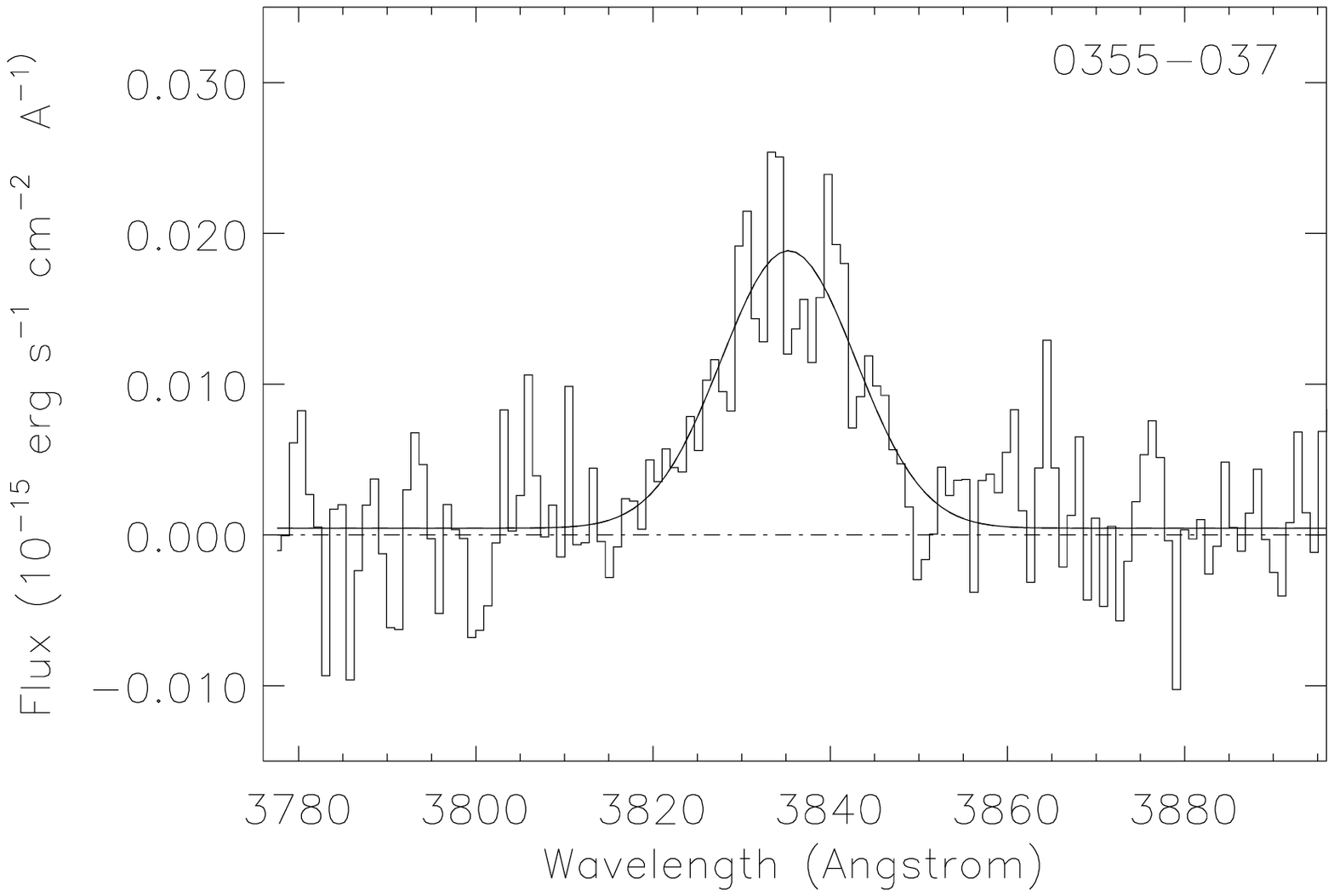,width=8cm,height=7cm}
}
\hbox{
\psfig{figure=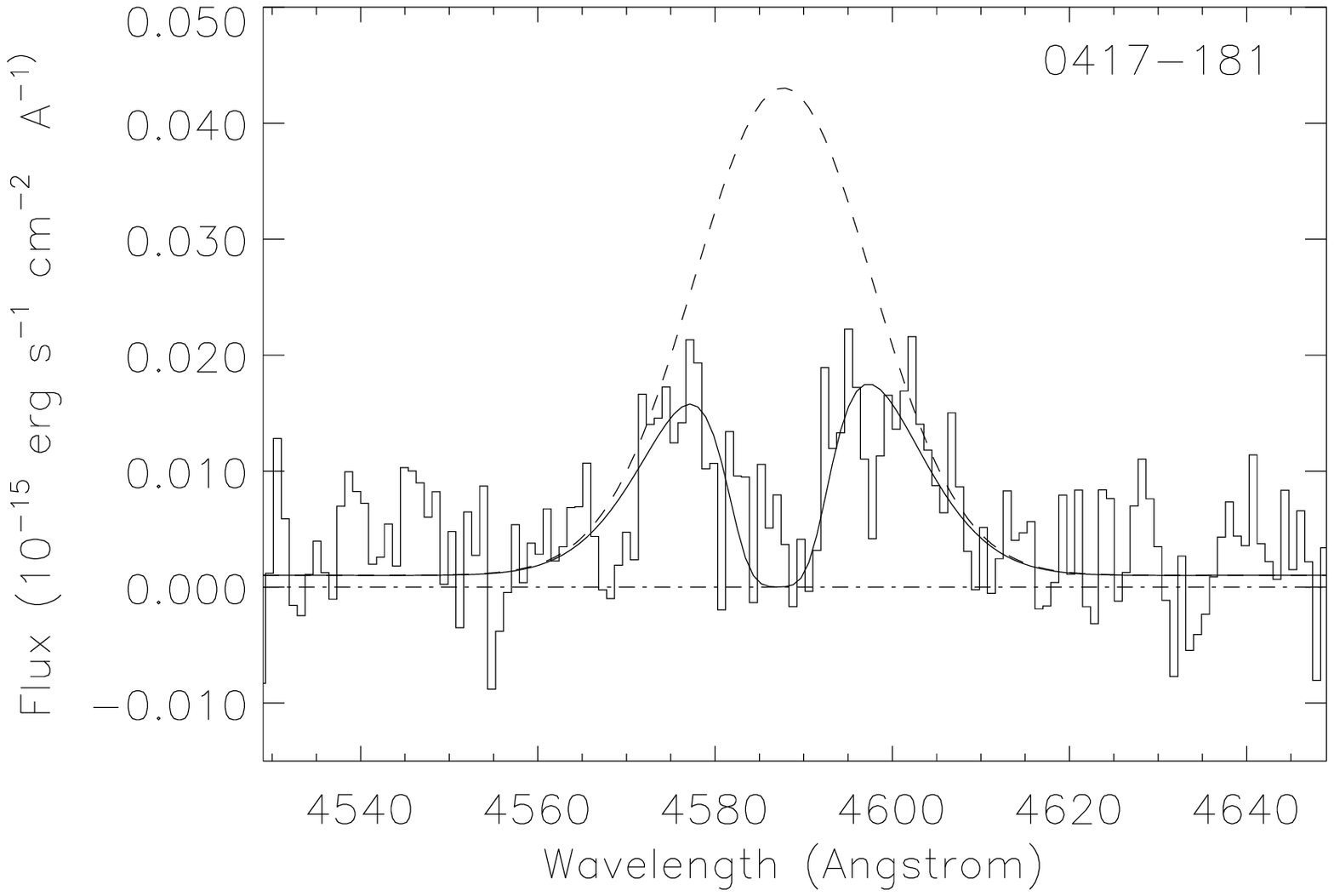,width=8cm,height=7cm}
\psfig{figure=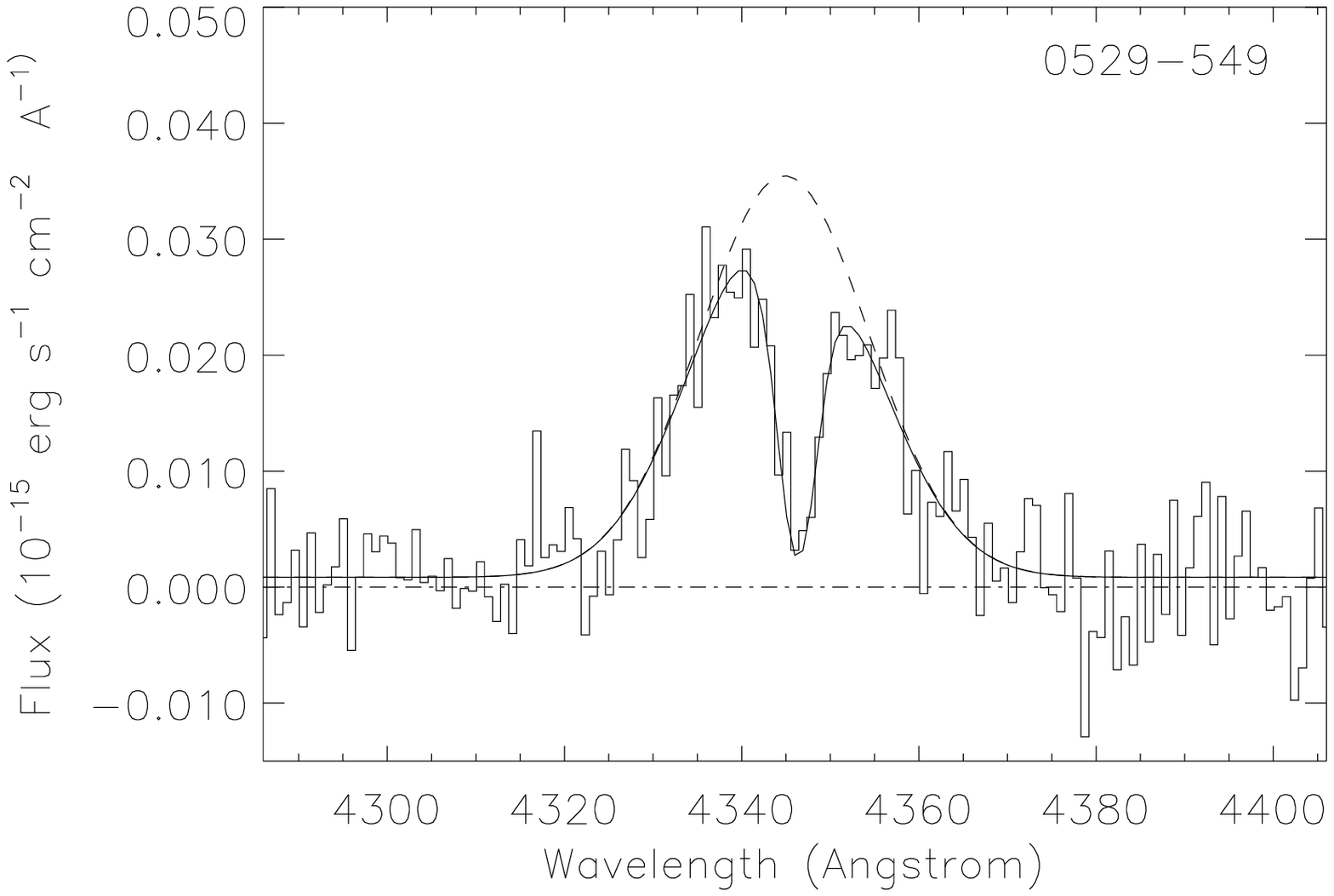,width=8cm,height=7cm}
}
\noindent 
{\bf Fig. 4.} Profiles of the Ly$\alpha$ emission lines.
The solid line is the resultant profile from the absorption model fitting,
and the dashed line is the original Gaussian emission line profile calculated
from the model fitting

\end{figure}

\clearpage \newpage

\begin{figure}[p]
\hbox{
\psfig{figure=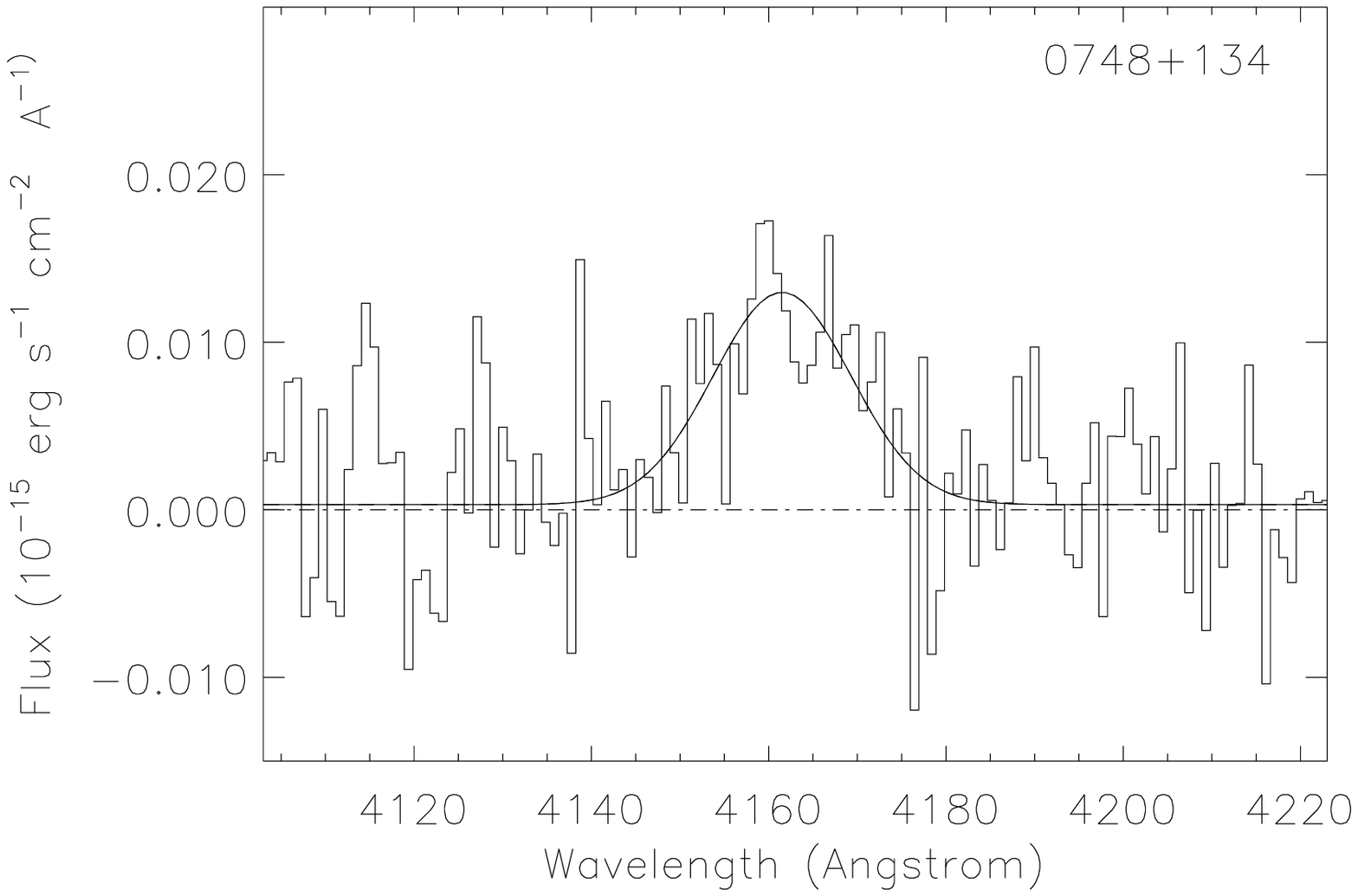,width=8cm,height=7cm}
\psfig{figure=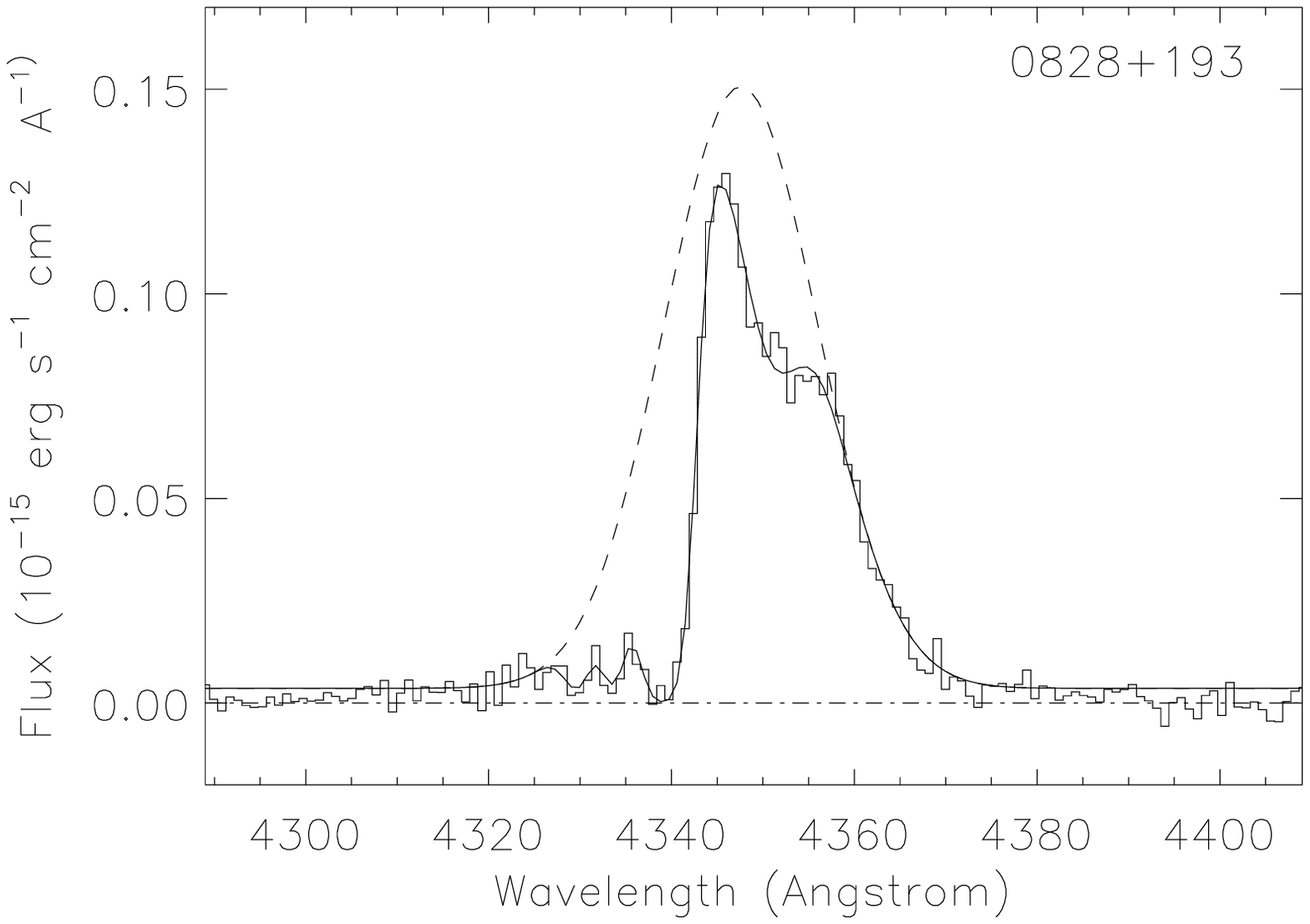,width=8cm,height=7cm}
}
\hbox{
\psfig{figure=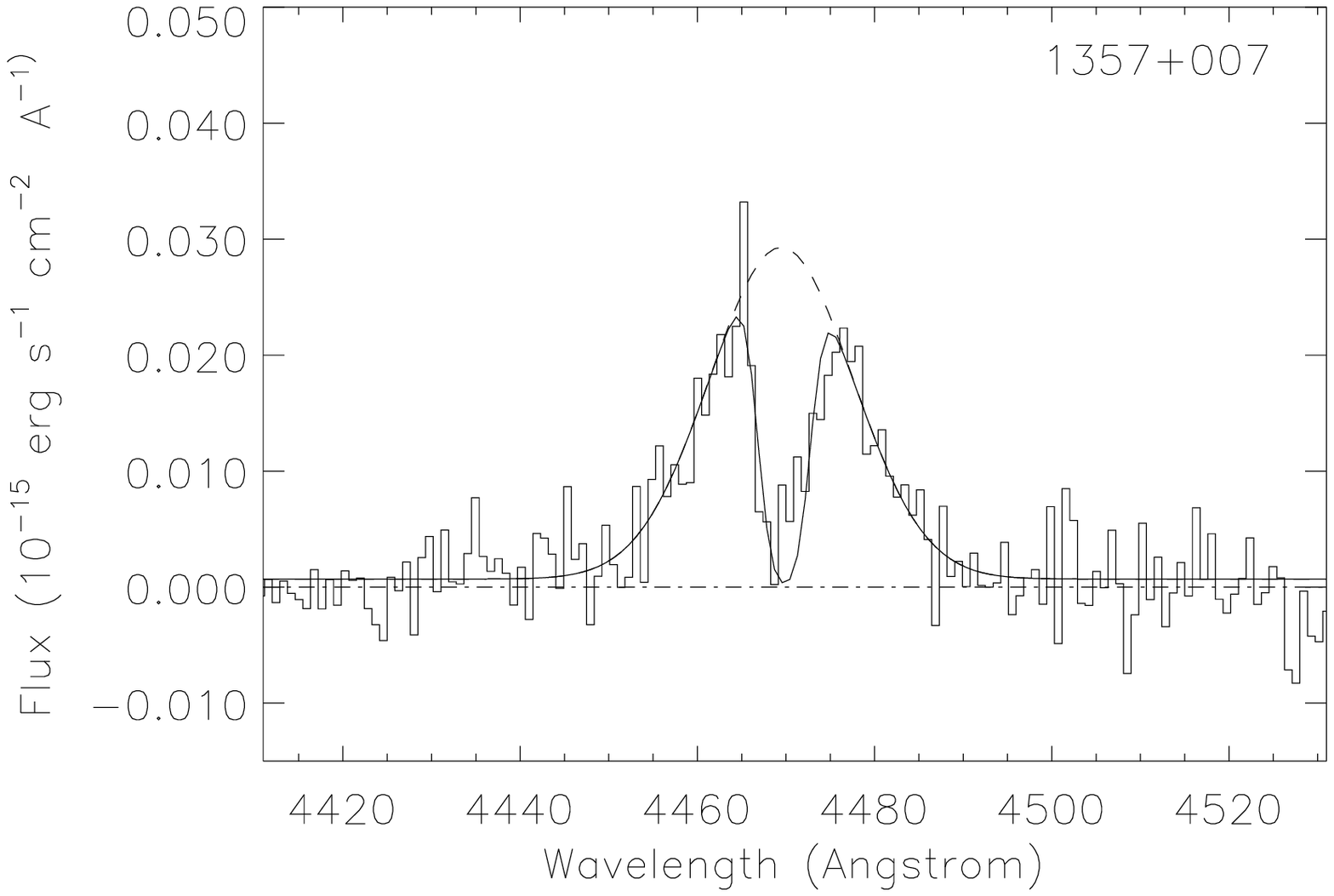,width=8cm,height=7cm}
\psfig{figure=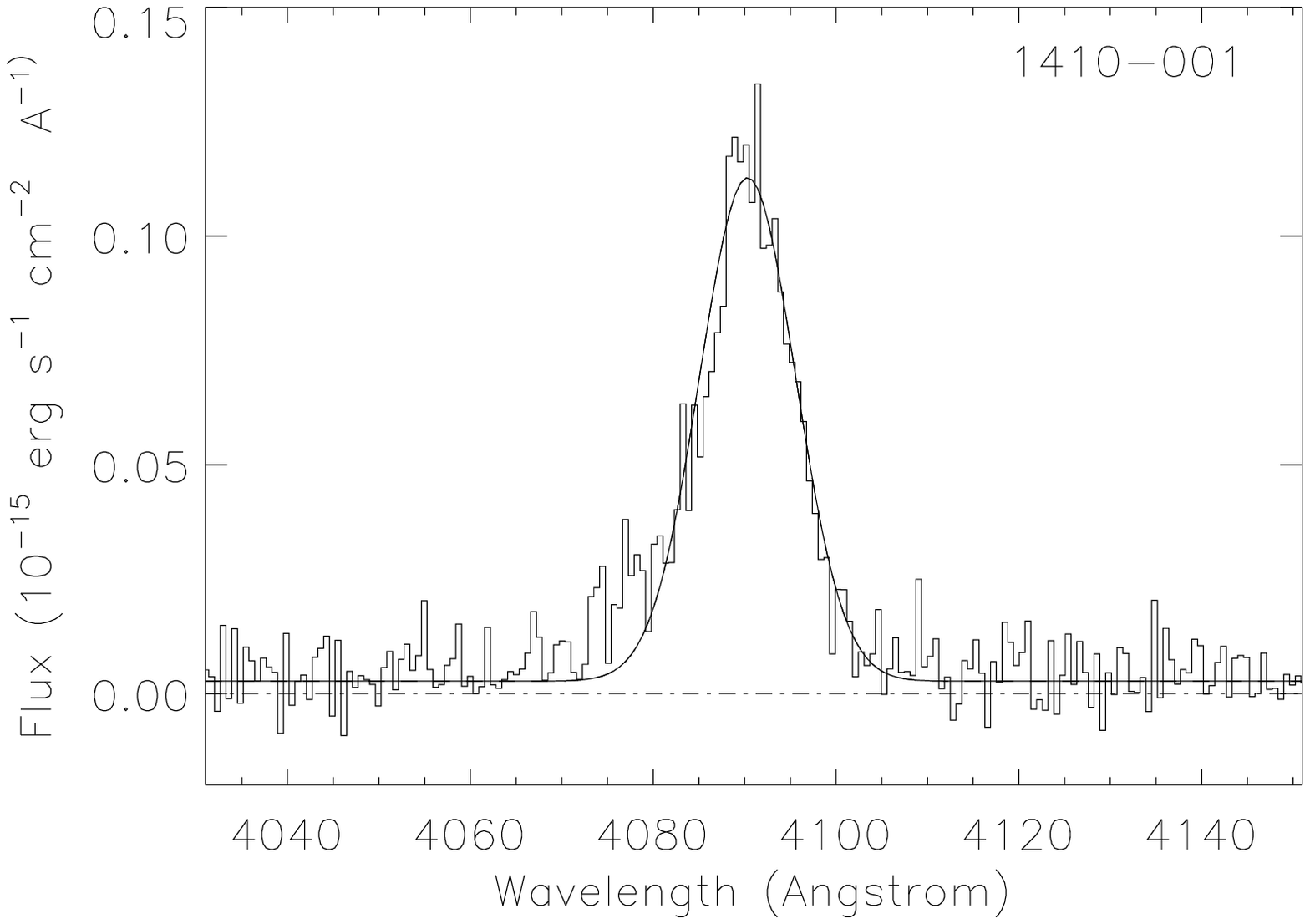,width=8cm,height=7cm}
}
\hbox{
\psfig{figure=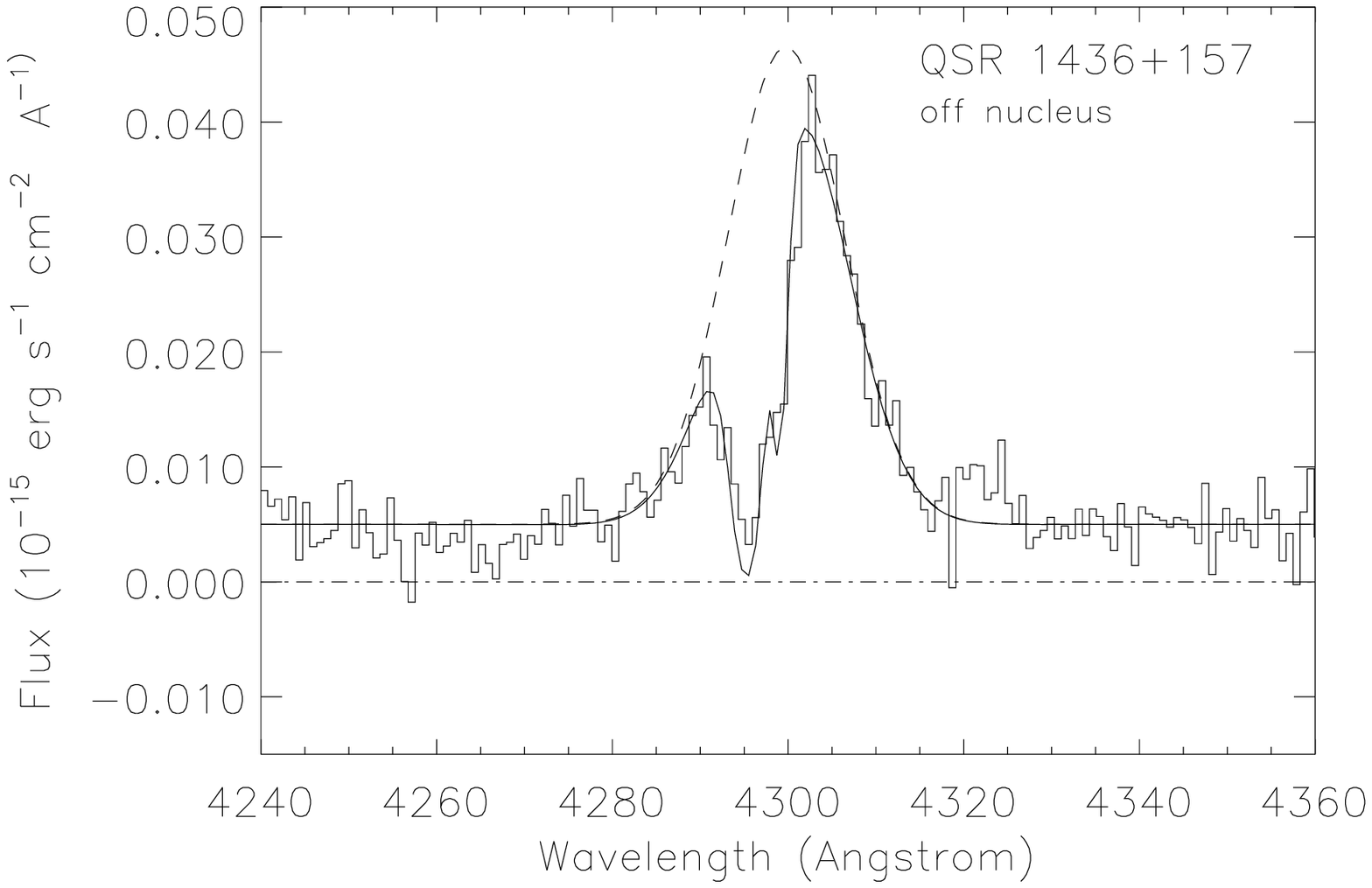,width=8cm,height=7cm}
\psfig{figure=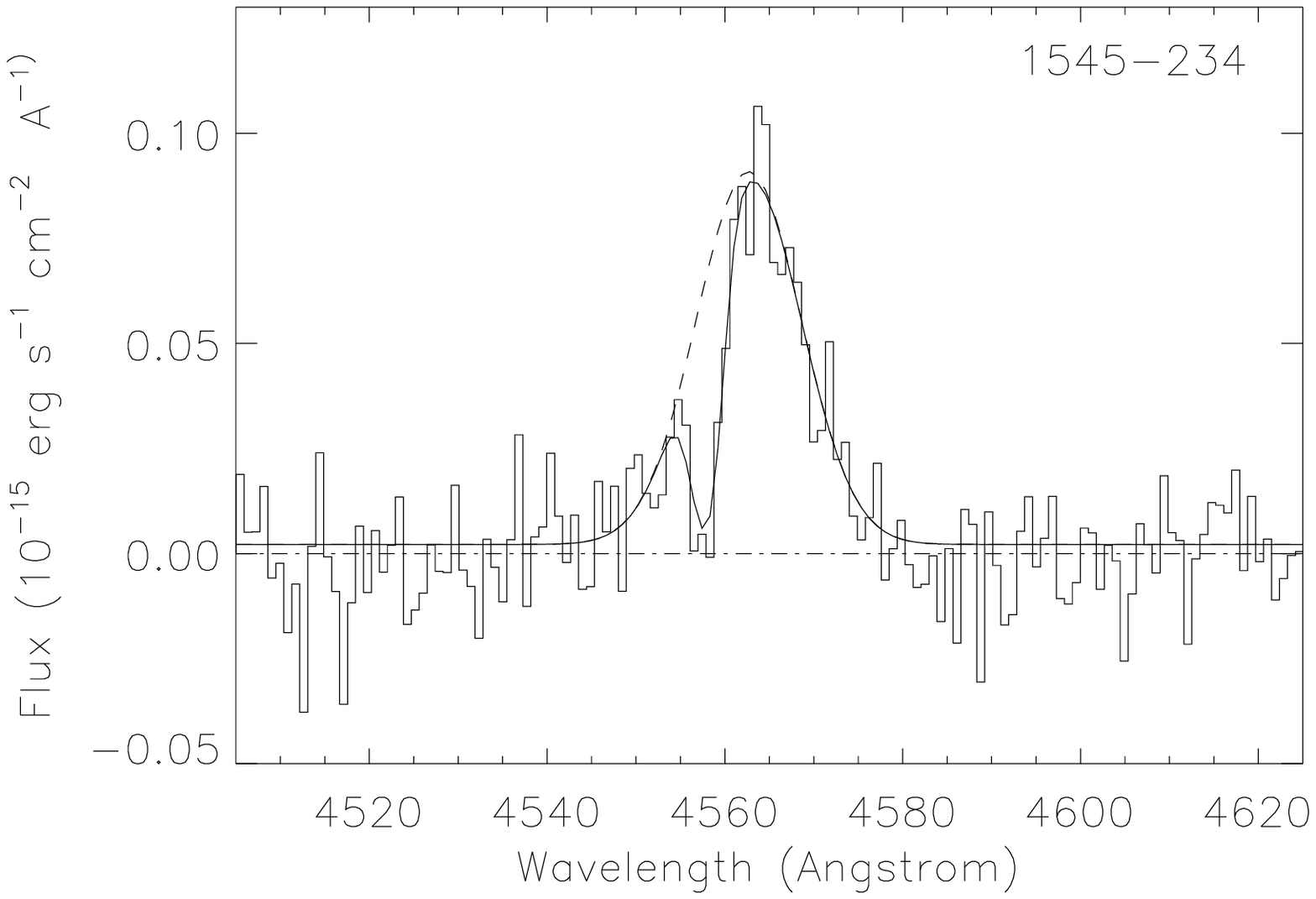,width=8cm,height=7cm}
}
\noindent {\bf Fig. 4.} -- continued --.
\end{figure}

\clearpage \newpage

\begin{figure}[t]
\hbox{
\psfig{figure=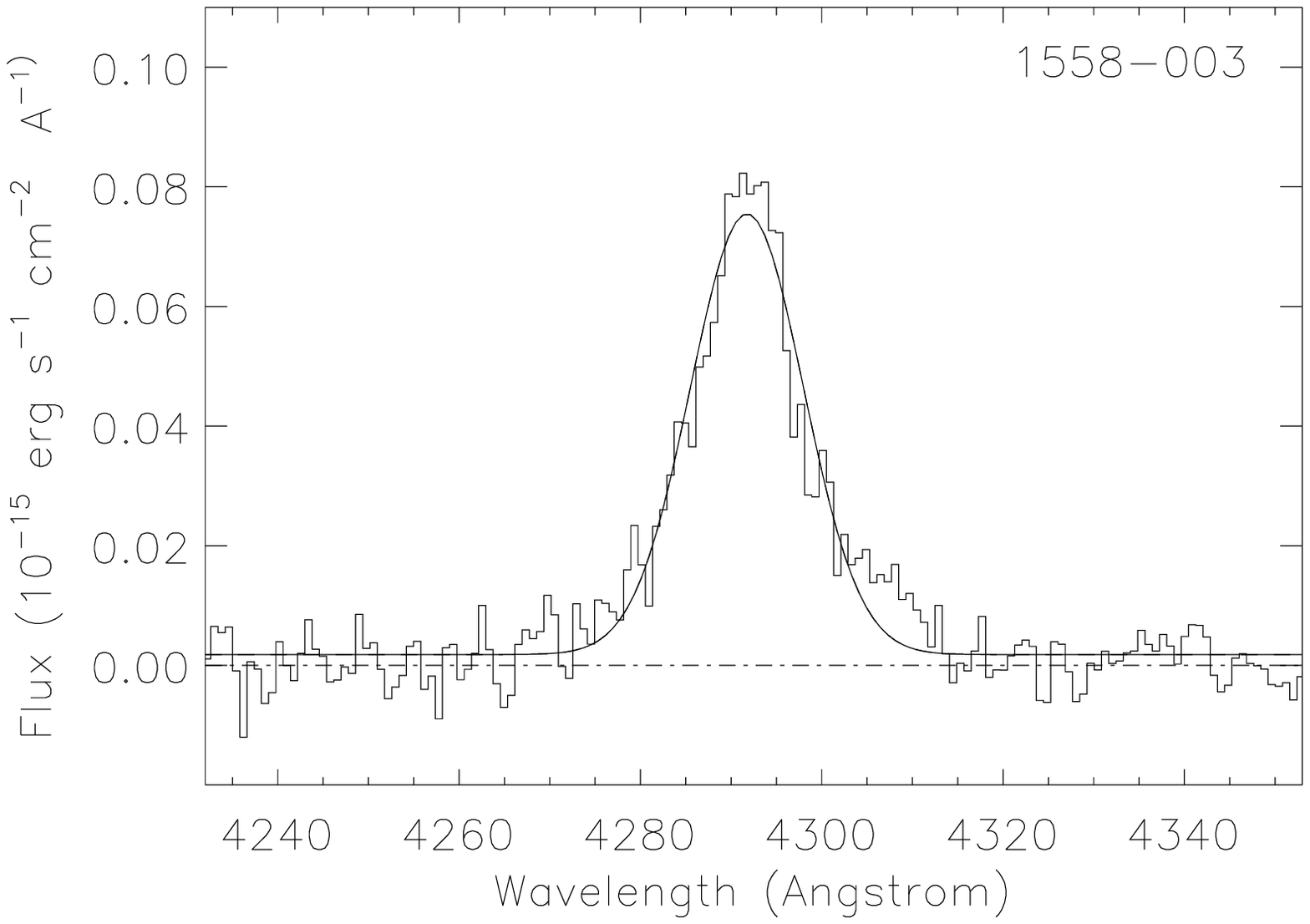,width=8cm,height=7cm}
\psfig{figure=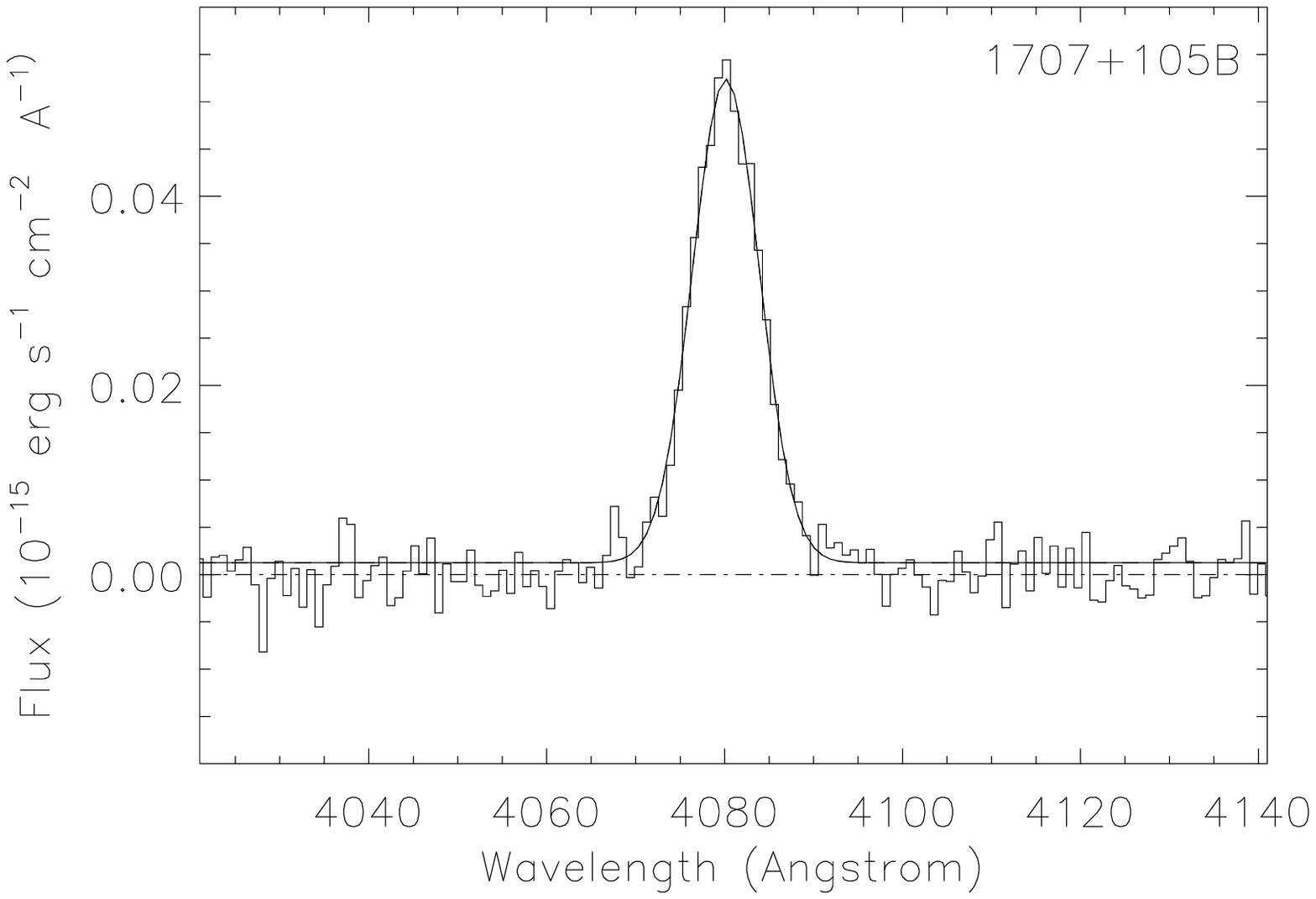,width=8cm,height=7cm}
}
\hbox{
\psfig{figure=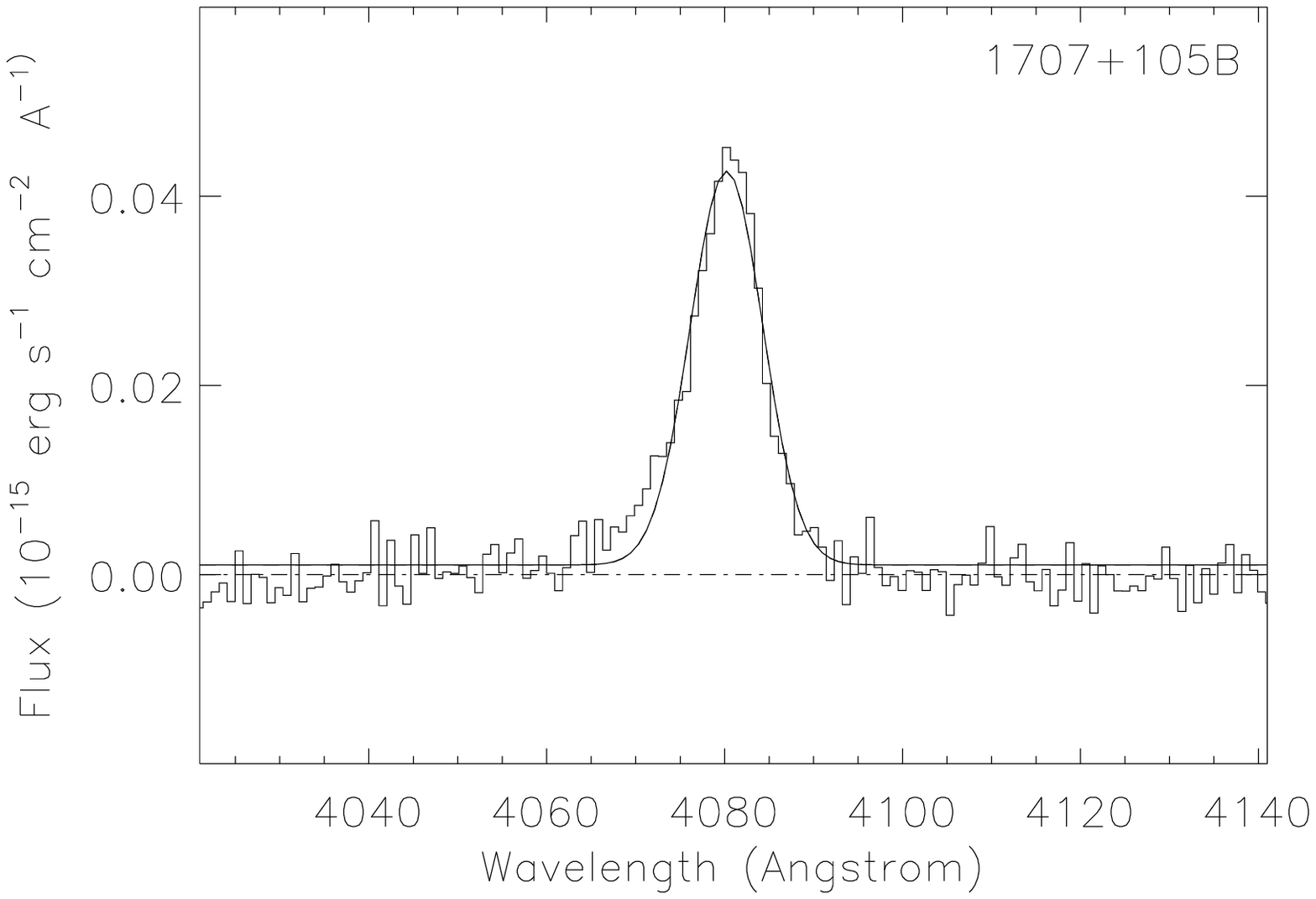,width=8cm,height=7cm}
\psfig{figure=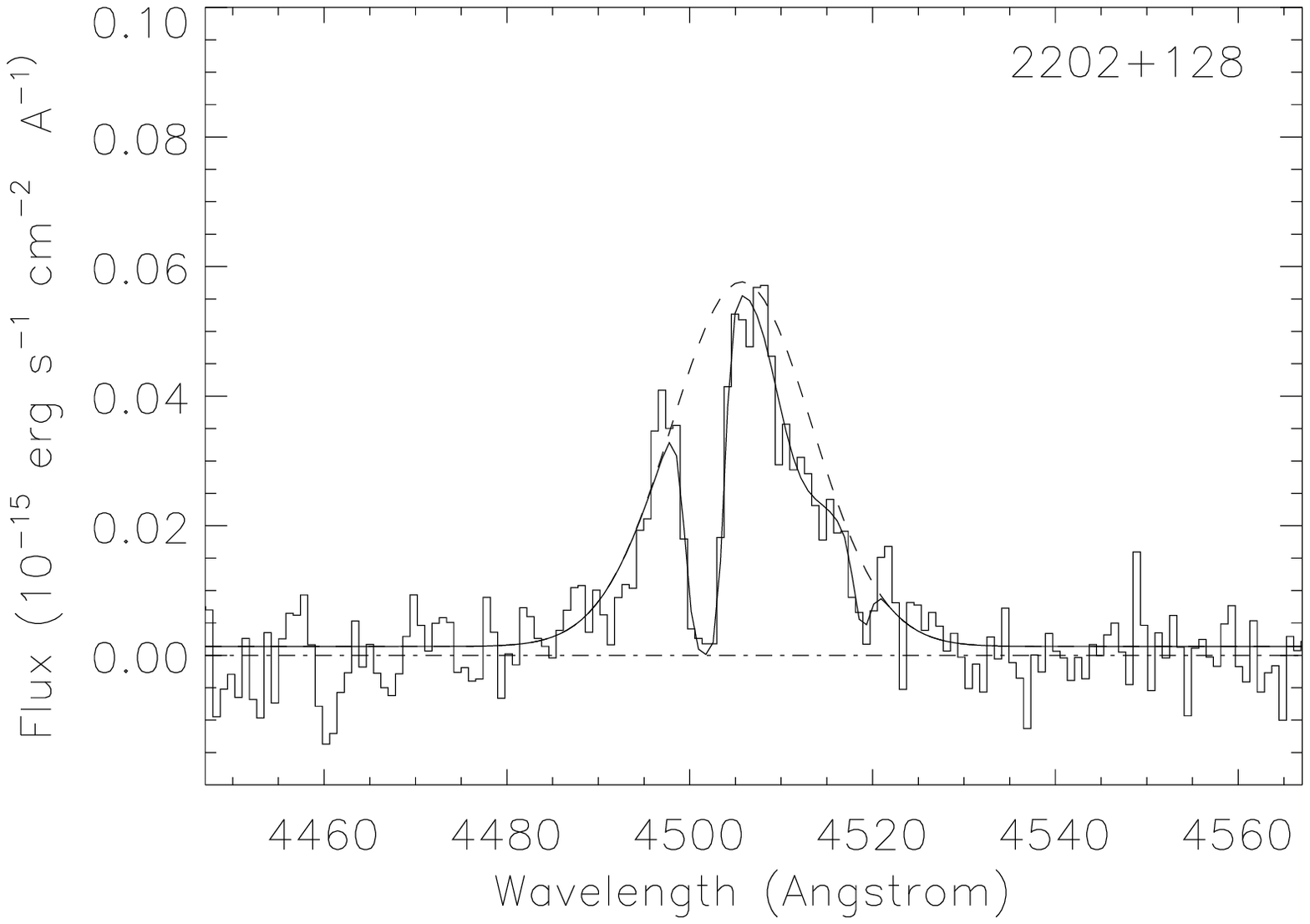,width=8cm,height=7cm}
}
\noindent {\bf Fig. 4.} -- continued --.
\end{figure}

\clearpage \newpage

\begin{figure}[p]
\hbox{
\psfig{figure=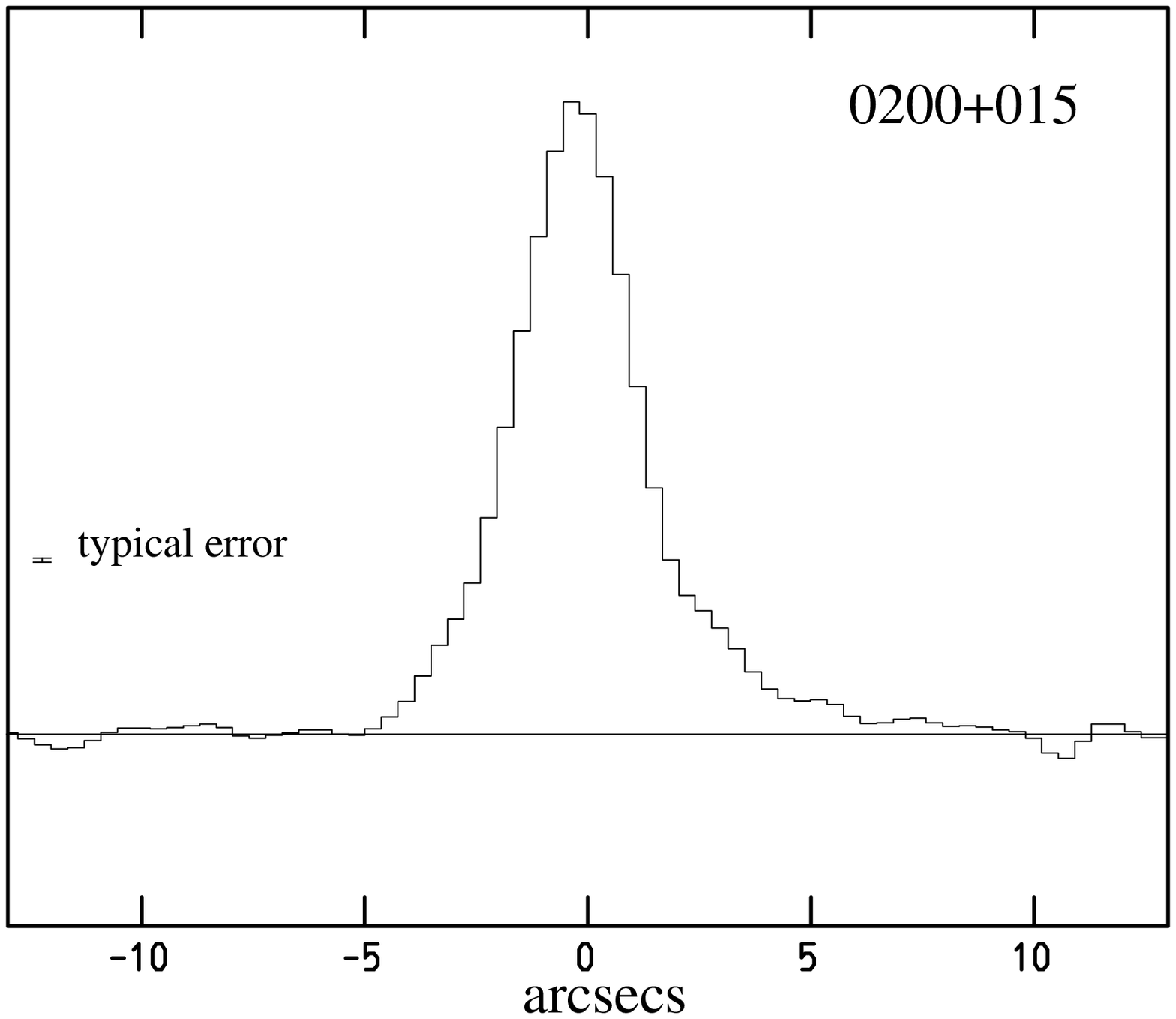,angle=90,width=8cm,height=7cm}
\psfig{figure=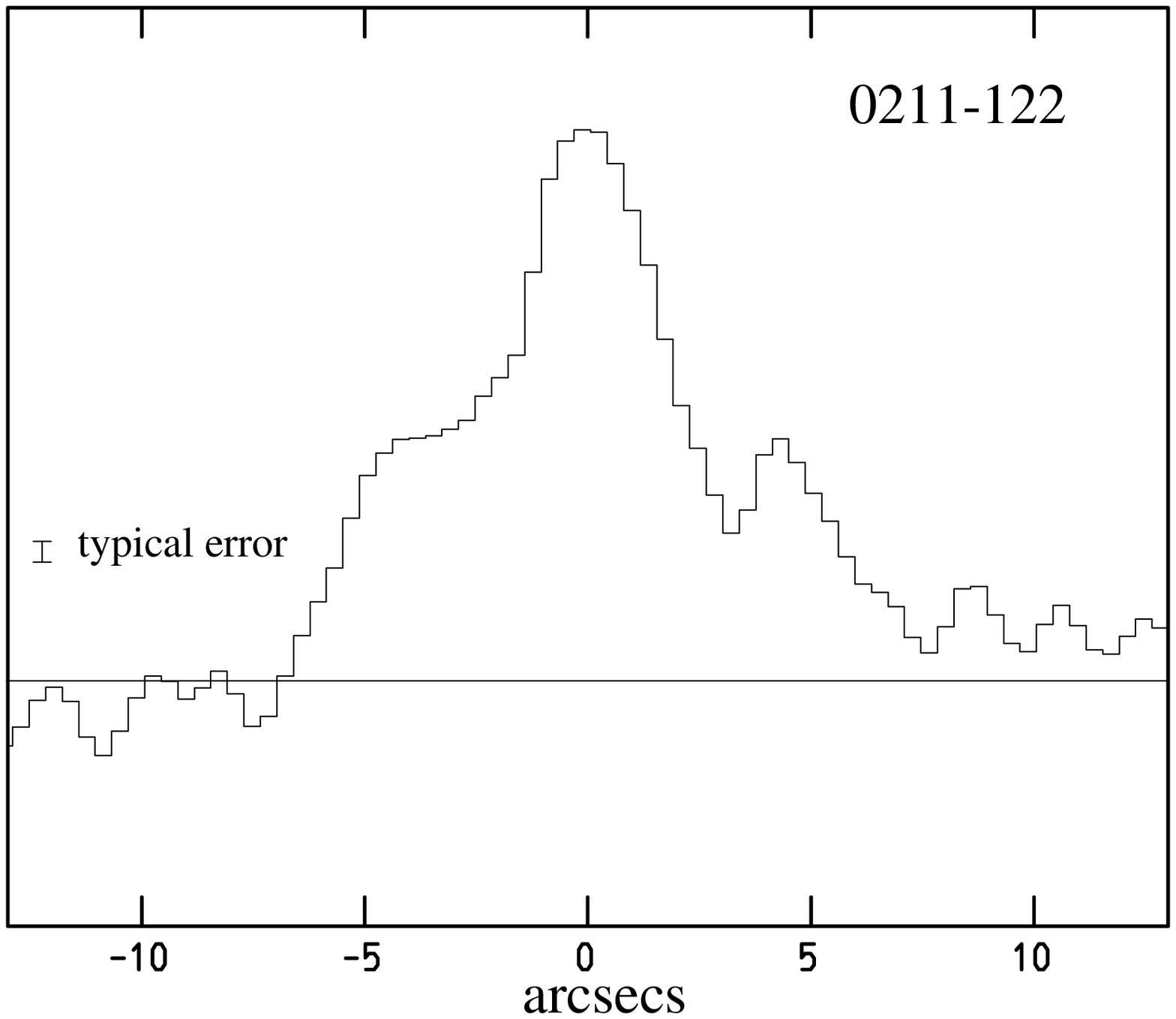,angle=90,width=8cm,height=7cm}
}
\hbox{
\psfig{figure=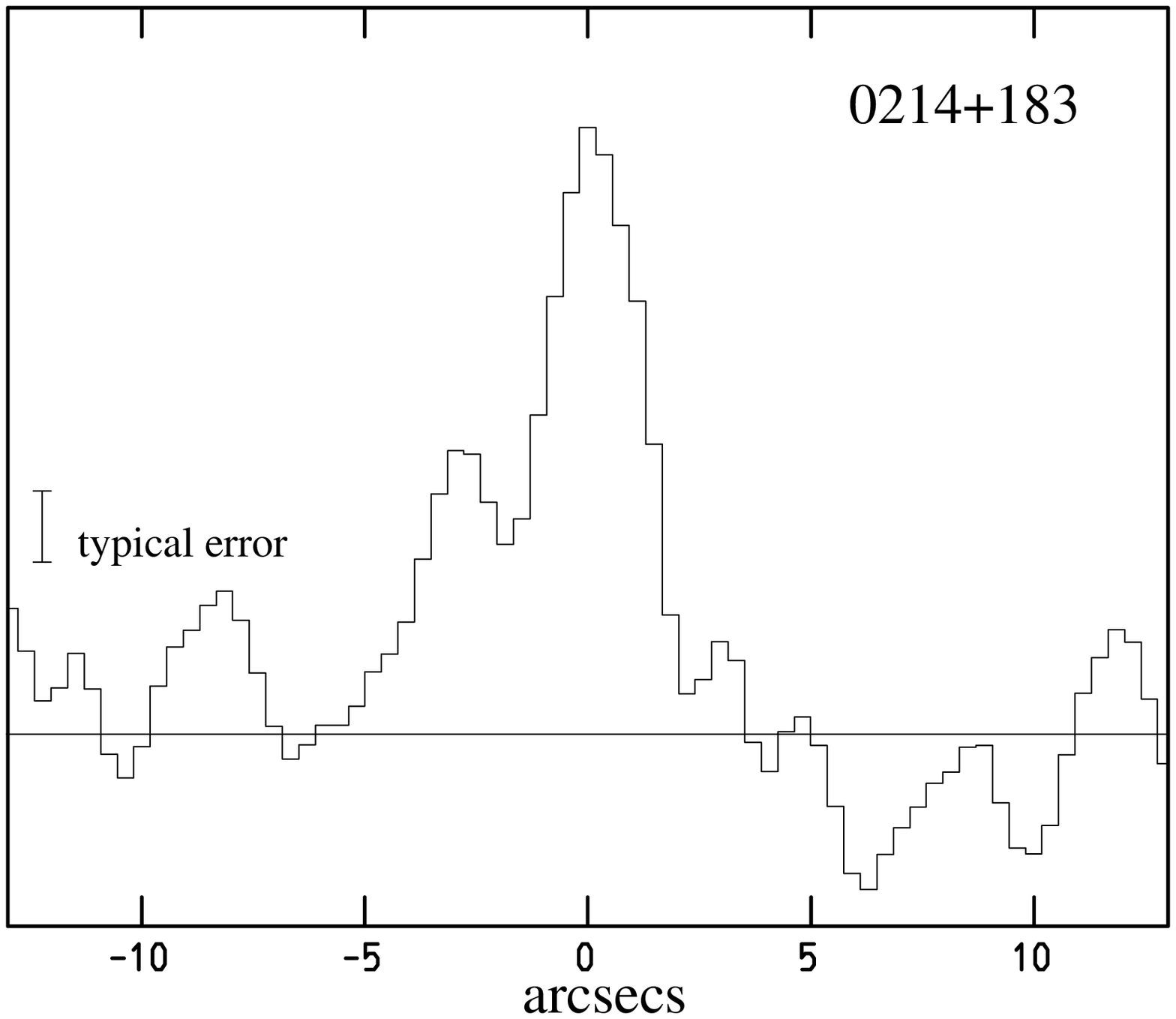,angle=90,width=8cm,height=7cm}
\psfig{figure=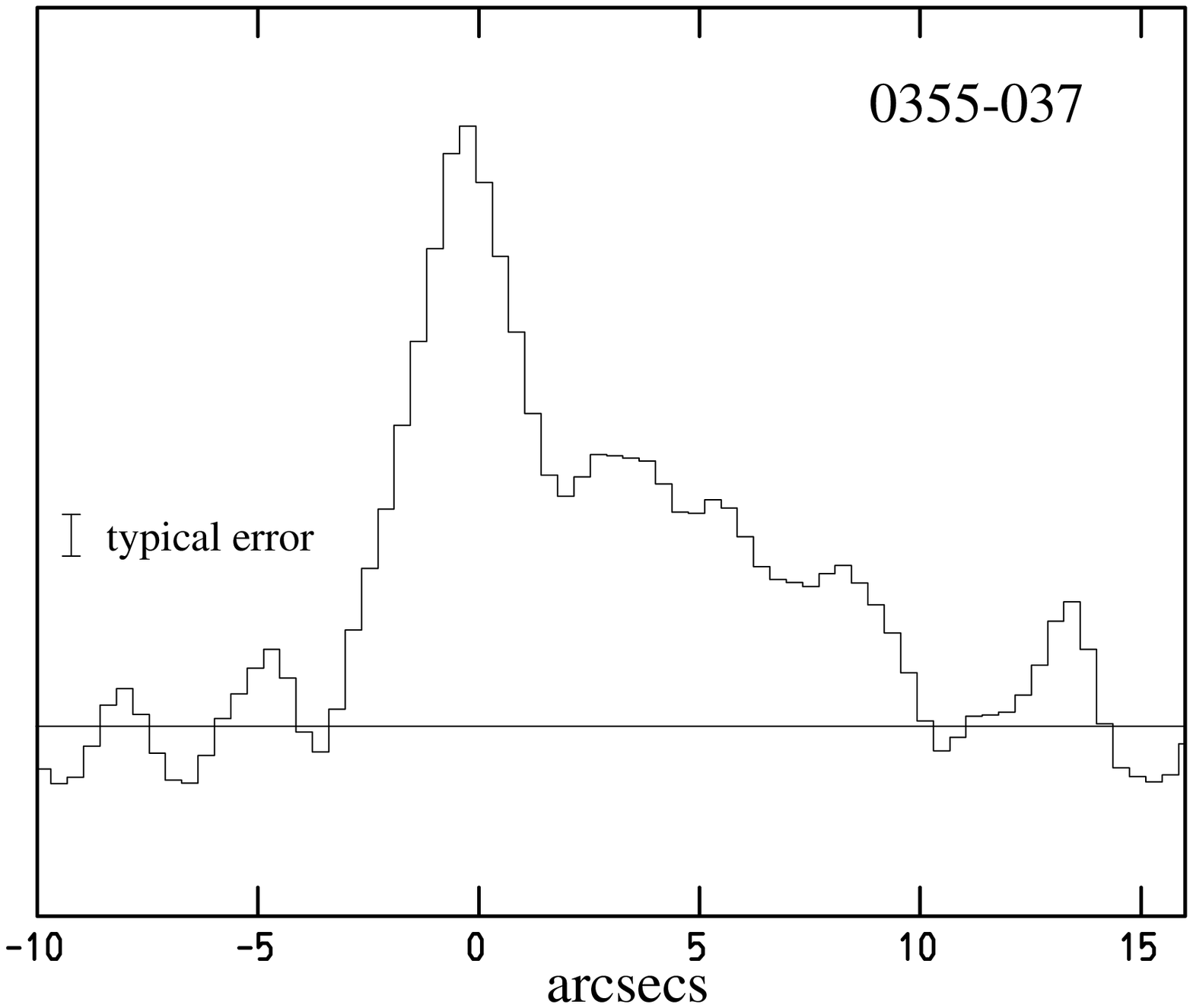,angle=90,width=8cm,height=7cm}
}
\hbox{
\psfig{figure=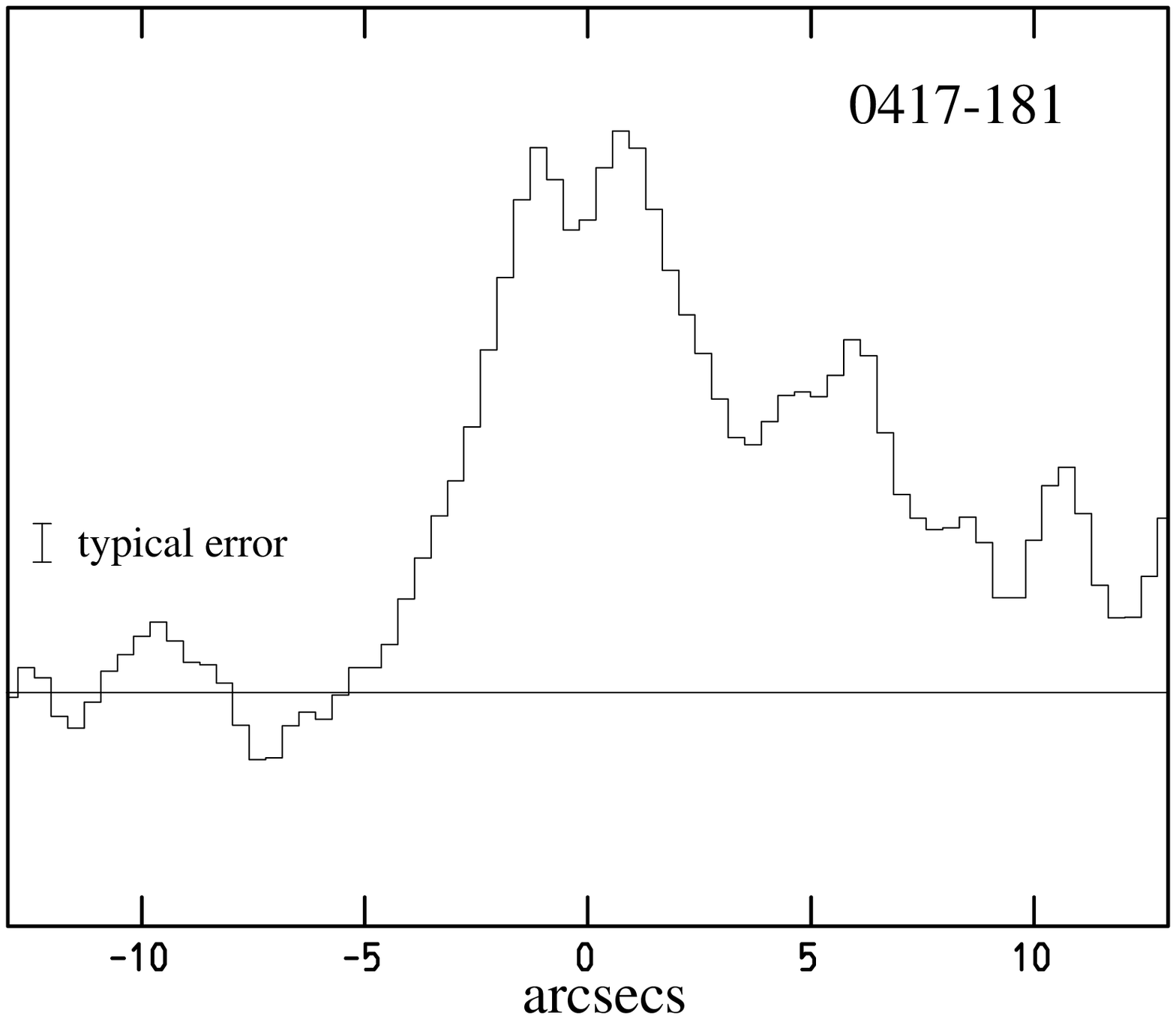,angle=90,width=8cm,height=7cm}
\psfig{figure=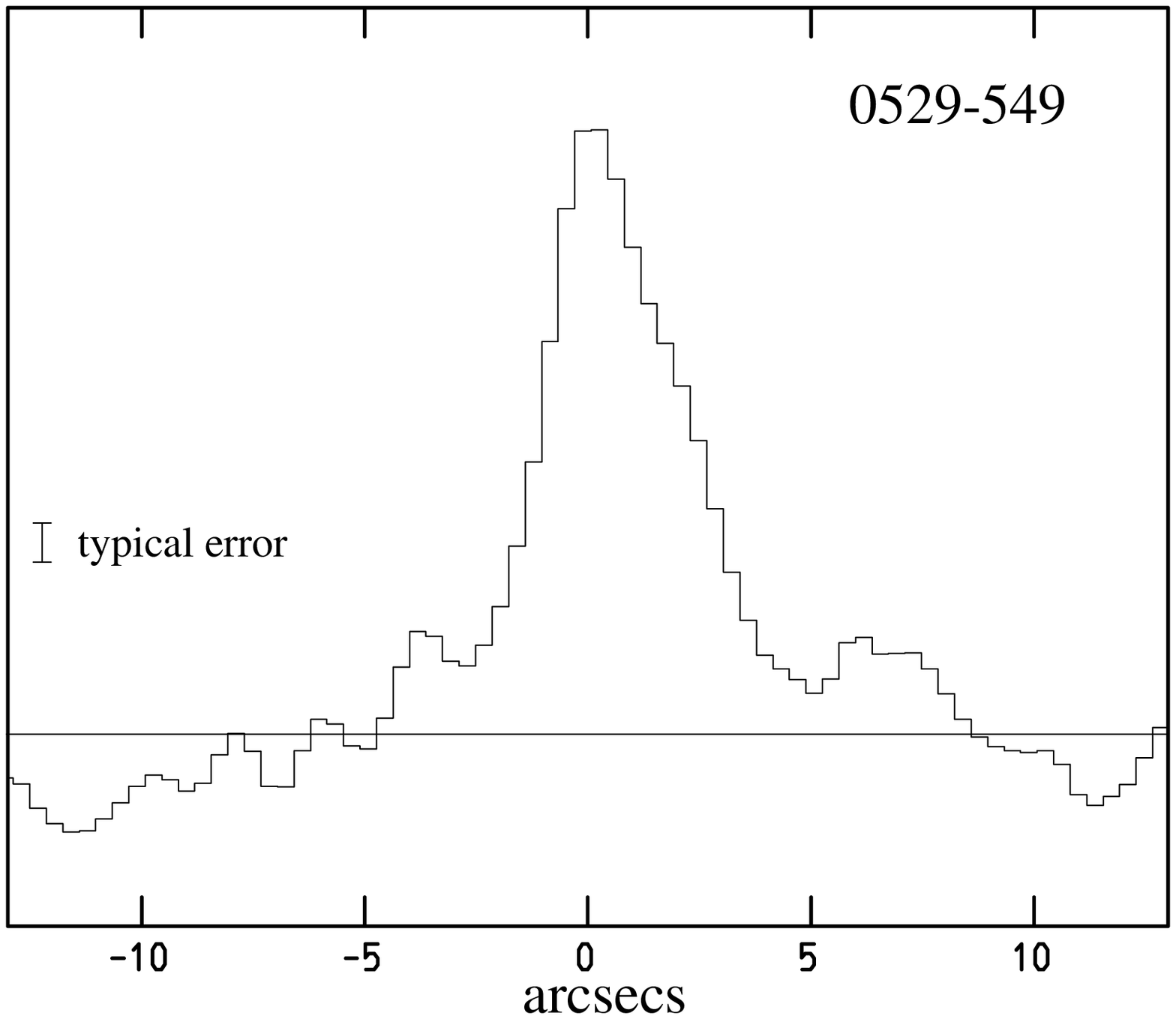,angle=90,width=8cm,height=7cm}
}

\noindent {\bf Fig. 5.} Spatial profiles of the Ly$\alpha$ 
emission, summed
over the full spectral range where Ly$\alpha$ emission was detected.
The profiles have been smoothed with a Gaussian function of $1''$ FWHM.
The plots have an arbitrary y-axis scale. Also indicated is the 1$\sigma$
error determined from the RMS background noise

\end{figure}

\clearpage \newpage

\begin{figure}[p]
\hbox{
\psfig{figure=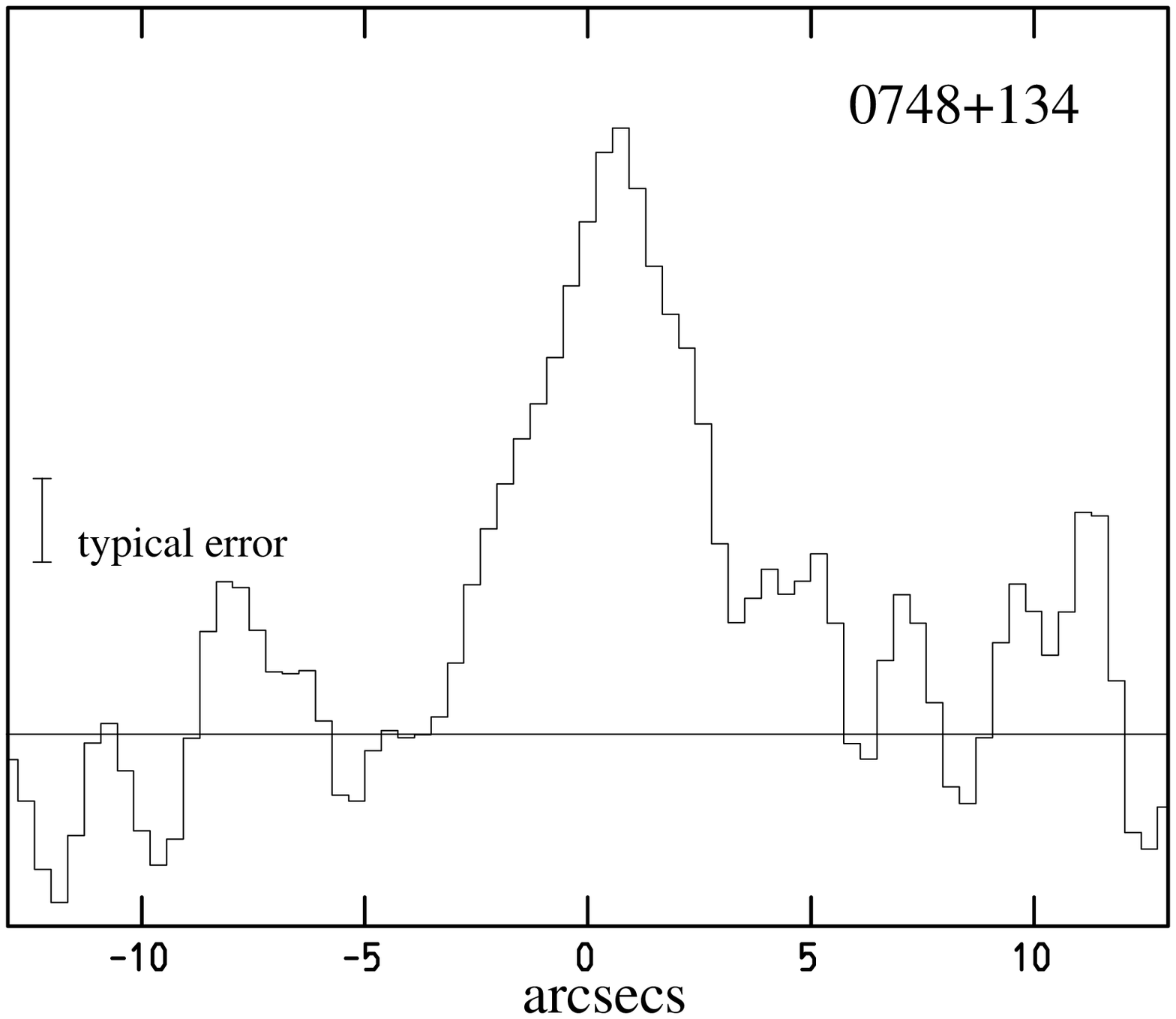,angle=90,width=8cm,height=7cm}
\psfig{figure=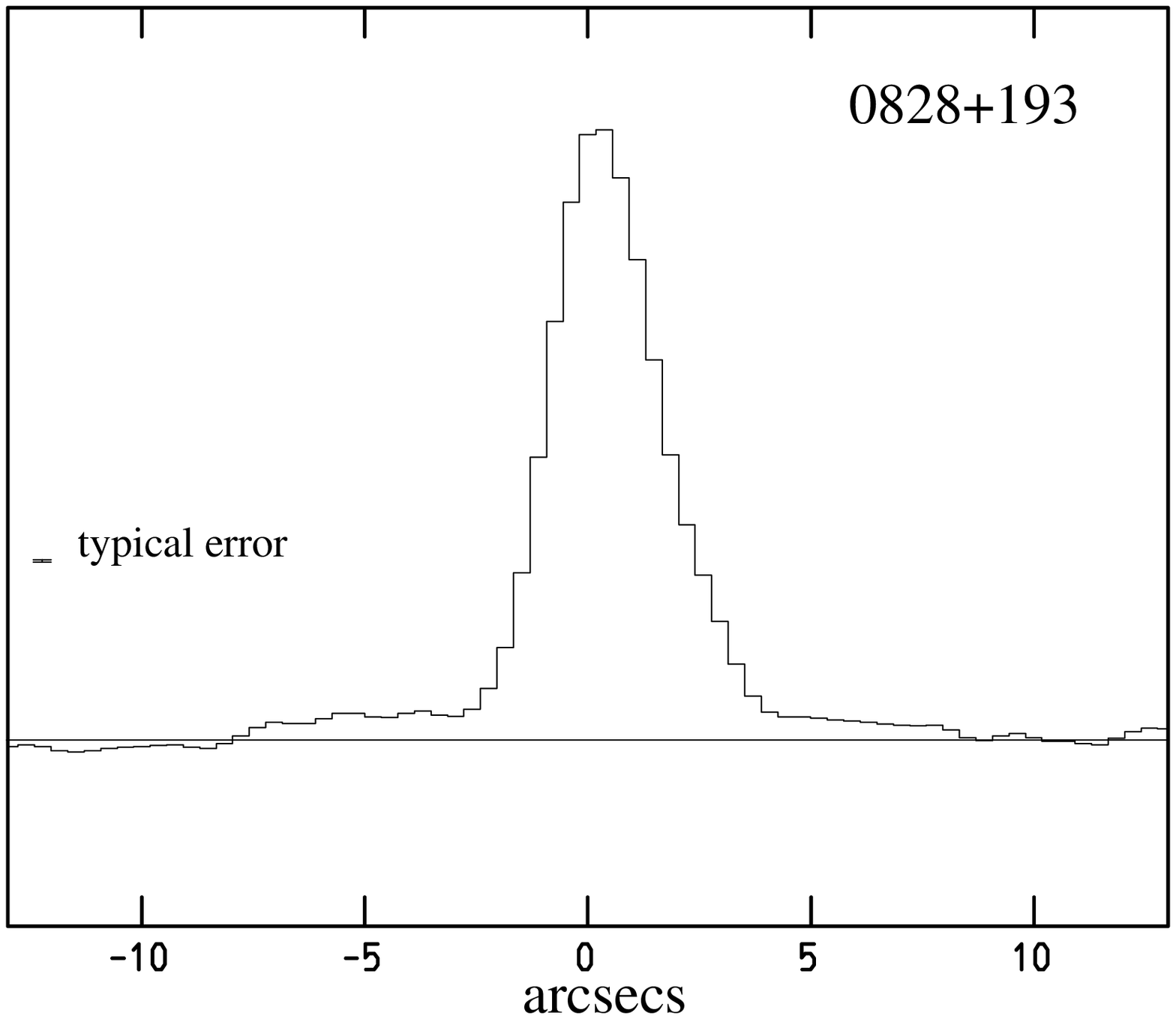,angle=90,width=8cm,height=7cm}
}
\hbox{
\psfig{figure=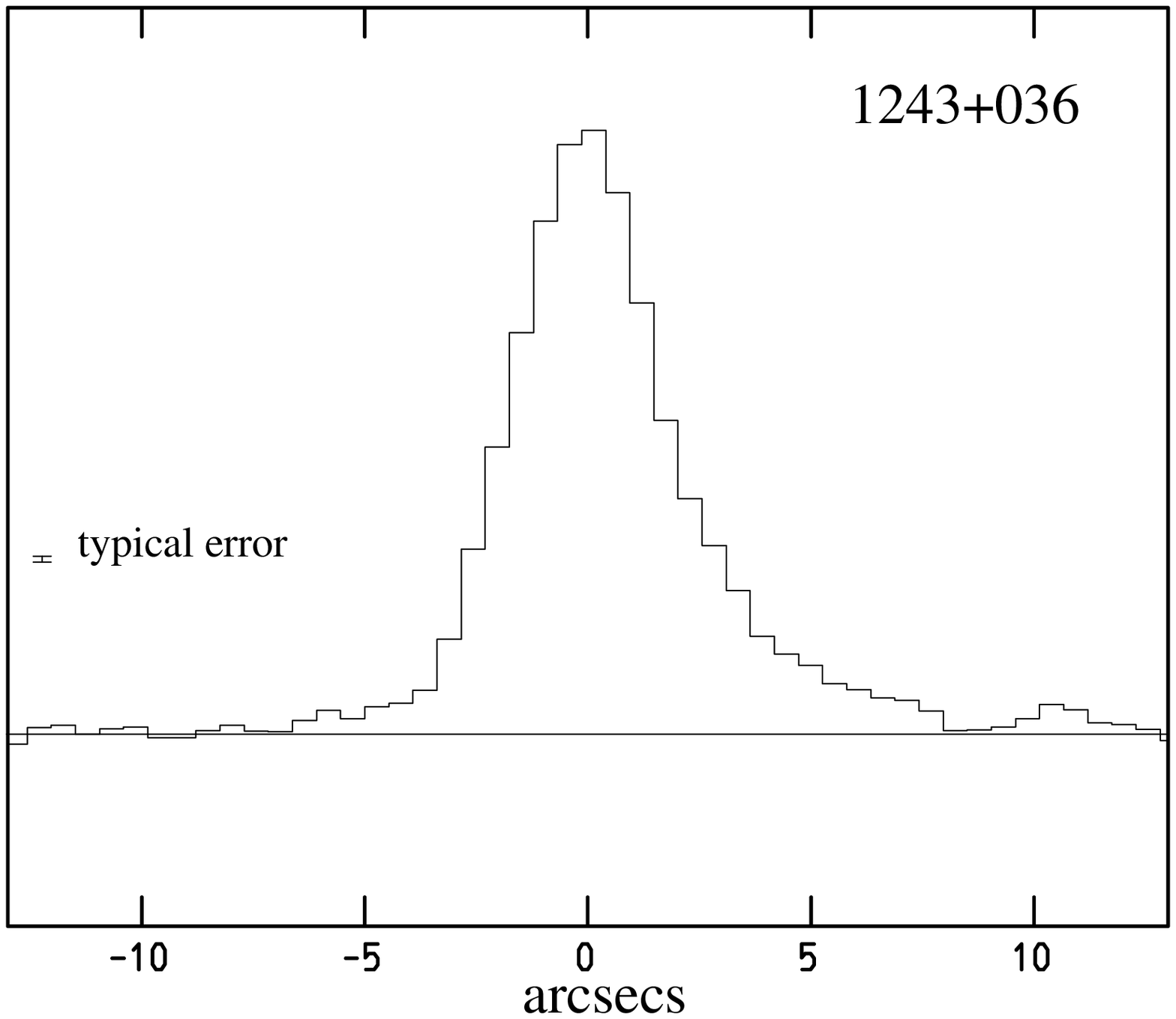,angle=90,width=8cm,height=7cm}
\psfig{figure=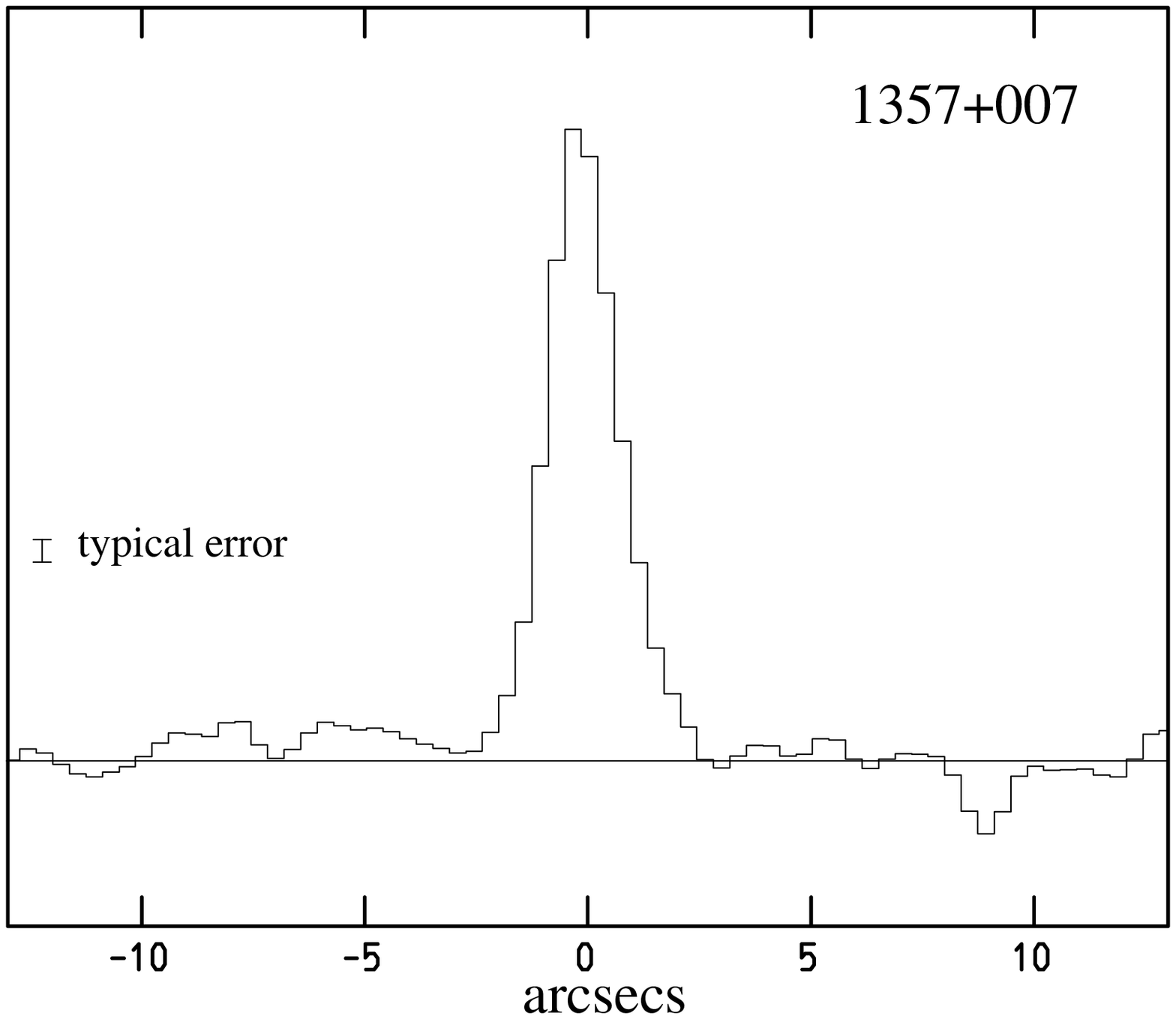,angle=90,width=8cm,height=7cm}
}
\hbox{
\psfig{figure=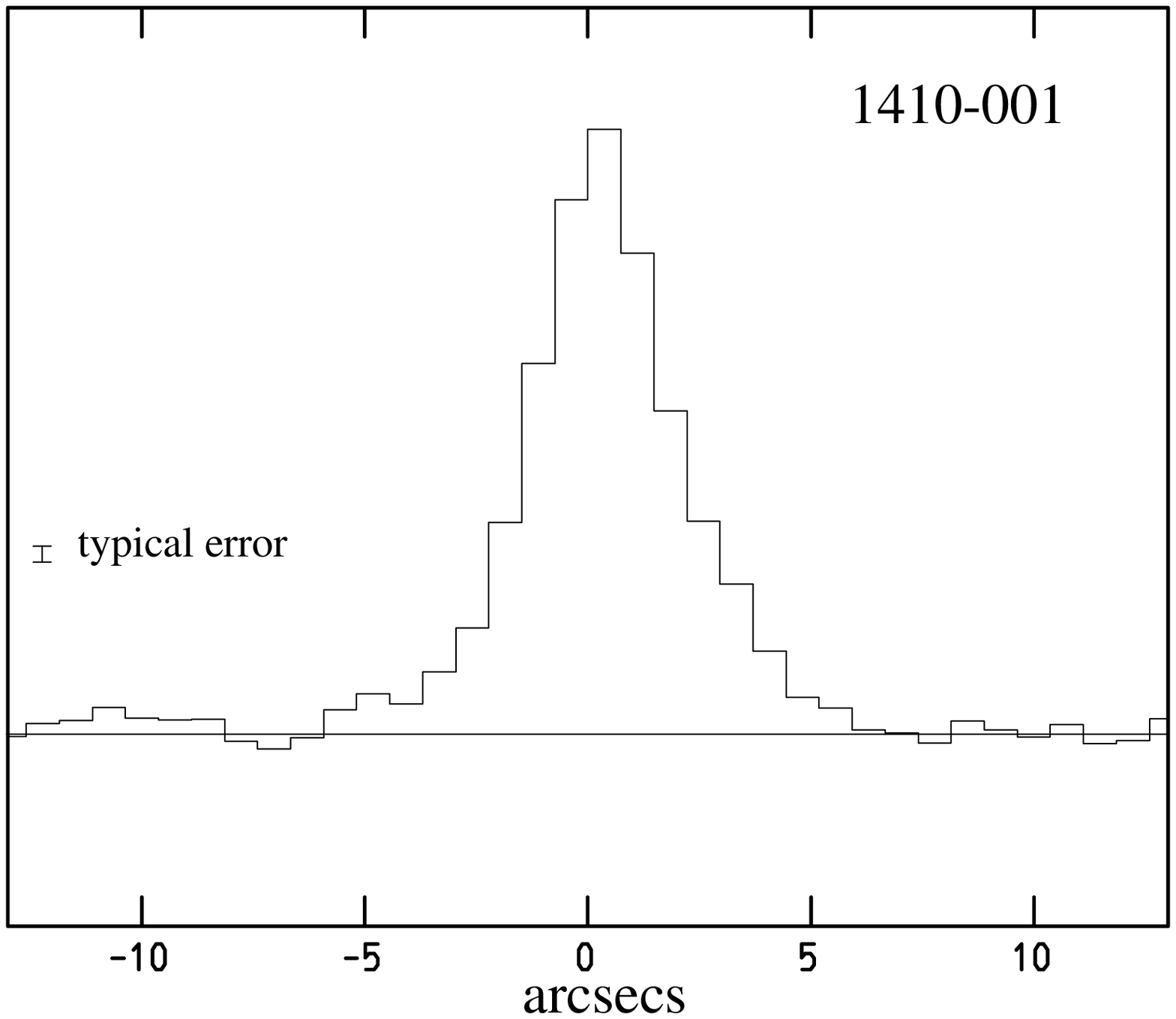,angle=90,width=8cm,height=7cm}
\psfig{figure=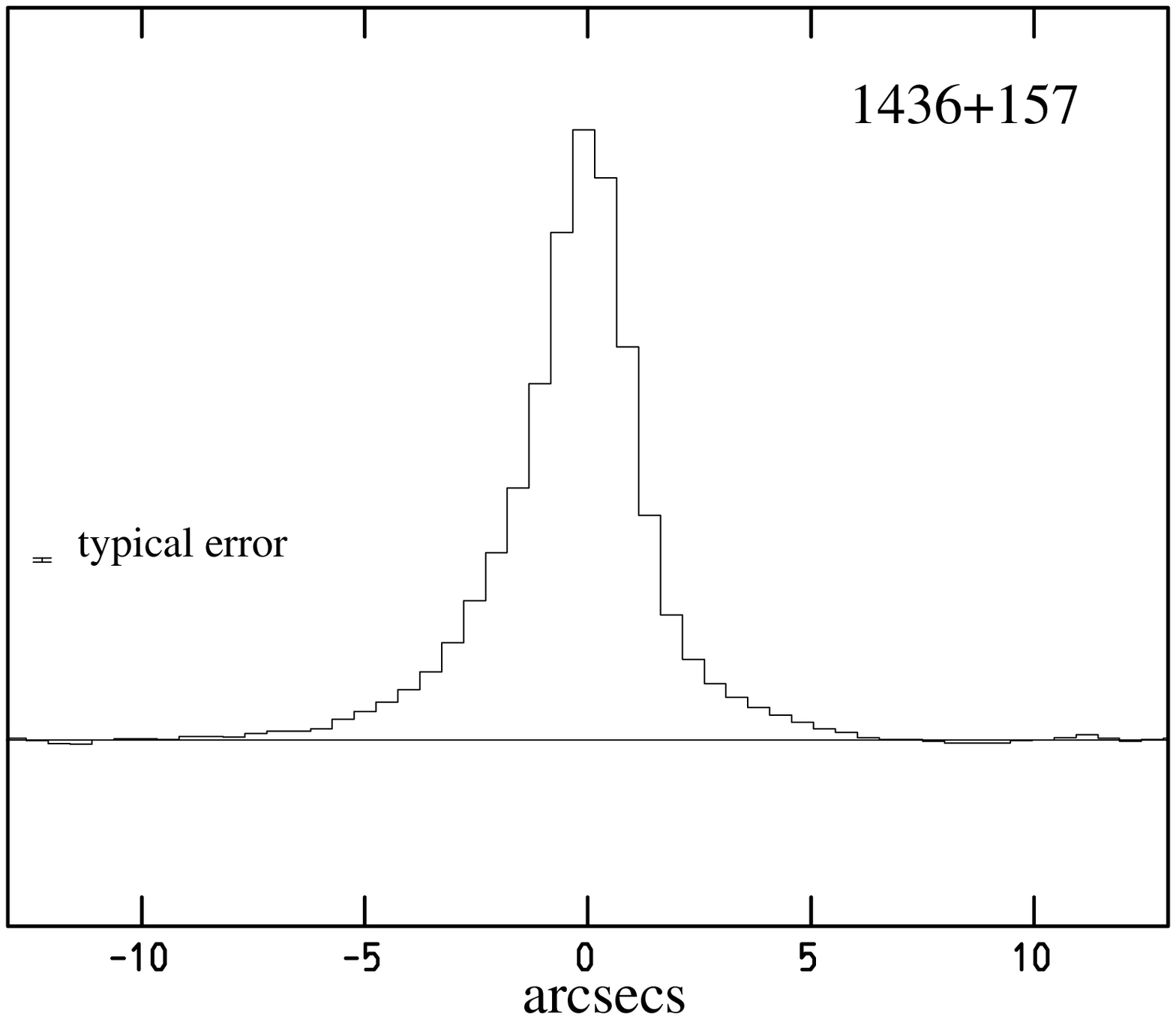,angle=90,width=8cm,height=7cm}
}

\noindent {\bf Fig. 5.} -- continued --.
\end{figure}

\clearpage \newpage

\begin{figure}[ht]
\hbox{
\psfig{figure=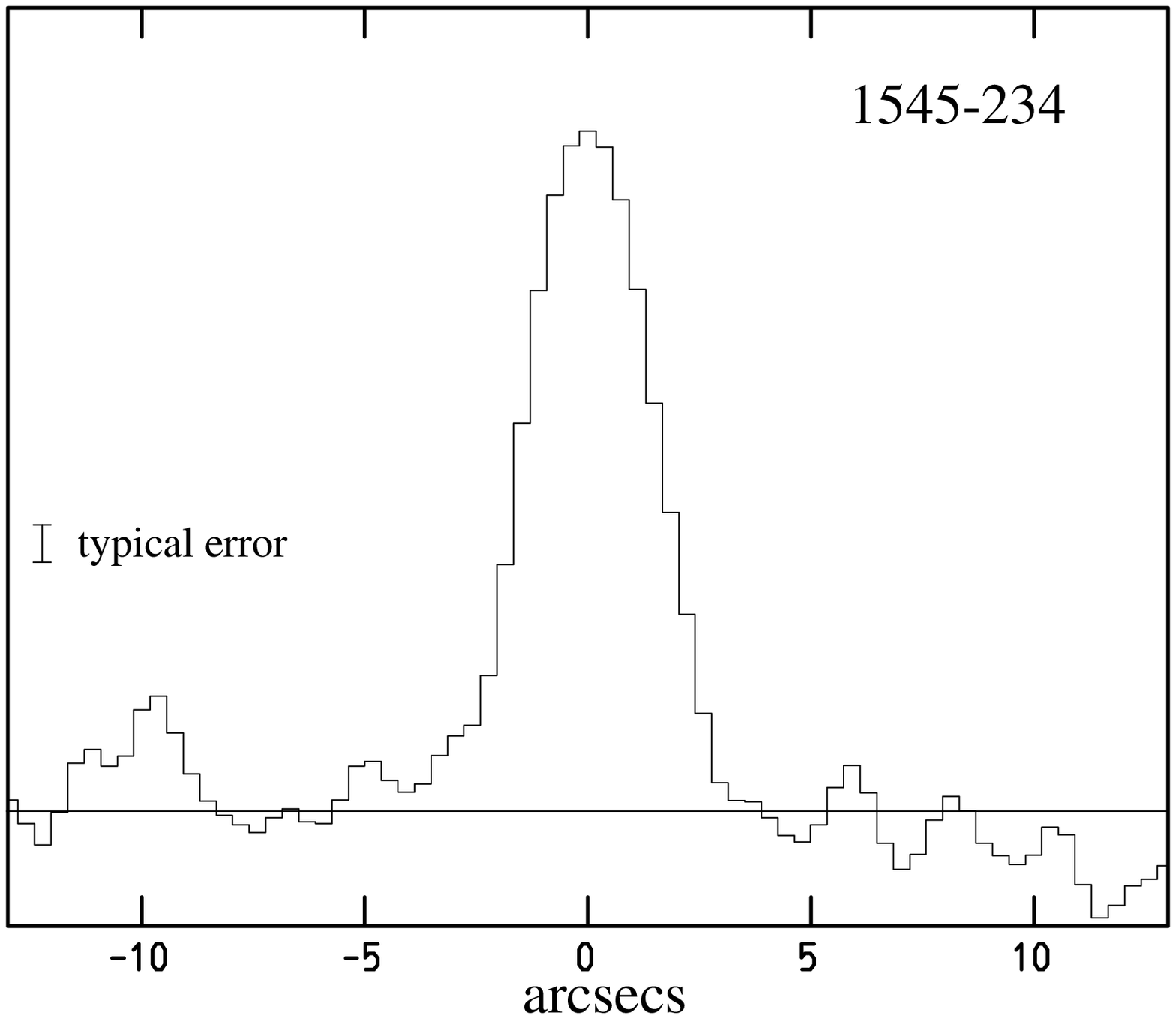,angle=90,width=8cm,height=7cm}
\psfig{figure=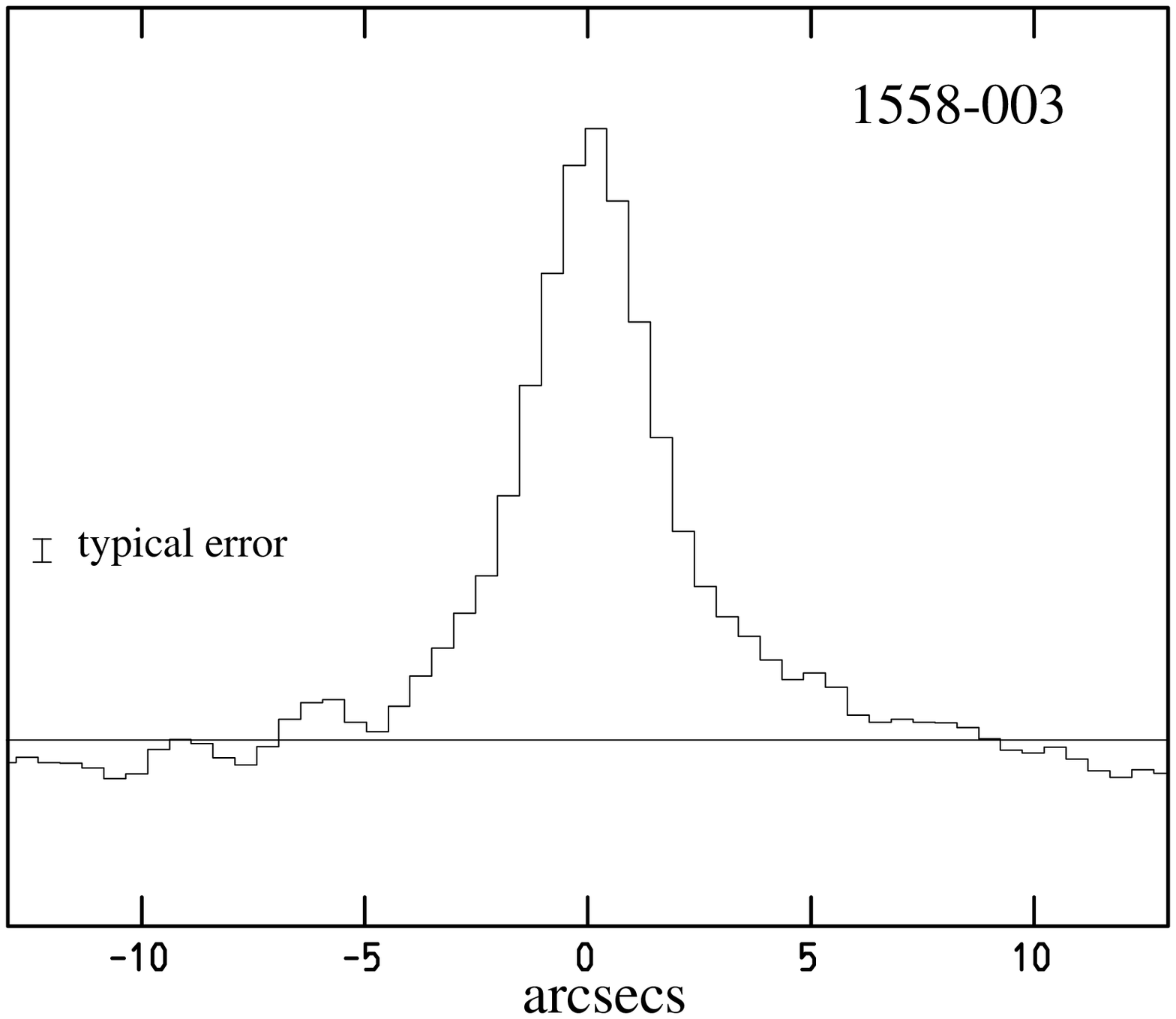,angle=90,width=8cm,height=7cm}
}
\hbox{
\psfig{figure=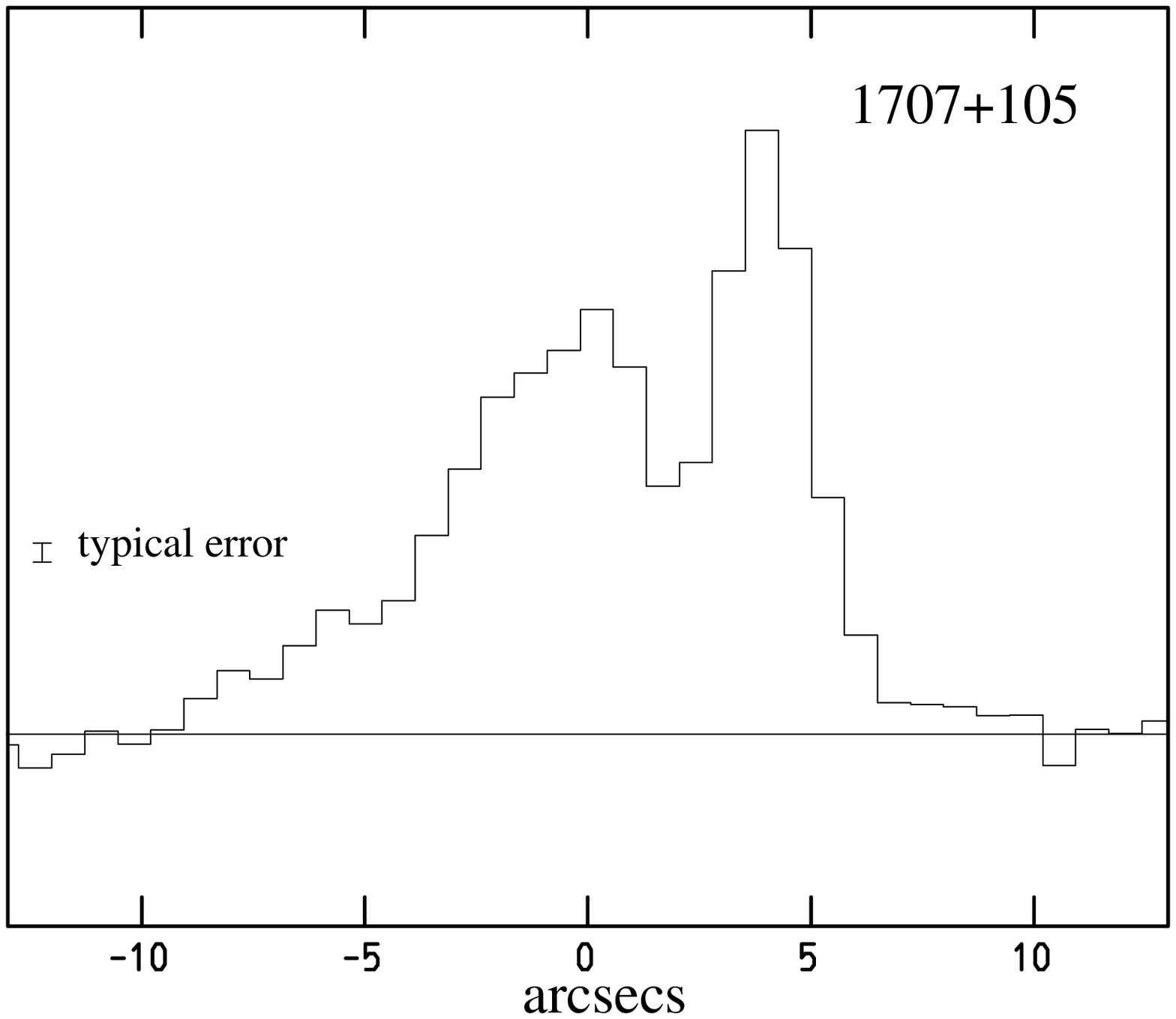,angle=90,width=8cm,height=7cm}
\psfig{figure=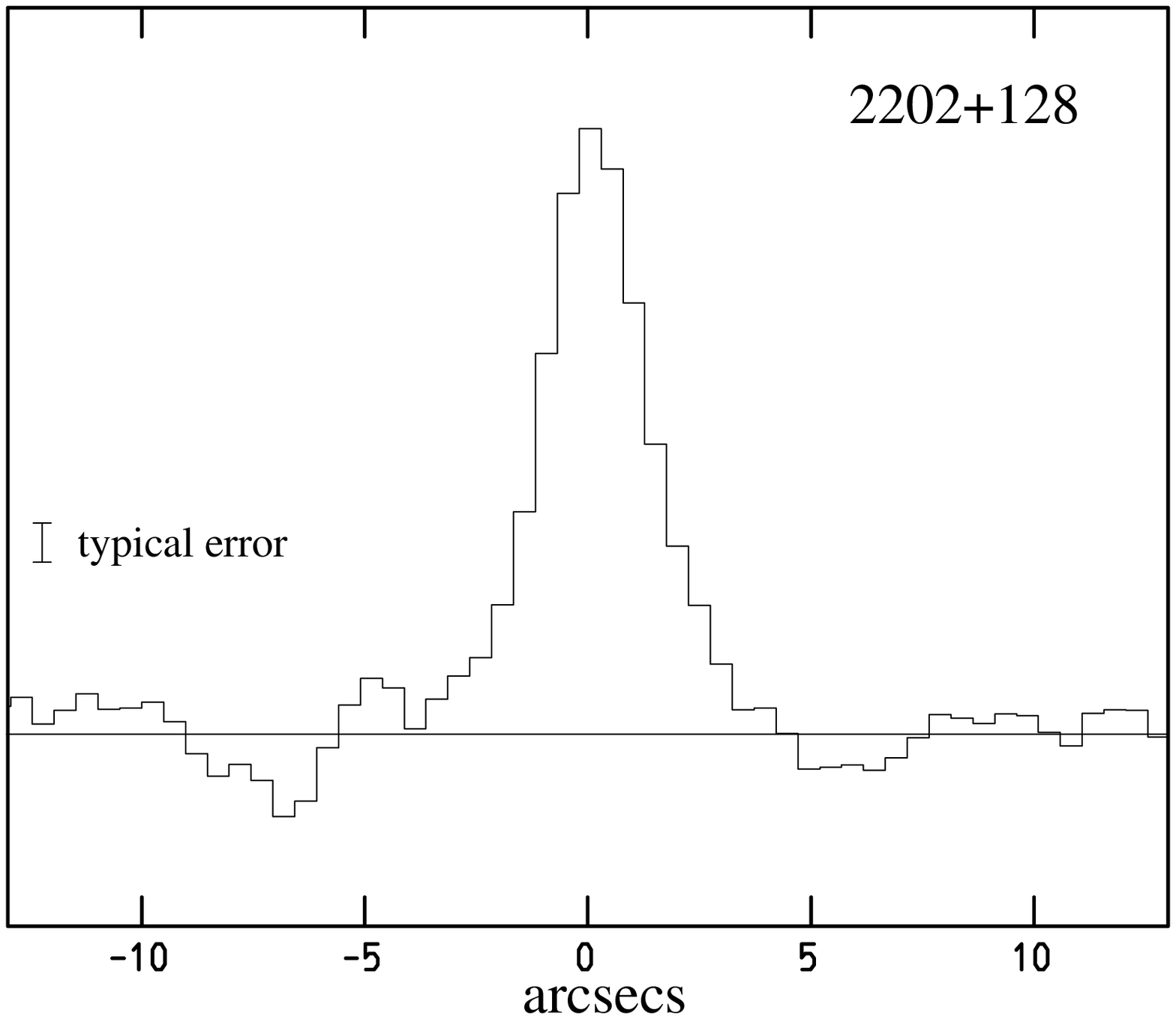,angle=90,width=8cm,height=7cm}
}
\noindent {\bf Fig. 5.} -- continued --.
\end{figure}

\clearpage \newpage

\begin{figure}[p]
\hbox{
\psfig{figure=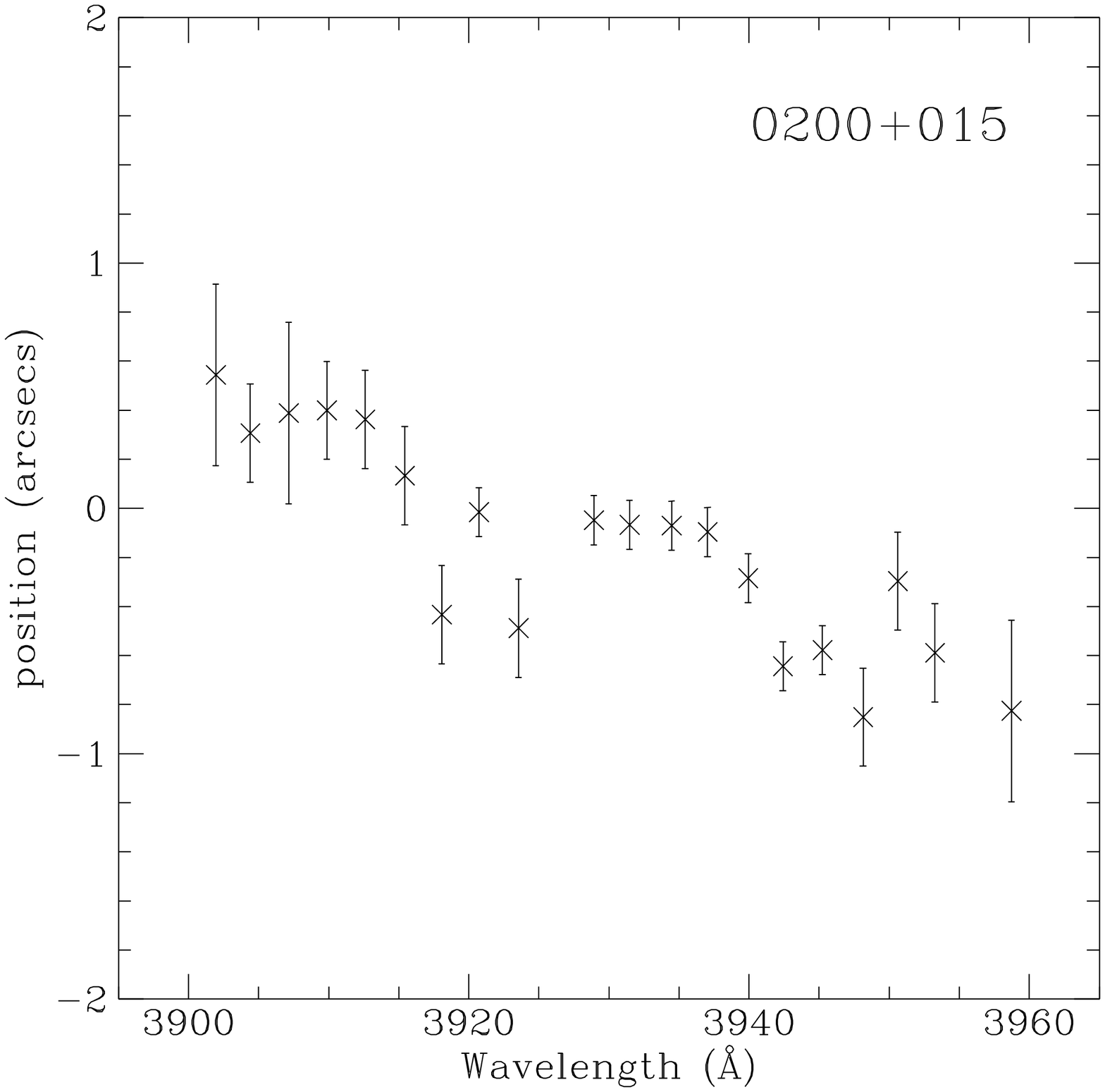,width=8cm,height=7cm}
\psfig{figure=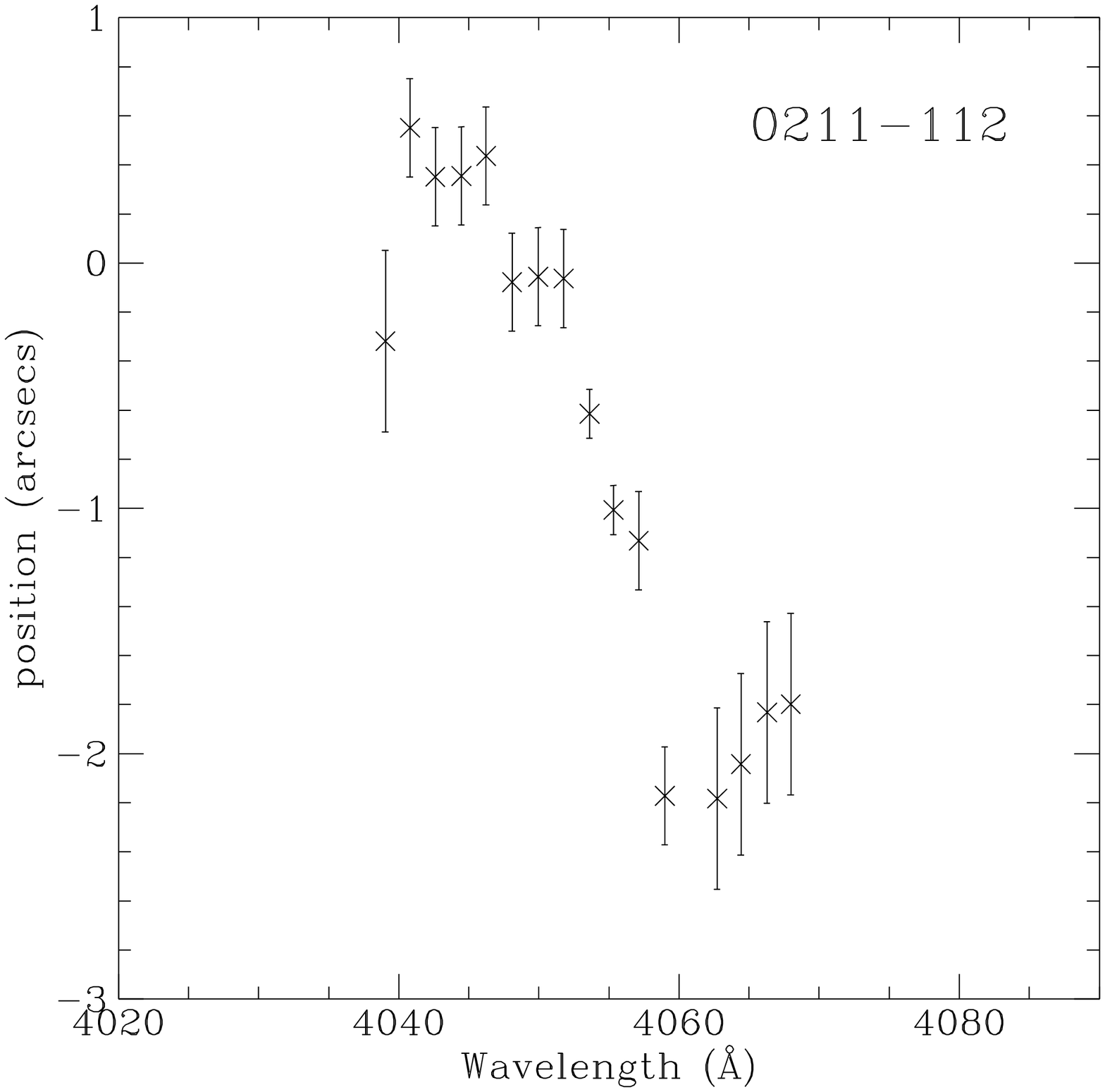,width=8cm,height=7cm}
}
\hbox{
\psfig{figure=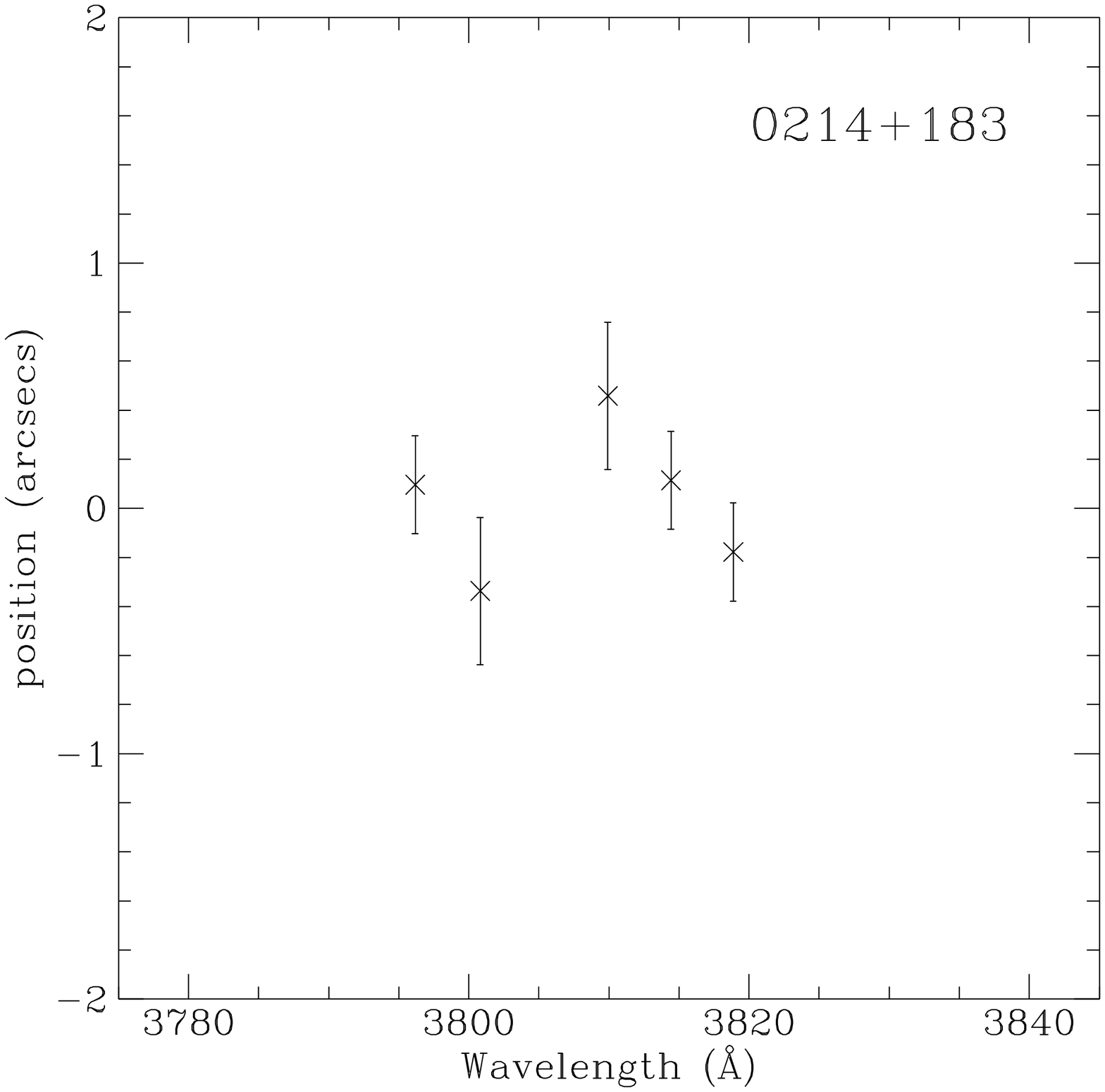,width=8cm,height=7cm}
\psfig{figure=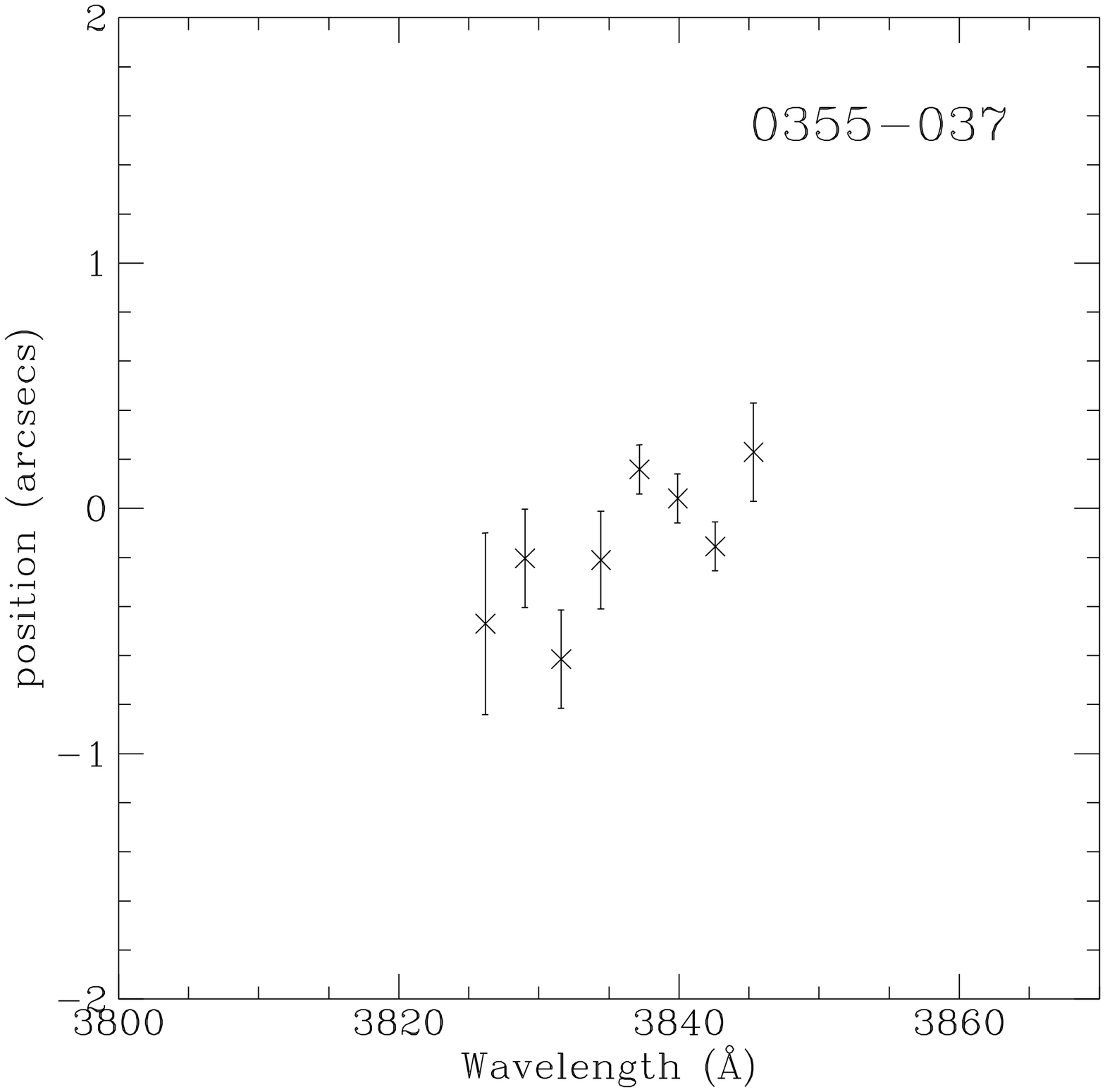,width=8cm,height=7cm}
}
\hbox{
\psfig{figure=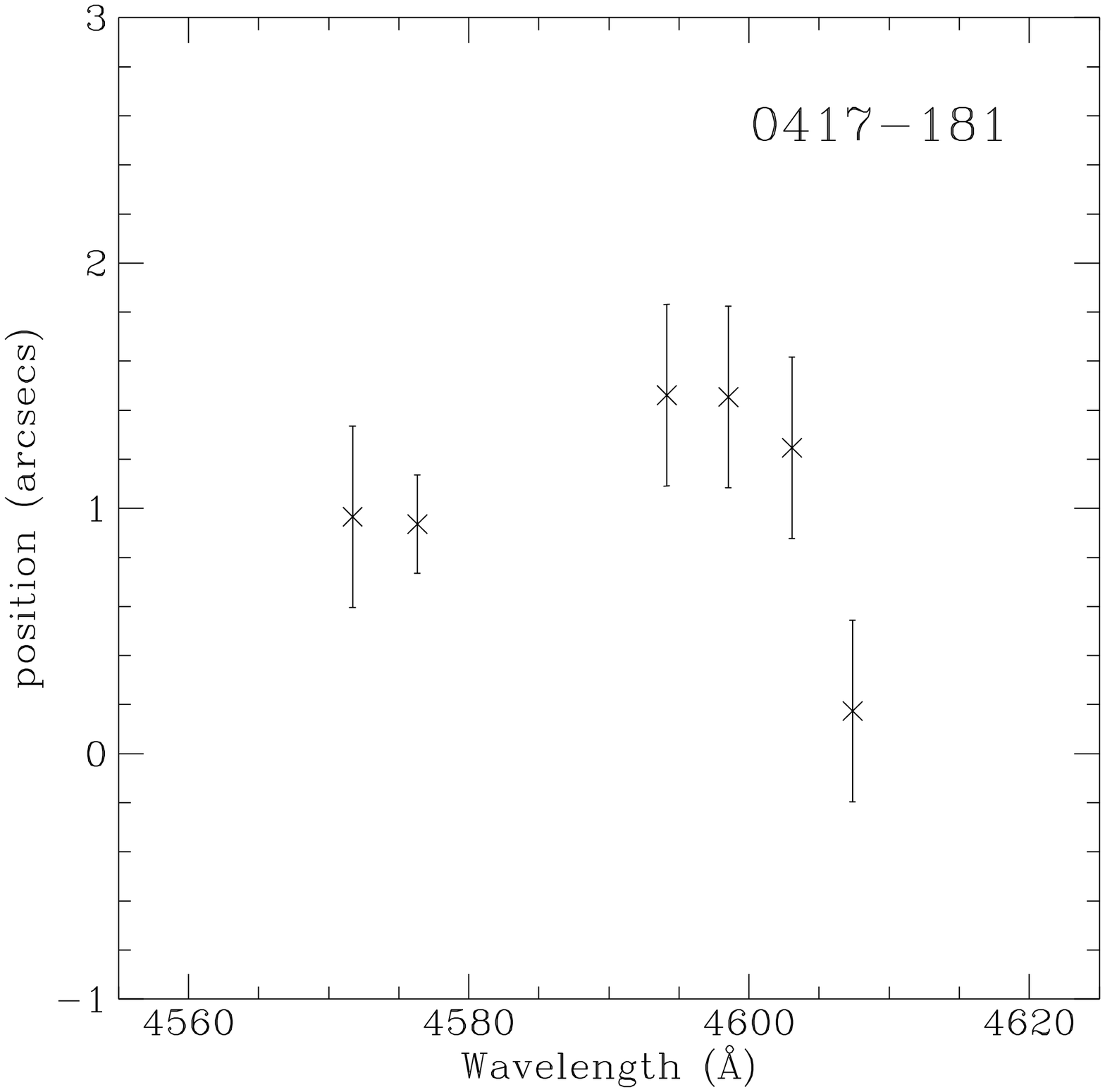,width=8cm,height=7cm}
\psfig{figure=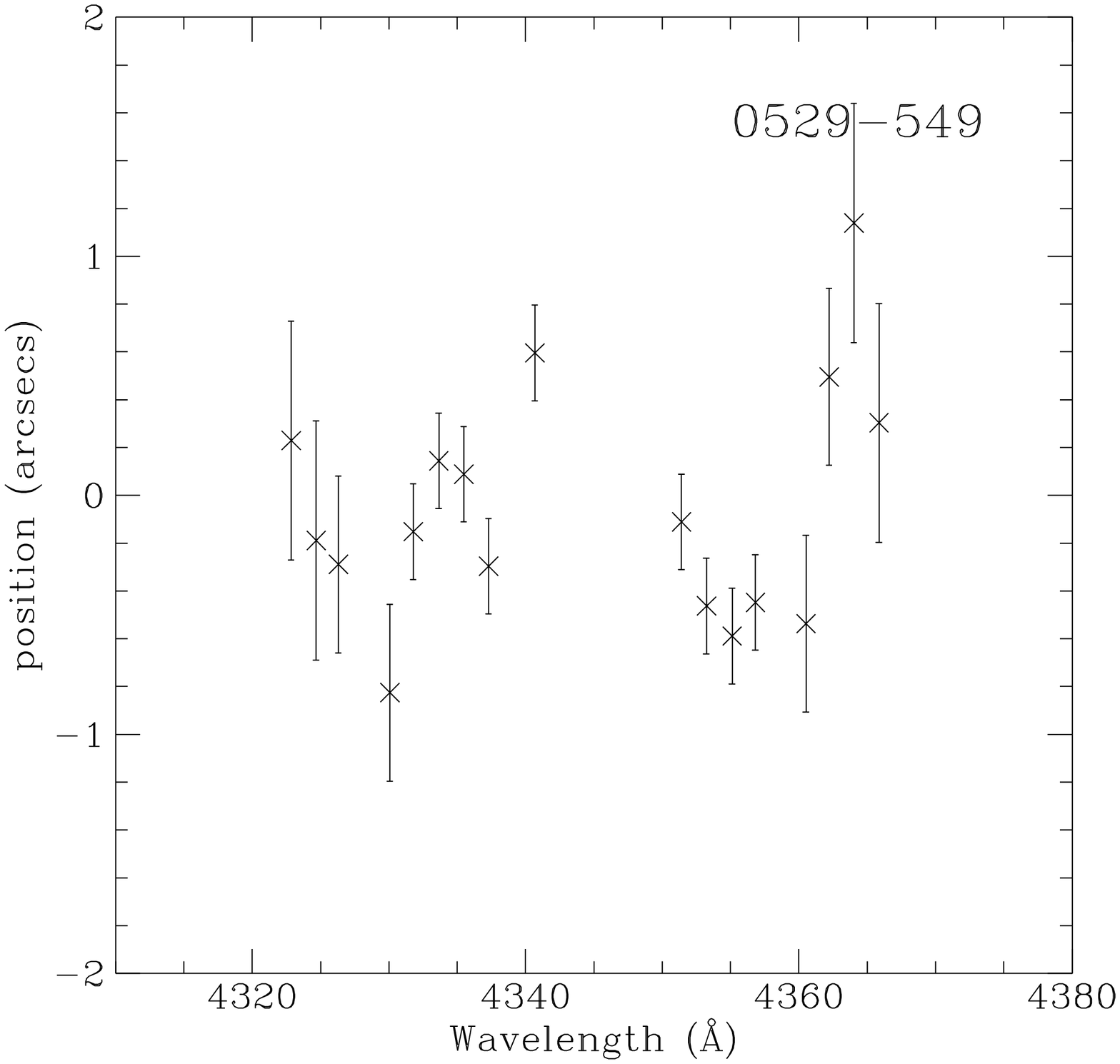,width=8cm,height=7cm}
}
\noindent {\bf Fig. 6.} Spatial positions of the maximum Ly$\alpha$ intensity
plotted against wavelength
\end{figure}

\clearpage \newpage

\begin{figure}[p]
\hbox{
\psfig{figure=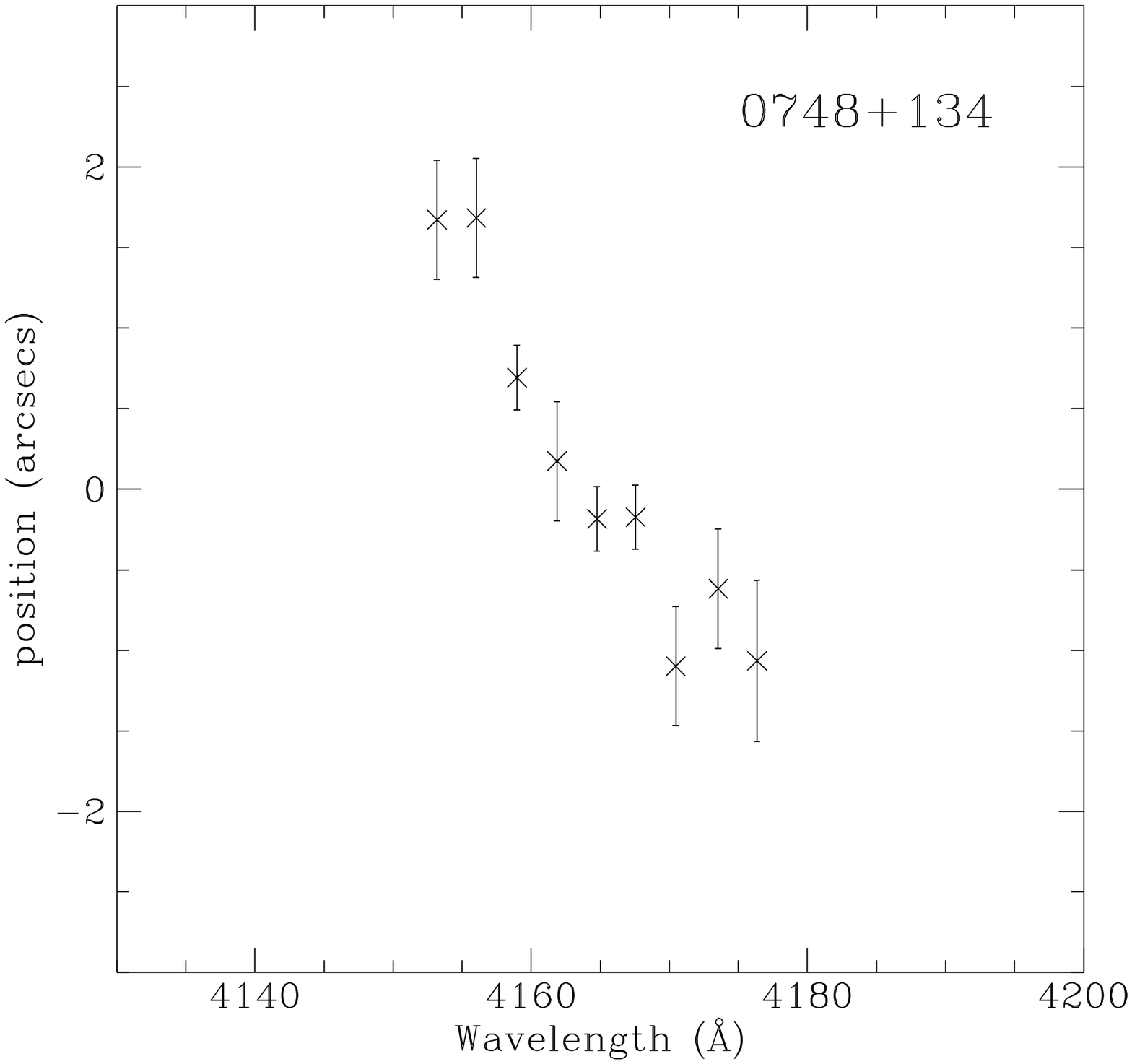,width=8cm,height=7cm}
\psfig{figure=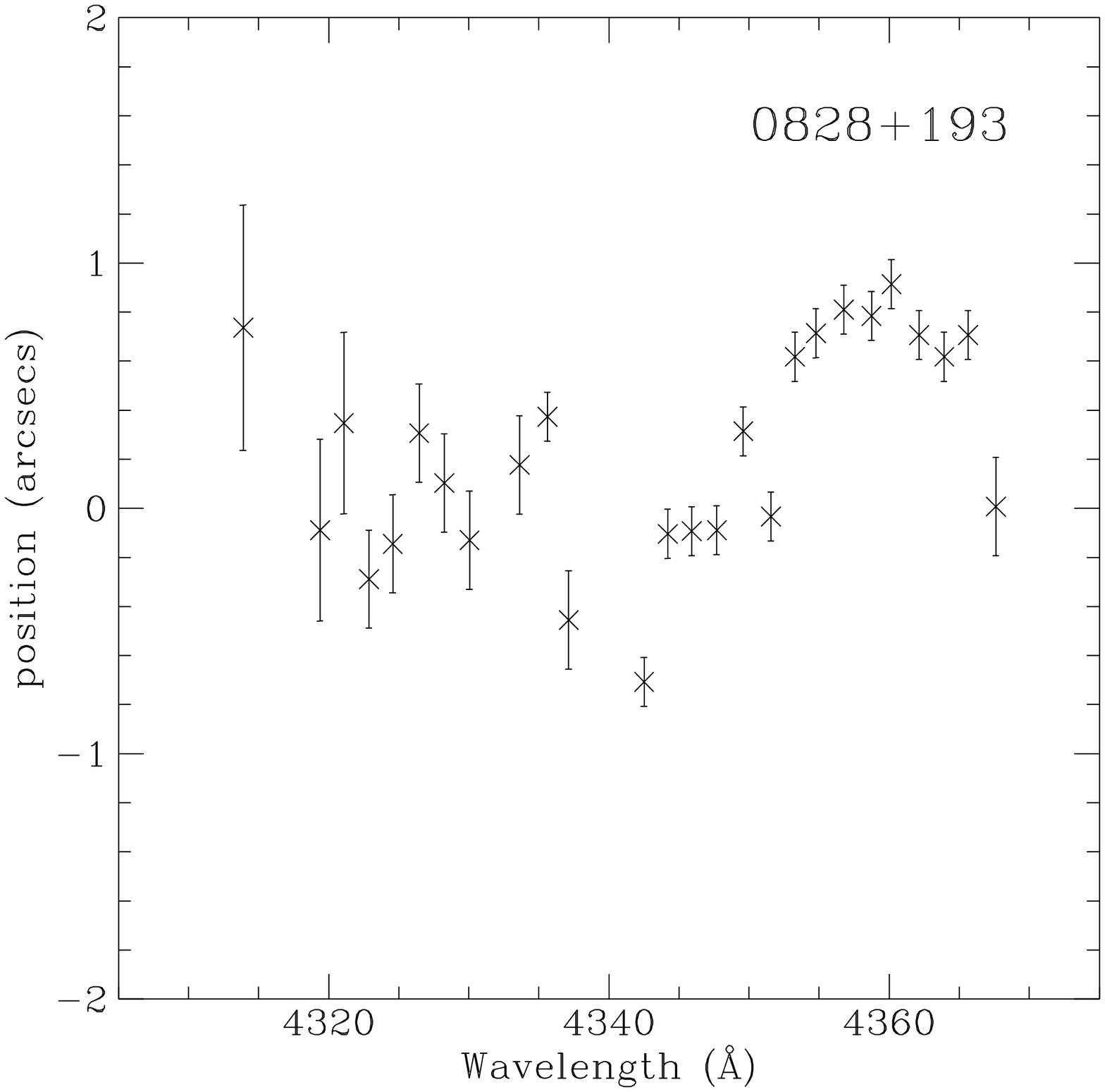,width=8cm,height=7cm}
}
\hbox{
\psfig{figure=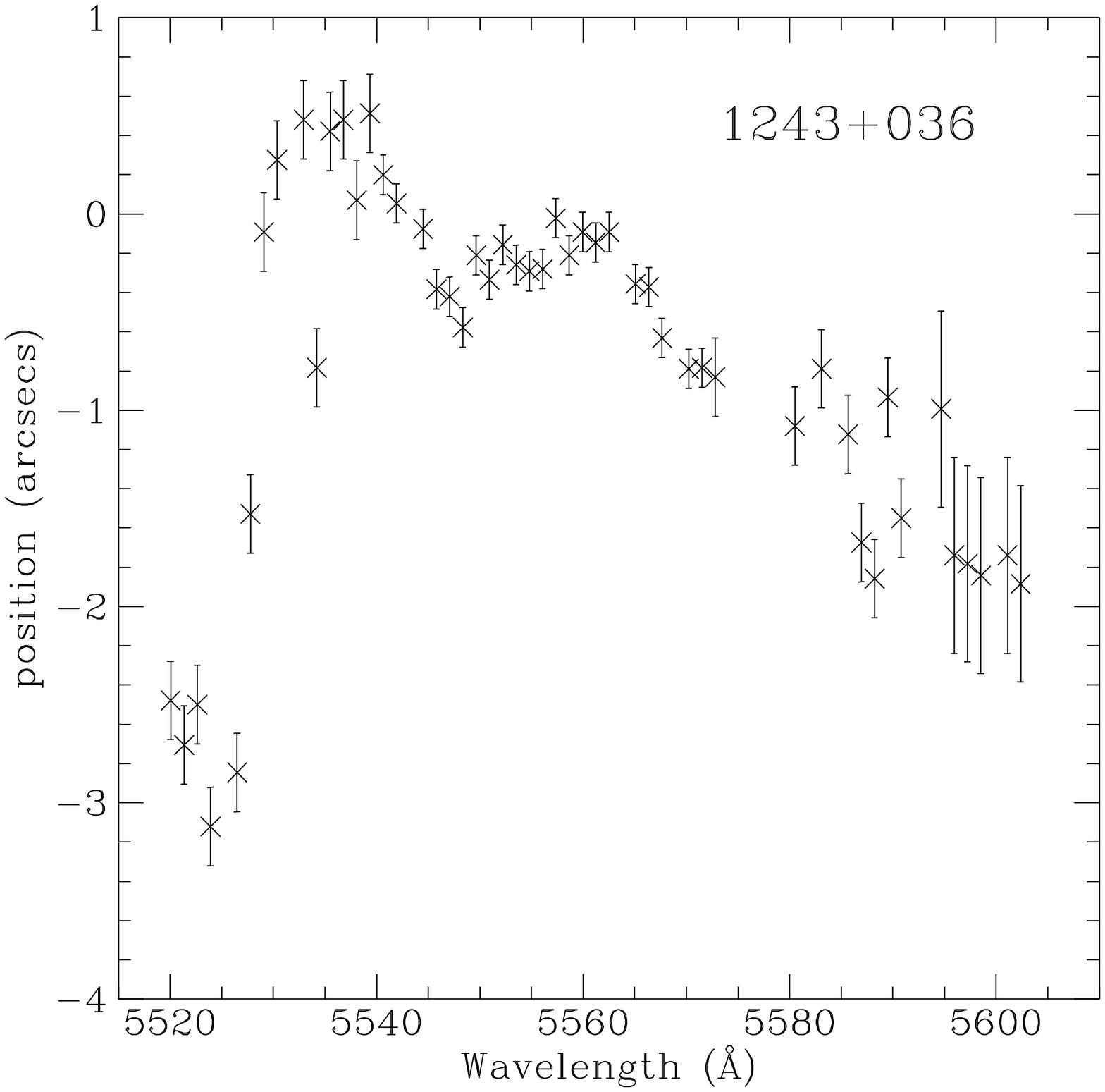,width=8cm,height=7cm}
\psfig{figure=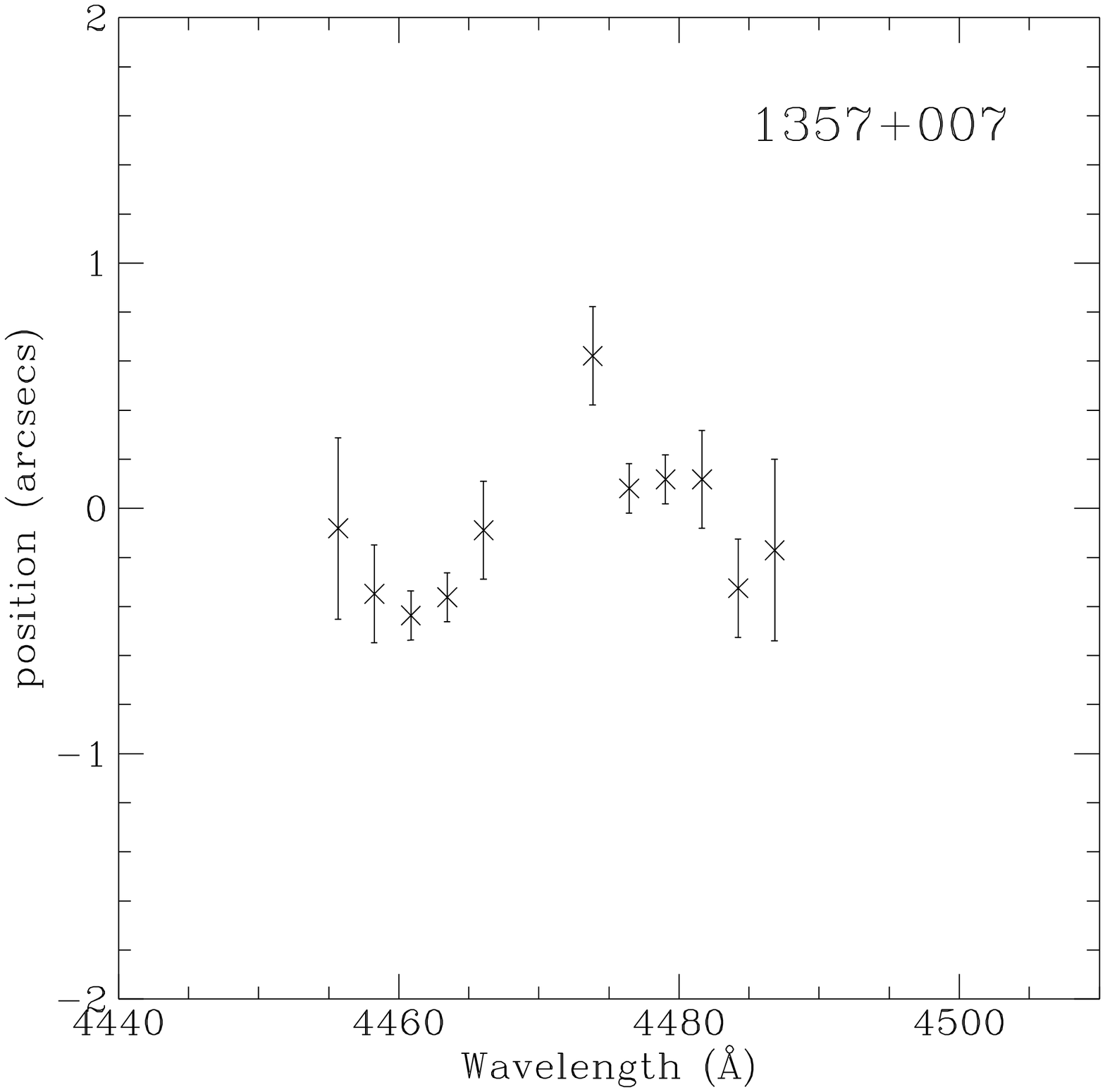,width=8cm,height=7cm}
}
\hbox{
\psfig{figure=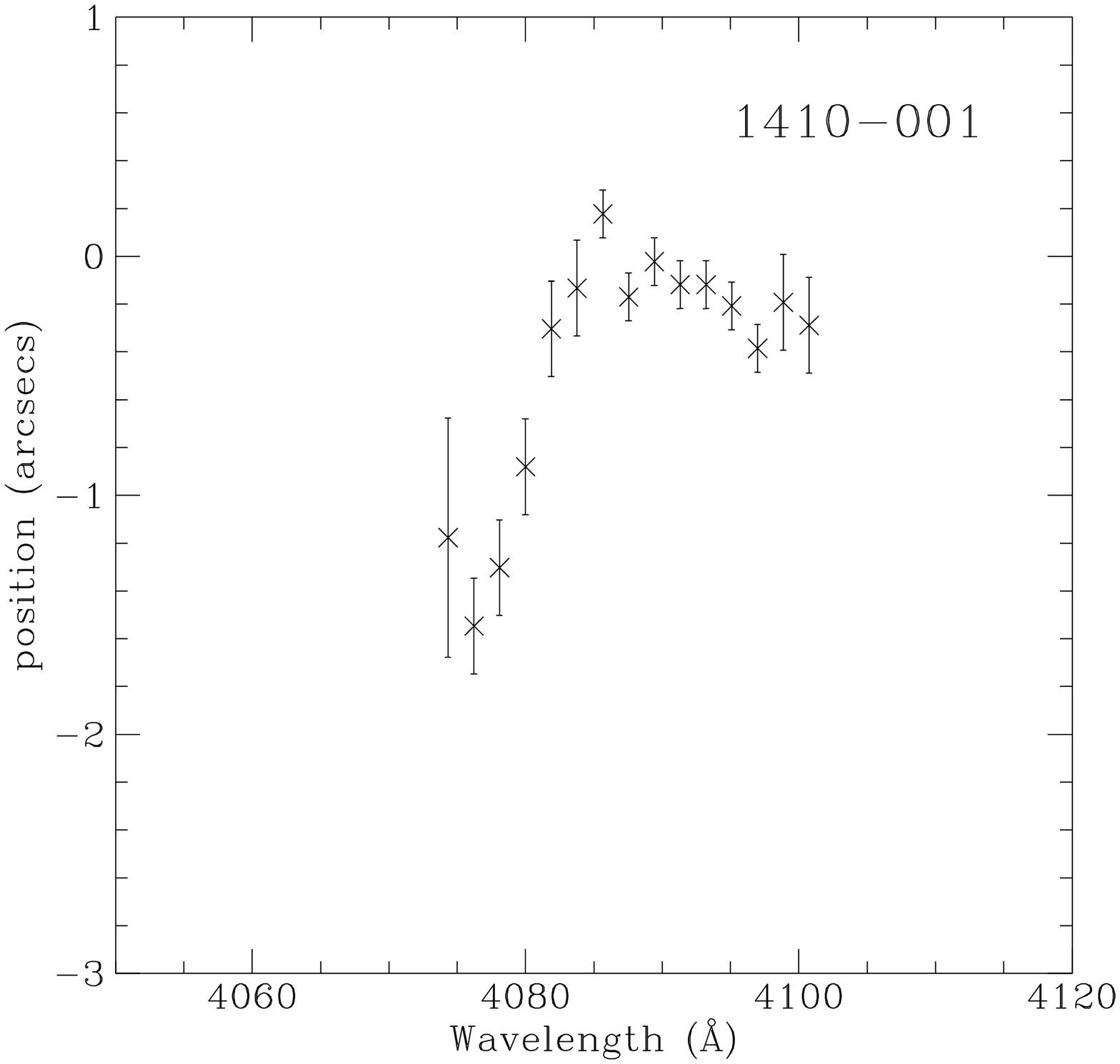,width=8cm,height=7cm}
\psfig{figure=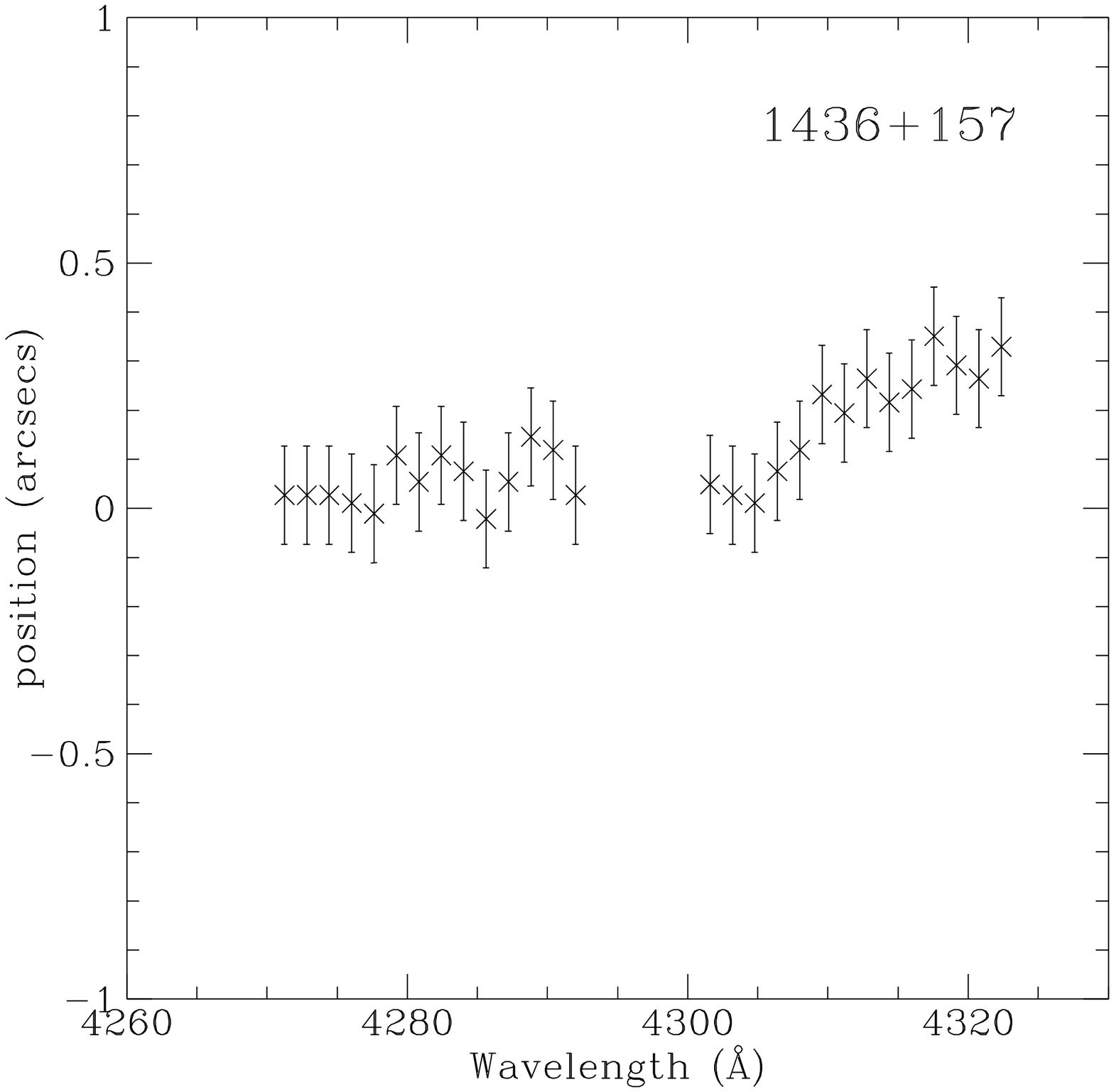,width=8cm,height=7cm}
}
\noindent {\bf Fig. 5.} -- continued --. 
\end{figure}

\clearpage \newpage

\begin{figure}[ht]
\hbox{
\psfig{figure=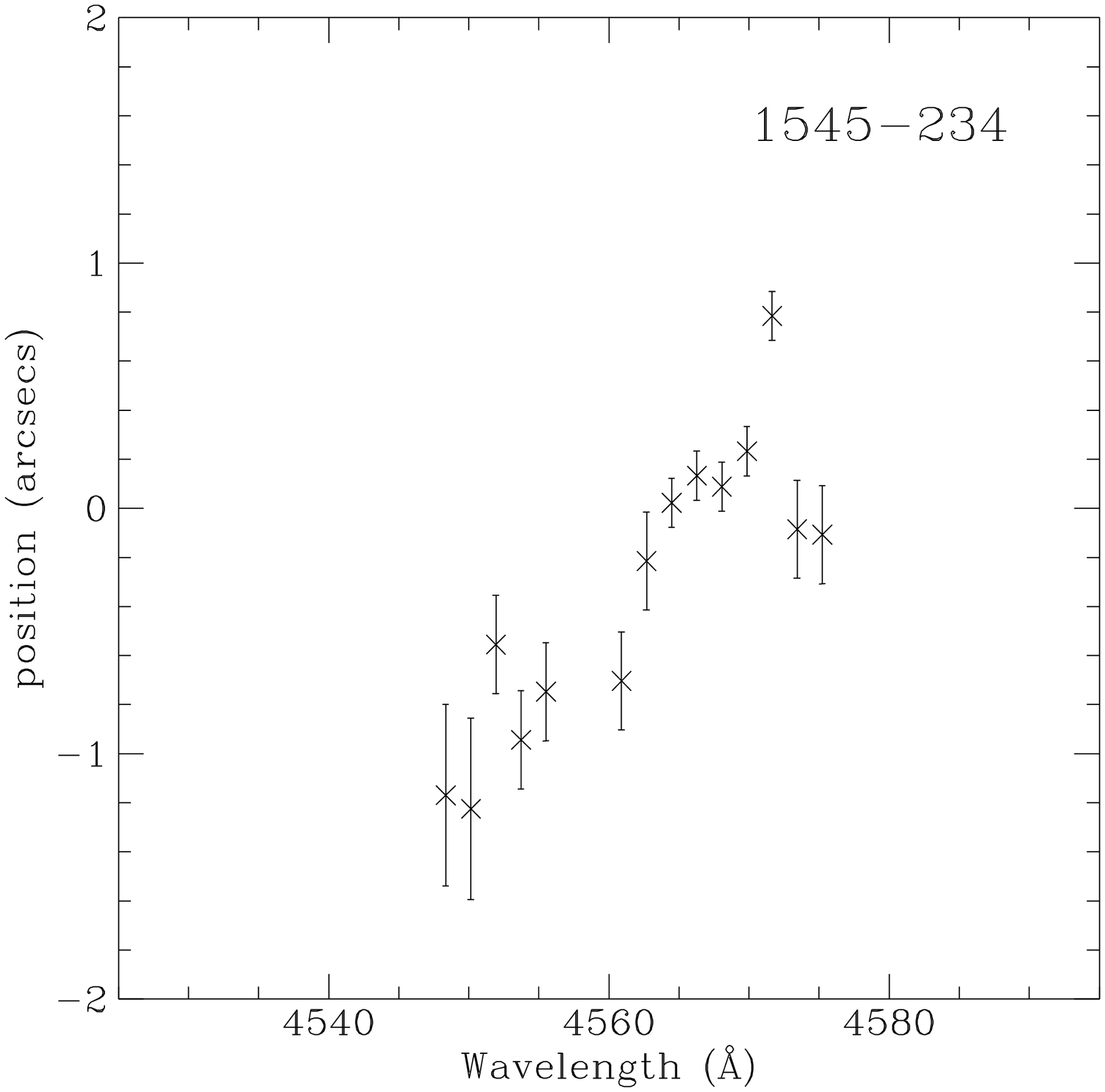,width=8cm,height=7cm}
\psfig{figure=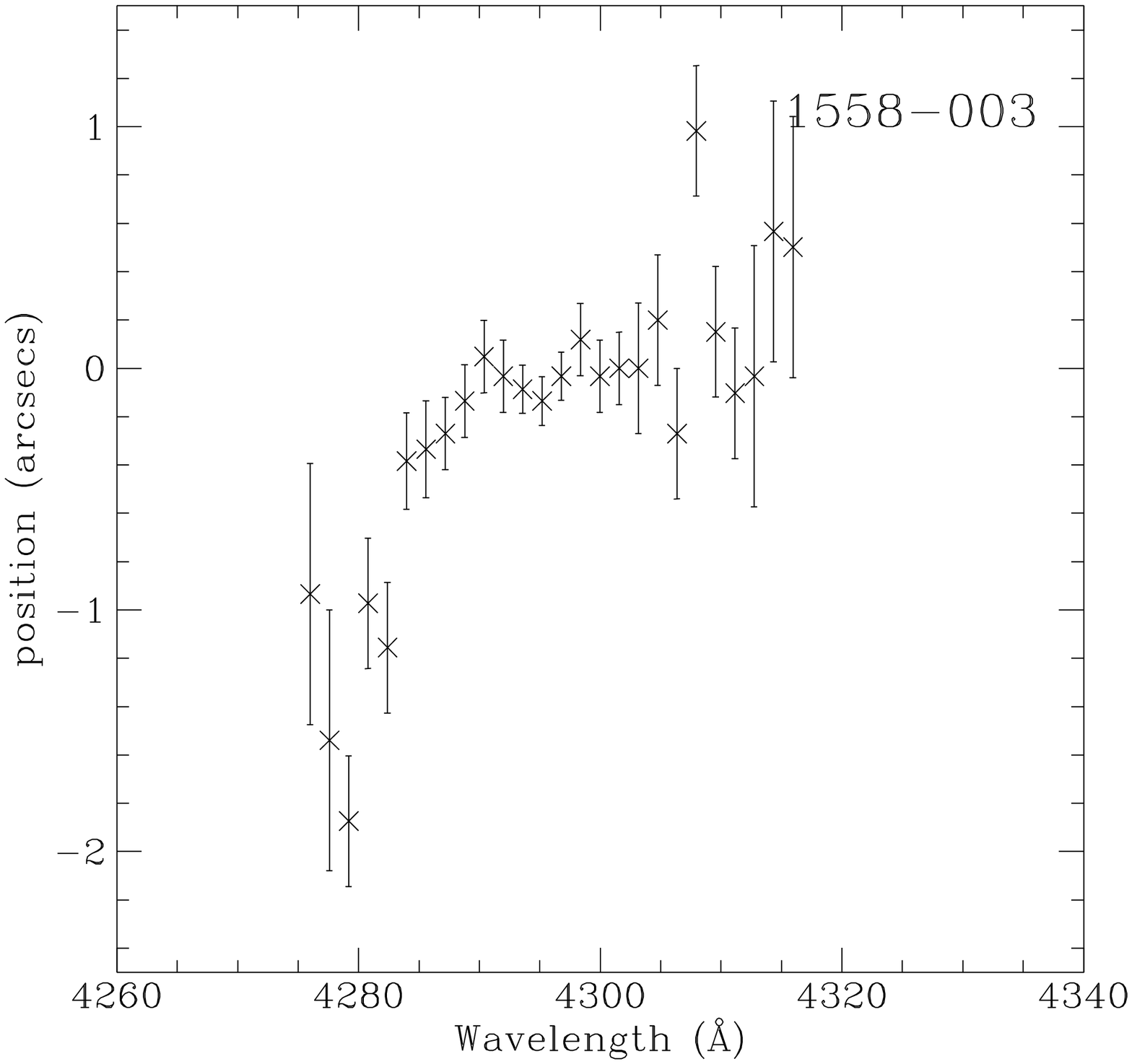,width=8cm,height=7cm}
}
\hbox{
\psfig{figure=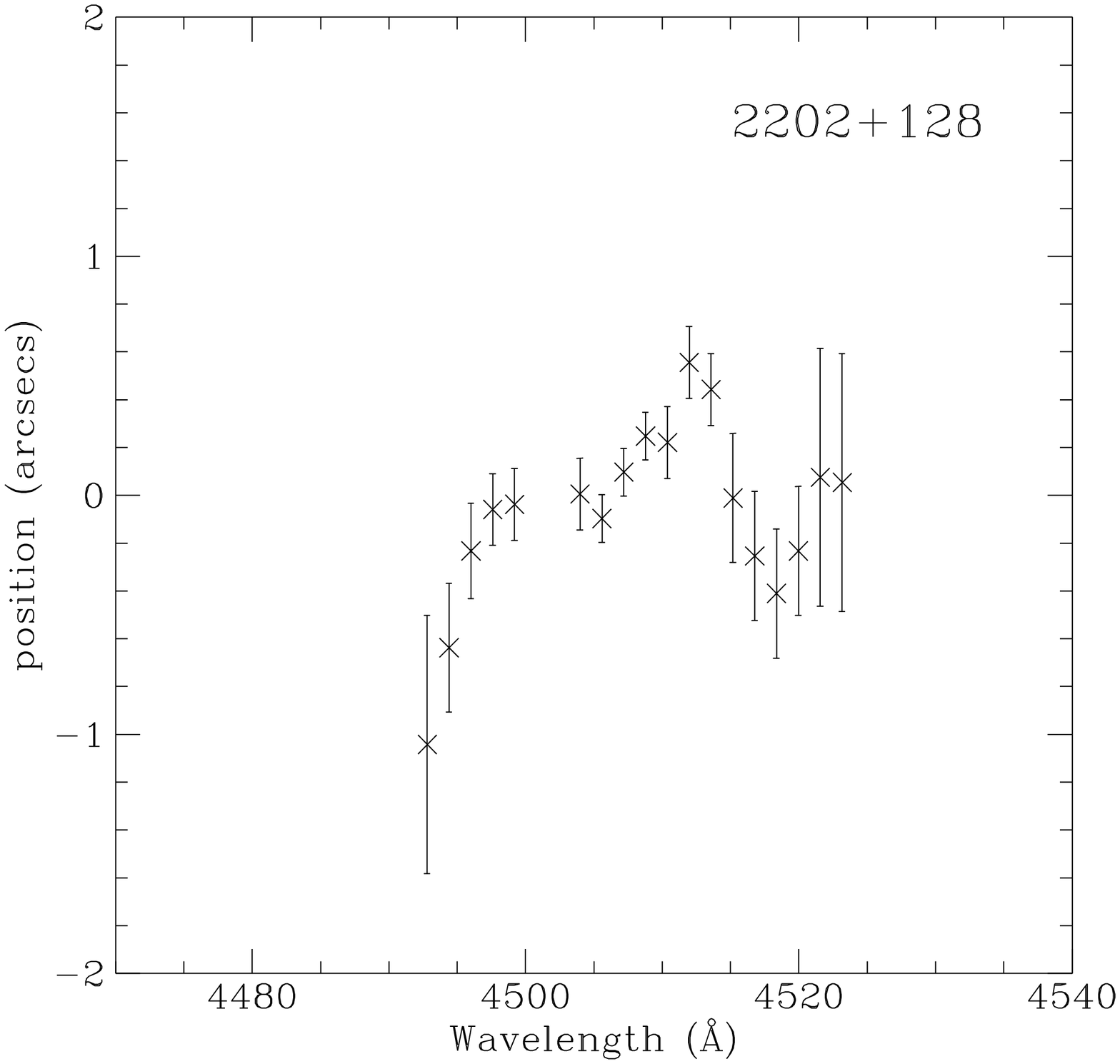,width=8cm,height=7cm}
}
\noindent 
\noindent {\bf Fig. 5.} -- continued --. 
\end{figure}

\clearpage \newpage

\begin{figure}
\hbox{
\psfig{figure=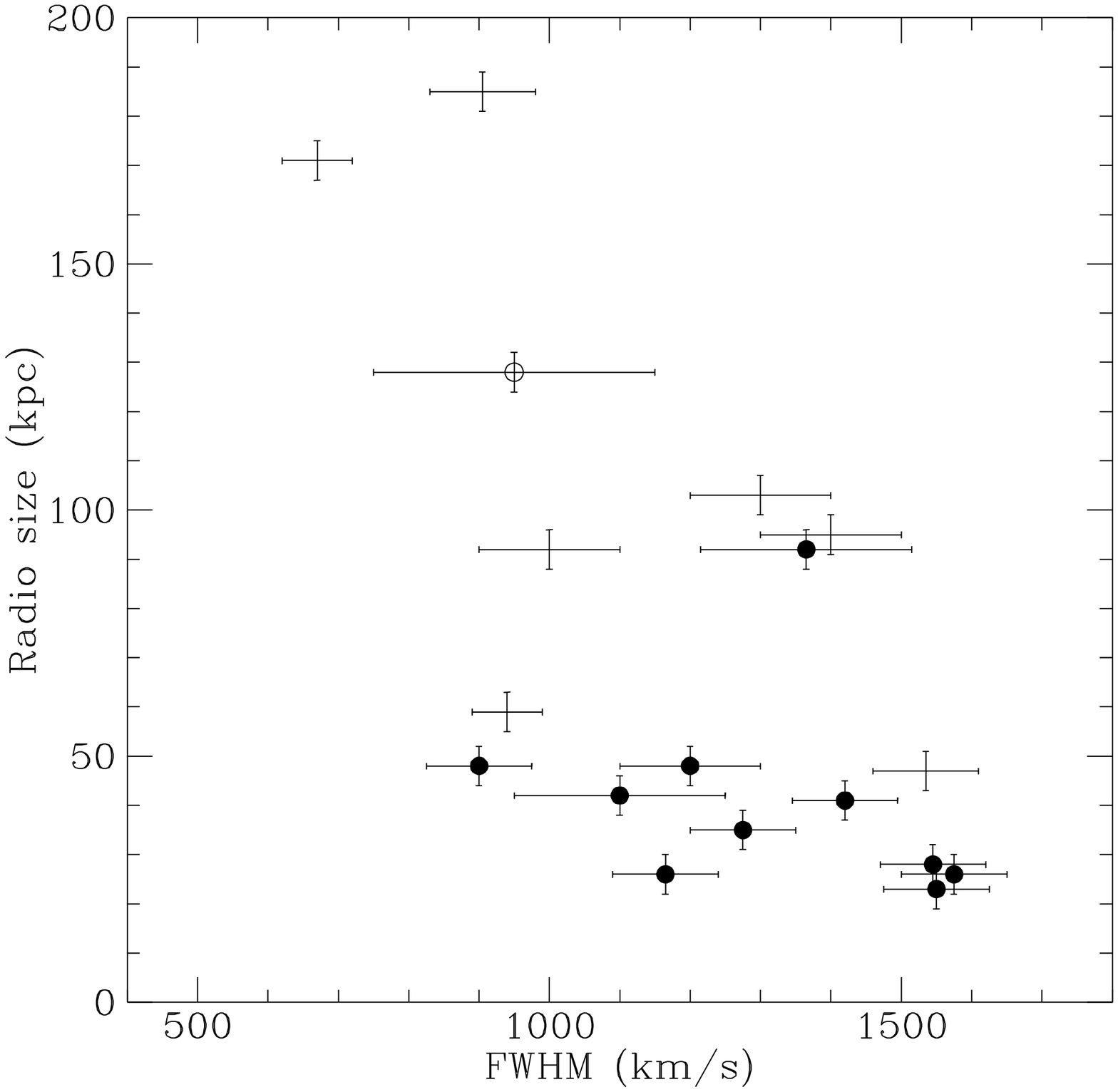,width=8cm}
\psfig{figure=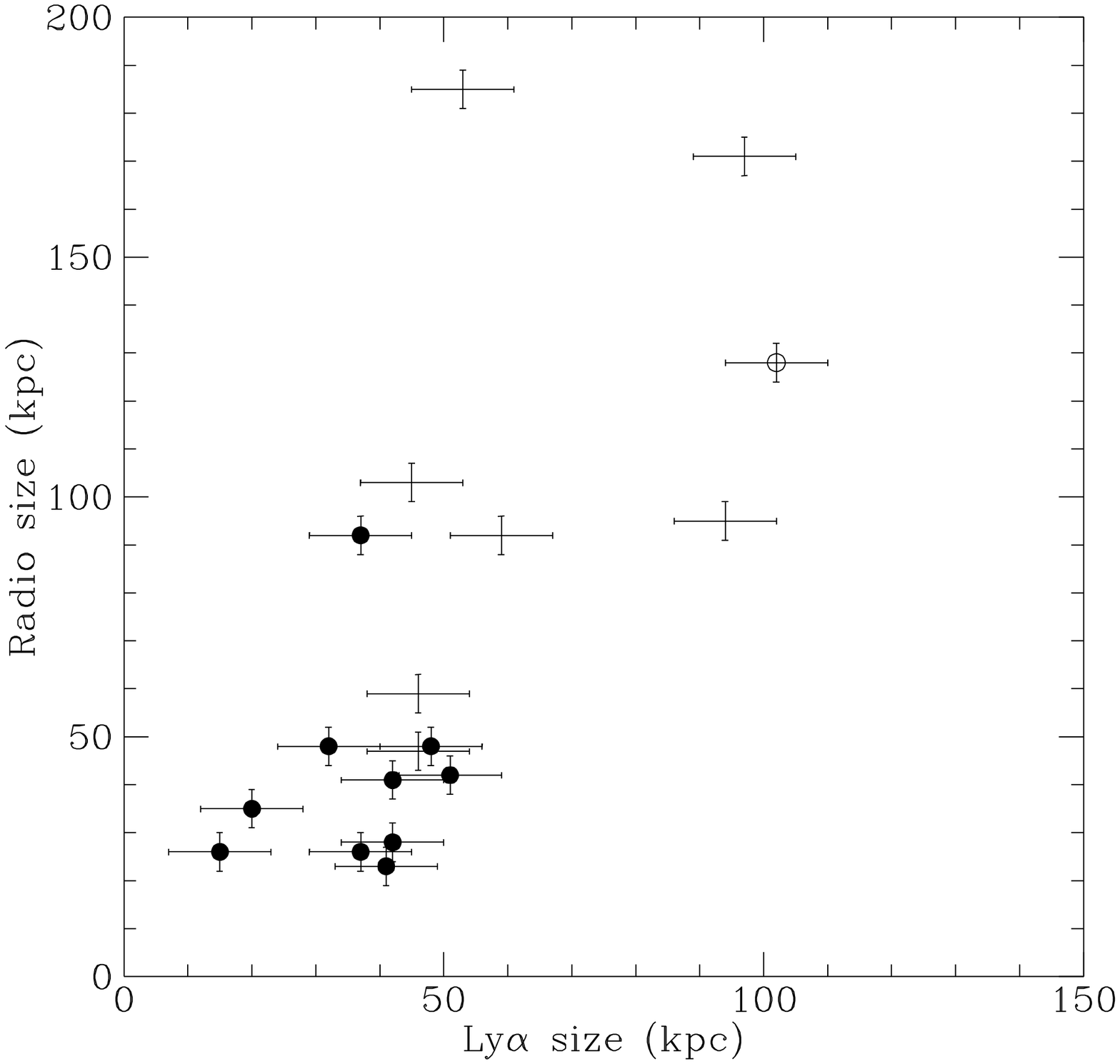,width=8cm}}
\noindent 
{\bf Fig. 7. a} Radio size plotted against Ly$\alpha$ width (FWHM).
{\bf b} Radio size plotted against Ly$\alpha$ size, D$_{Ly\alpha}^{20\%}$.
The objects marked with black dots are the ones with strong HI absorption
(N(HI)$>10^{18}$ cm$^{-2}$)
in the Ly$\alpha$ profile. The object marked with an open circle is 0211$-$122
whose absorption is likely to be strongly influenced by dust (see text)

\end{figure}

\clearpage \newpage

\begin{figure}
\hbox{
\psfig{figure=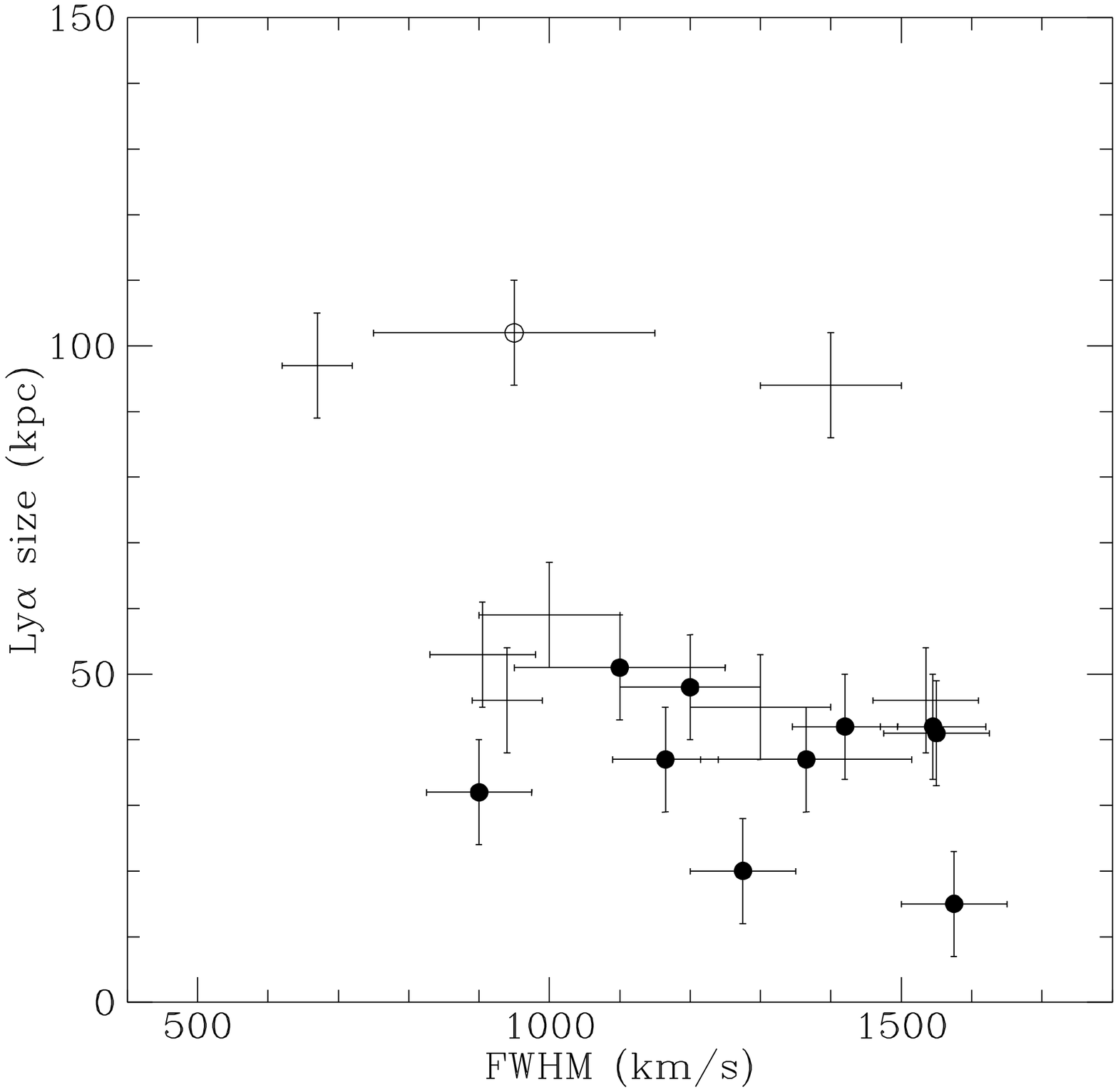,width=8cm}
\psfig{figure=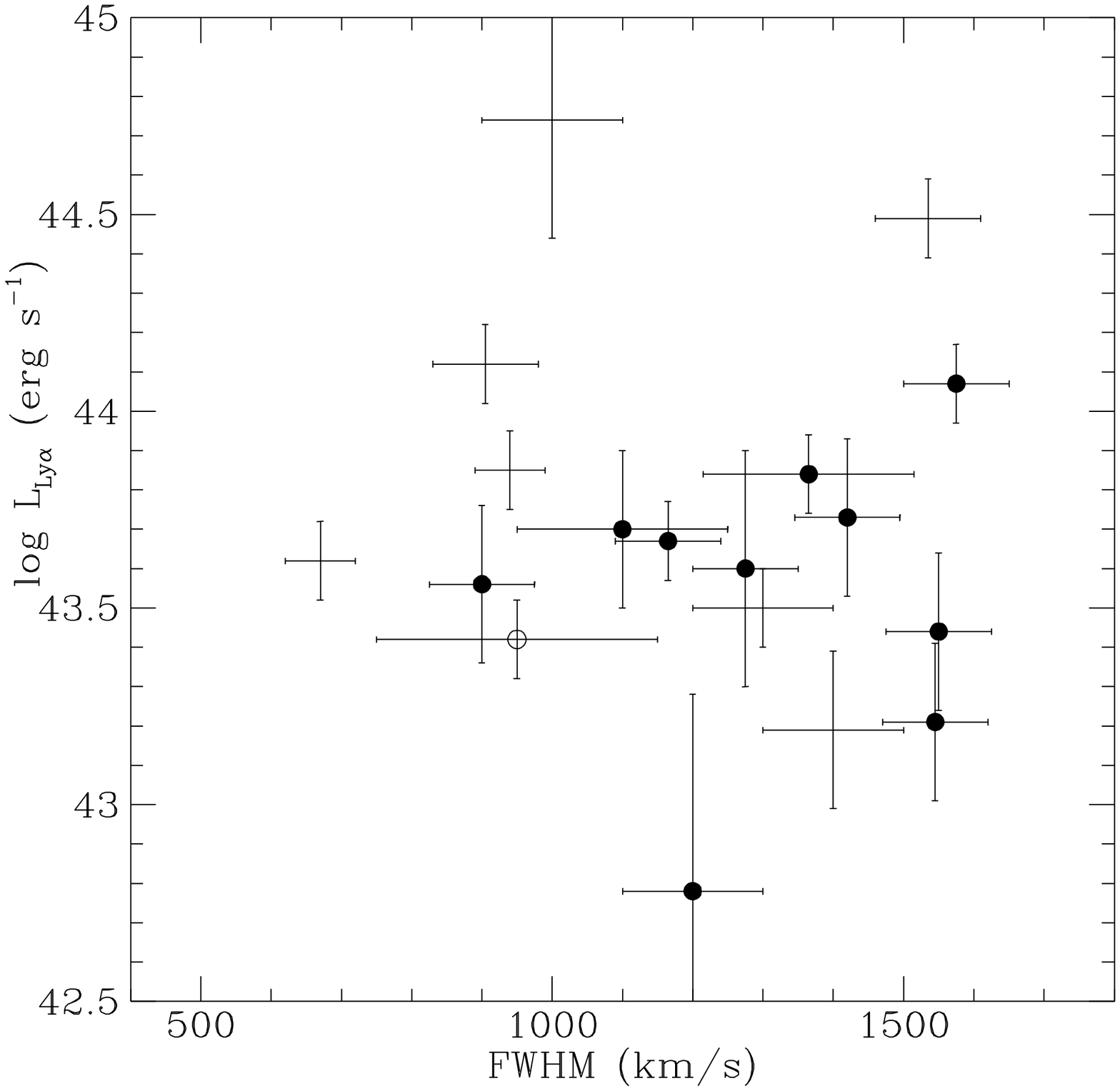,width=8cm}}

\noindent 
{\bf Fig. 8. a} Ly$\alpha$ size, D$_{Ly\alpha}^{20\%}$, 
plotted against Ly$\alpha$ width (FWHM).
{\bf b} Ly$\alpha$ luminosity plotted against Ly$\alpha$ width (FWHM).
The objects marked with black dots are the ones with strong HI absorption
in the Ly$\alpha$ profile. The object marked with an open circle is 0211$-$122

\end{figure}

\clearpage \newpage

\begin{figure}
\centerline{
\psfig{figure=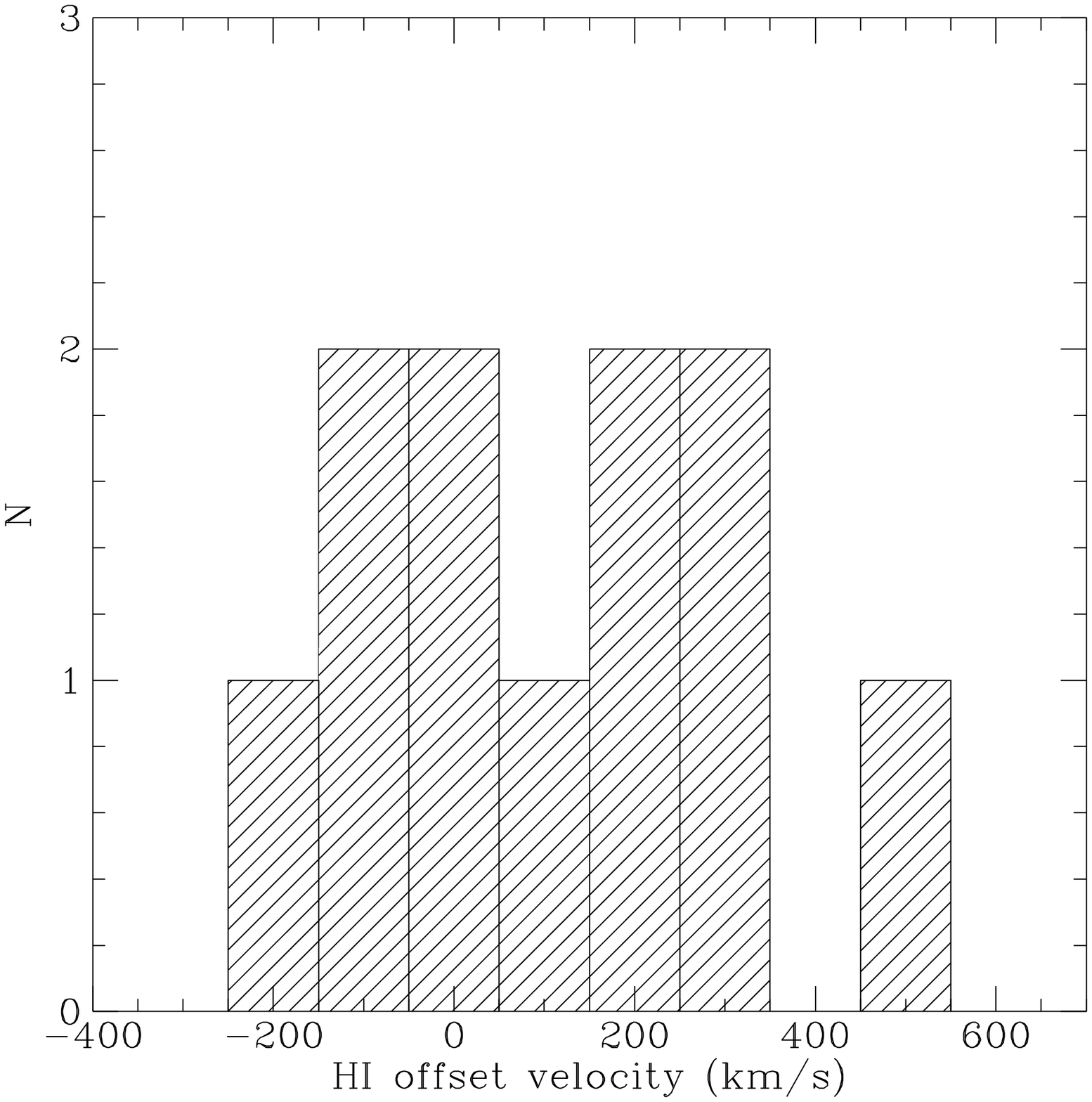,width=8cm}}
\noindent {\bf Fig. 9.} 
Histogram of the relative velocities of the strong HI absorption systems
(N(HI)$>10^{18}$ cm$^{-2}$)
with respect to the peak of the Gaussian fitted to the Ly$\alpha$ emission.
Negative velocities are redshifted with respect to the Ly$\alpha$ emission

\end{figure}

\clearpage \newpage

\begin{figure}
\hbox{
\psfig{figure=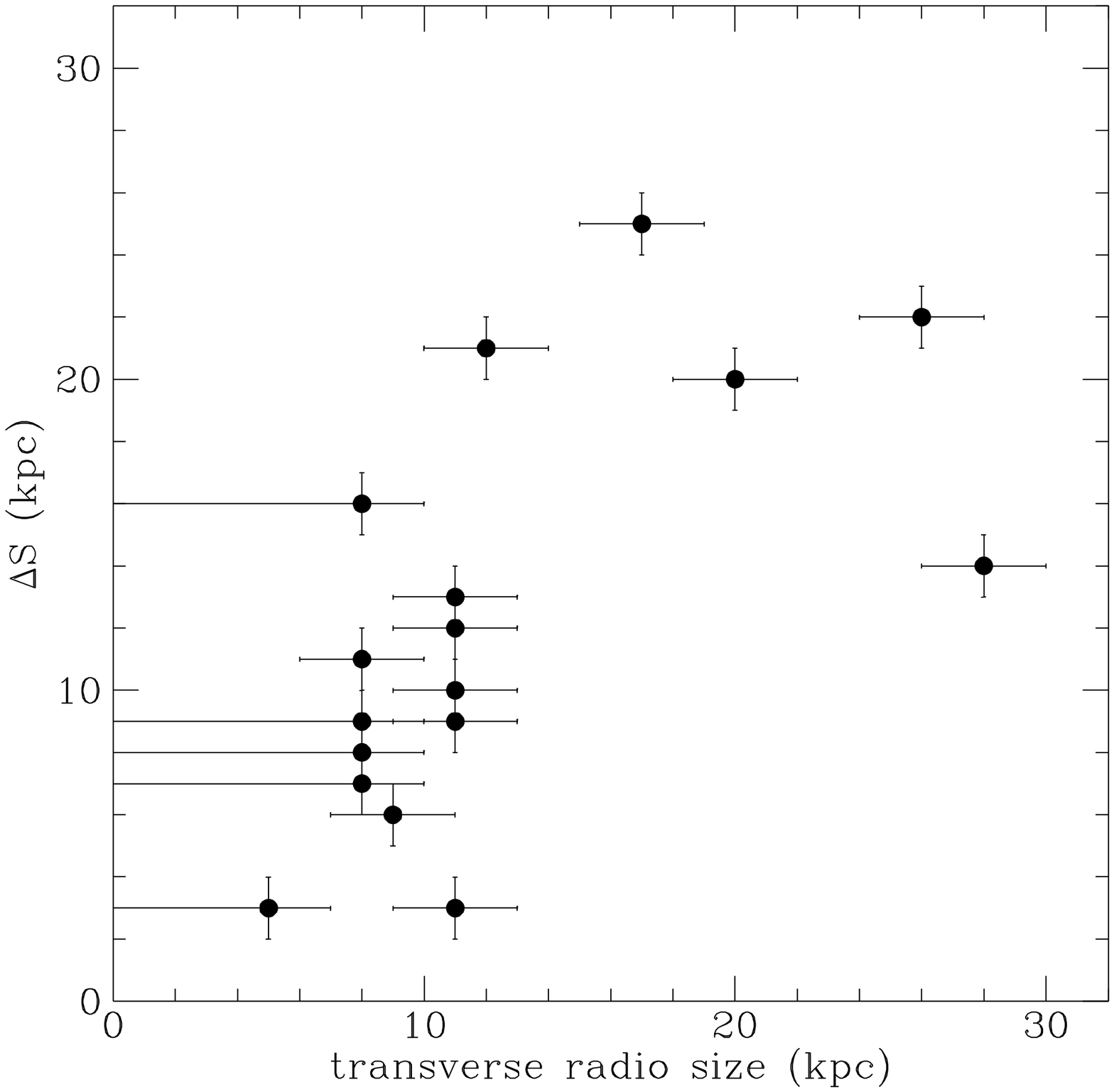,width=8cm}
\psfig{figure=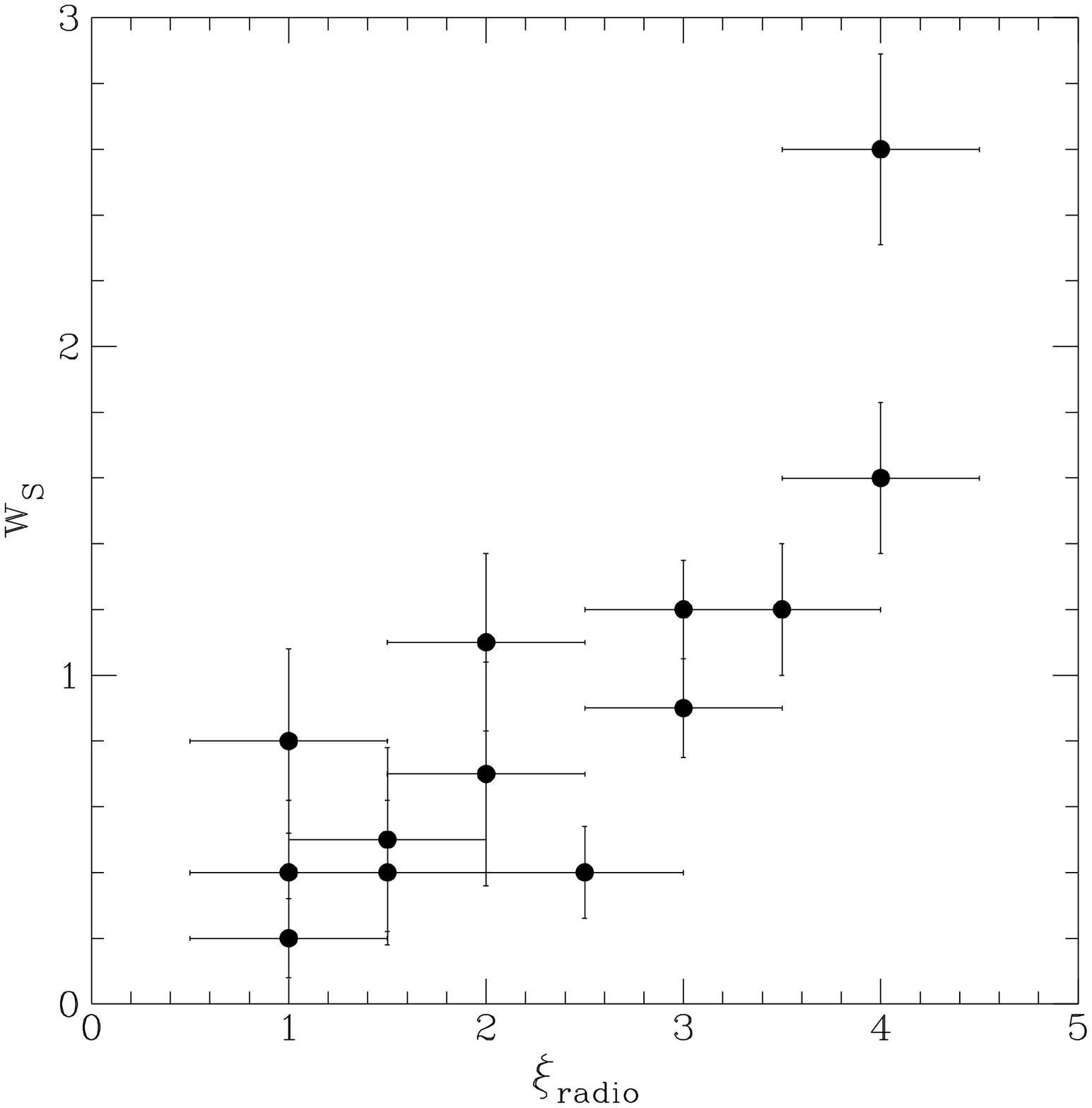,width=8cm}}

\noindent {\bf Fig. 10.  a} Maximum shift of Ly$\alpha$ peak position 
($\Delta$S) plotted against the transverse radio size.
{\bf b} ``Wiggling index'' of the Ly$\alpha$ position ($w_S$) 
plotted against the radio distortion index ($\xi_{radio}$)
\end{figure}

\clearpage \newpage

\begin{figure}
\centerline{
\psfig{figure=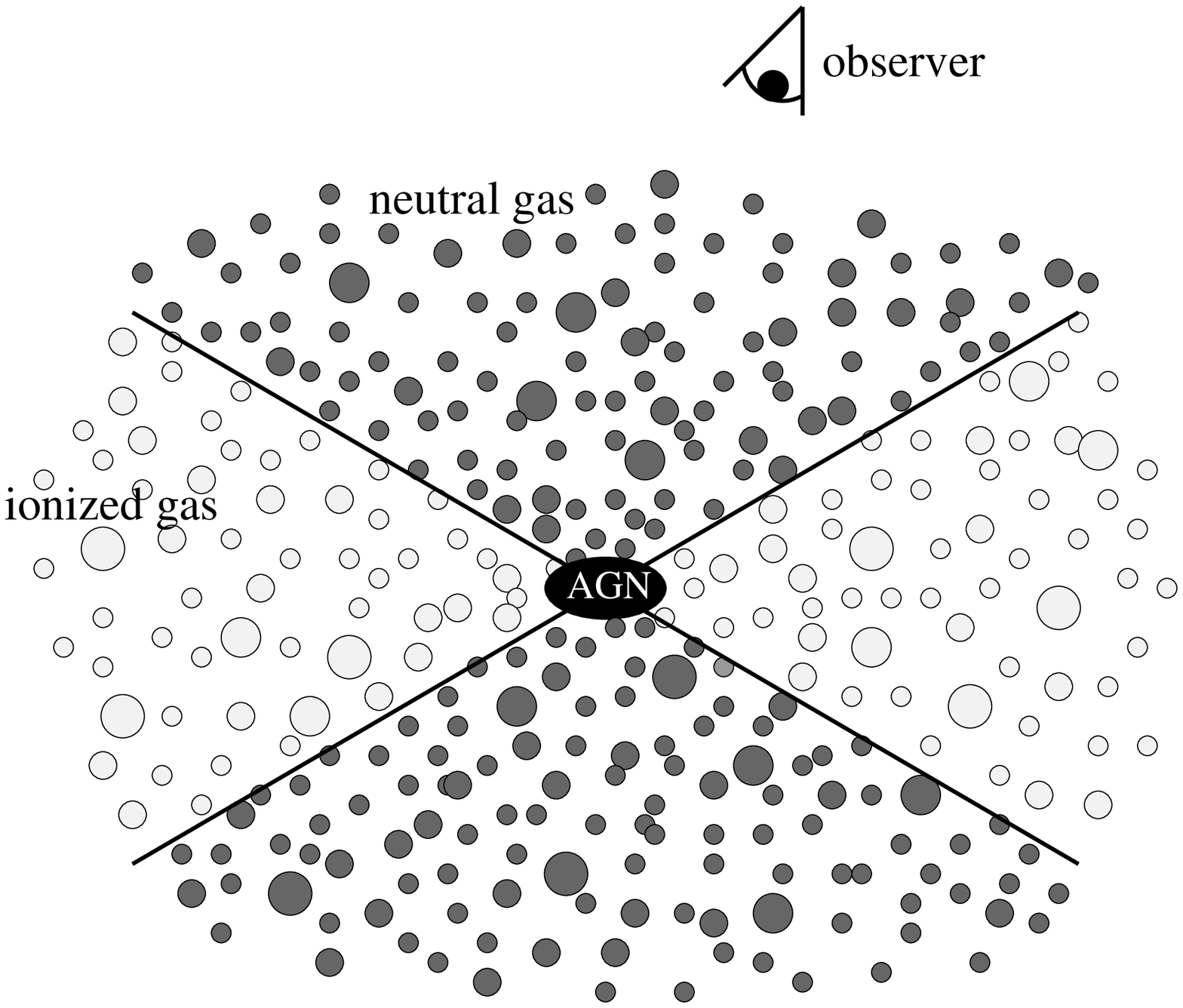,angle=270,width=12cm}}
\noindent {\bf Fig. 11.} Possible location of the absorbing gas in a radio
galaxy with extended emission line gas according to
an orientation-unification scenario. Gas in the ionization cone is photoionized
by the AGN, while the gas outside the ionization cone is predominantly
neutral and will cause absorption of Ly$\alpha$ emission when this
is observed through the neutral gas. The velocity dispersion of the gas 
inside the ionization cone is increased by interaction
with the radio jet. The gas outside the ionization cone is more quiescent
and produces a narrow absorption feature in the Ly$\alpha$ emission profile

\end{figure}
\newpage 

\section{Tables}

\begin{table}
\begin{center}
{\bf Table 1.} High Resolution Spectroscopic Observations
\begin{tabular}{llcccc}  \hline\hline
Session & Date        &  Telescope/Instrument & Resolution & Average Seeing & Photometric\\ \hline
1 & Nov 1992    & ESO-NTT / EMMI        & 2.8 \AA\ & $1''$   &  yes \\
2 & Apr 1993    & AAT                   & 1.5 \AA\ & $2''$   &  no  \\
3 & Nov 1993    & ESO-NTT / EMMI        & 2.8 \AA\ & $1''$   &  yes \\
4 & Apr 1994    & ESO-NTT EMMI          & 2.8 \AA\ & $1''$   &  yes \\
5 & Jun 1994    & WHT / ISIS            & 1.7 \AA\ & 1.5$''$ &  no  \\
6 & Jan 1995    & ESO-NTT / EMMI        & 2.8 \AA\ & 0.9$''$ &  yes \\
7 & Apr 1995    & ESO-NTT / EMMI        & 2.8 \AA\ & $1''$   &  no  \\ \hline\hline
\end{tabular}
\end{center}
\end{table}

\begin{table}[p]
\begin{center}
{\bf Table 2.} Log of the observations
\begin{tabular}{lcccc}  \hline\hline
Source   & $z$   & Session & Total exposure time & Slit P.A.($^{\circ}$) \\ \hline
0200+015 & 2.230 & 6       &  7200 s & 155      \\
0211$-$122 & 2.336 & 3       &  5400 s & 102     \\
0214+183 & 2.131 & 1,6     &  7200 s   & 160    \\
0355$-$037 & 2.151 & 6       &  8100 s & 120      \\
0417$-$181 & 2.770 & 3       & 10800 s & 168      \\
0529$-$549 & 2.572 & 1,3     & 10447 s & 0      \\
0748+134 & 2.410 & 6       &  5400 s & 153      \\
0828+193 & 2.572 & 4       &  5400 s  & 46      \\
1357+007 & 2.671 & 4       &  7200 s  & 136     \\
1410$-$001 & 2.359 & 2       &  8400 s  & 132      \\
1436+157 & 2.550 & 5       &  6300 s & 140      \\
1545$-$234 & 2.754 & 7     &  9000 s & 159       \\
1558$-$003 & 2.520 & 5       &  7200 s & 85      \\
1707+105 & 2.345 & 4,5     & 18000 s & 59,149      \\
2202+128 & 2.704 & 5       &  5400 s & 86      \\ \hline
\end{tabular}
\end{center}
\end{table}

\begin{table}[p]
\vspace{-2.5cm}
\centerline{
\psfig{figure=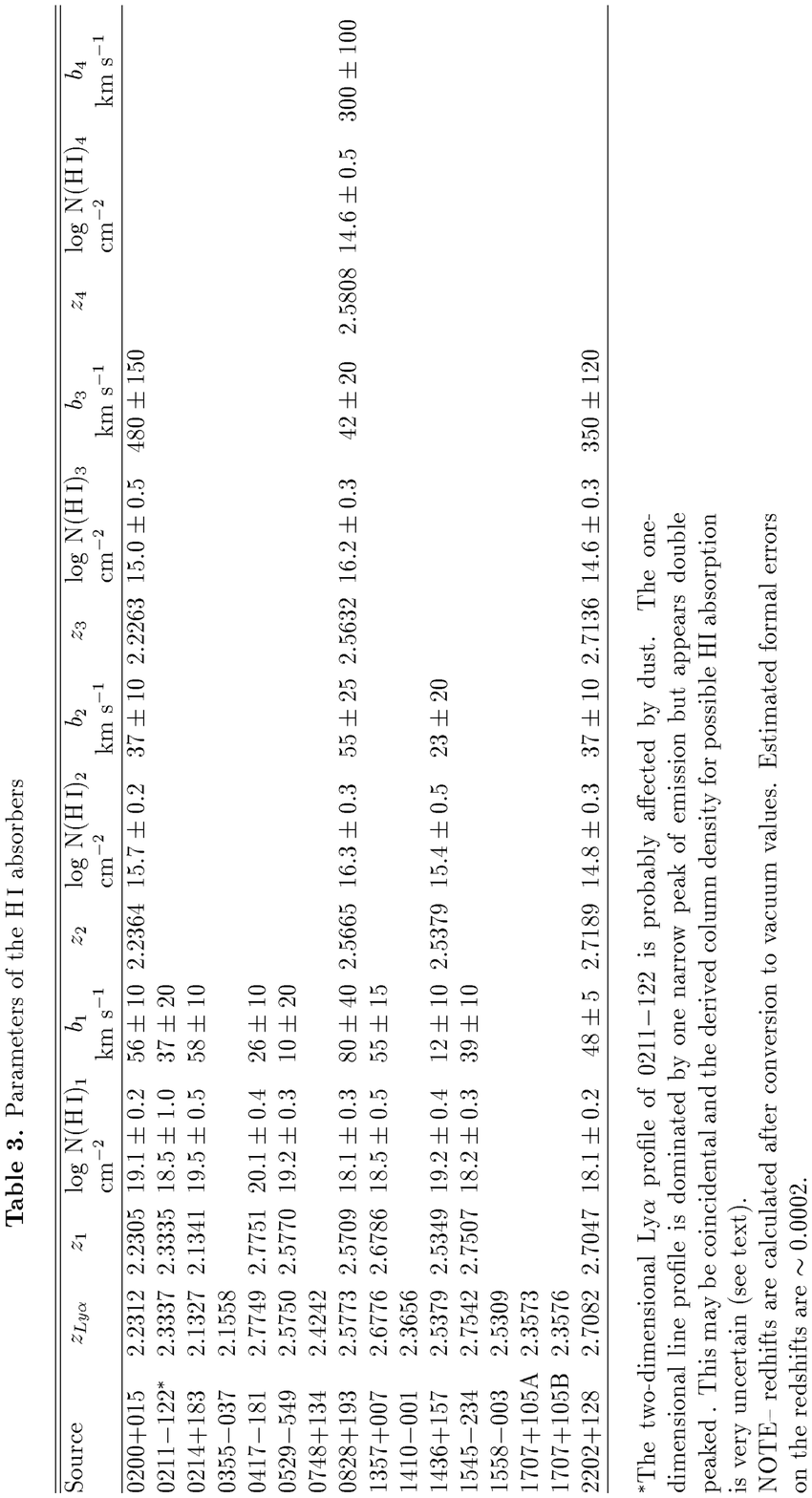,angle=180,height=27cm}
}
\end{table}

\begin{table}[p]
\vspace{-5cm}
\centerline{
\psfig{figure=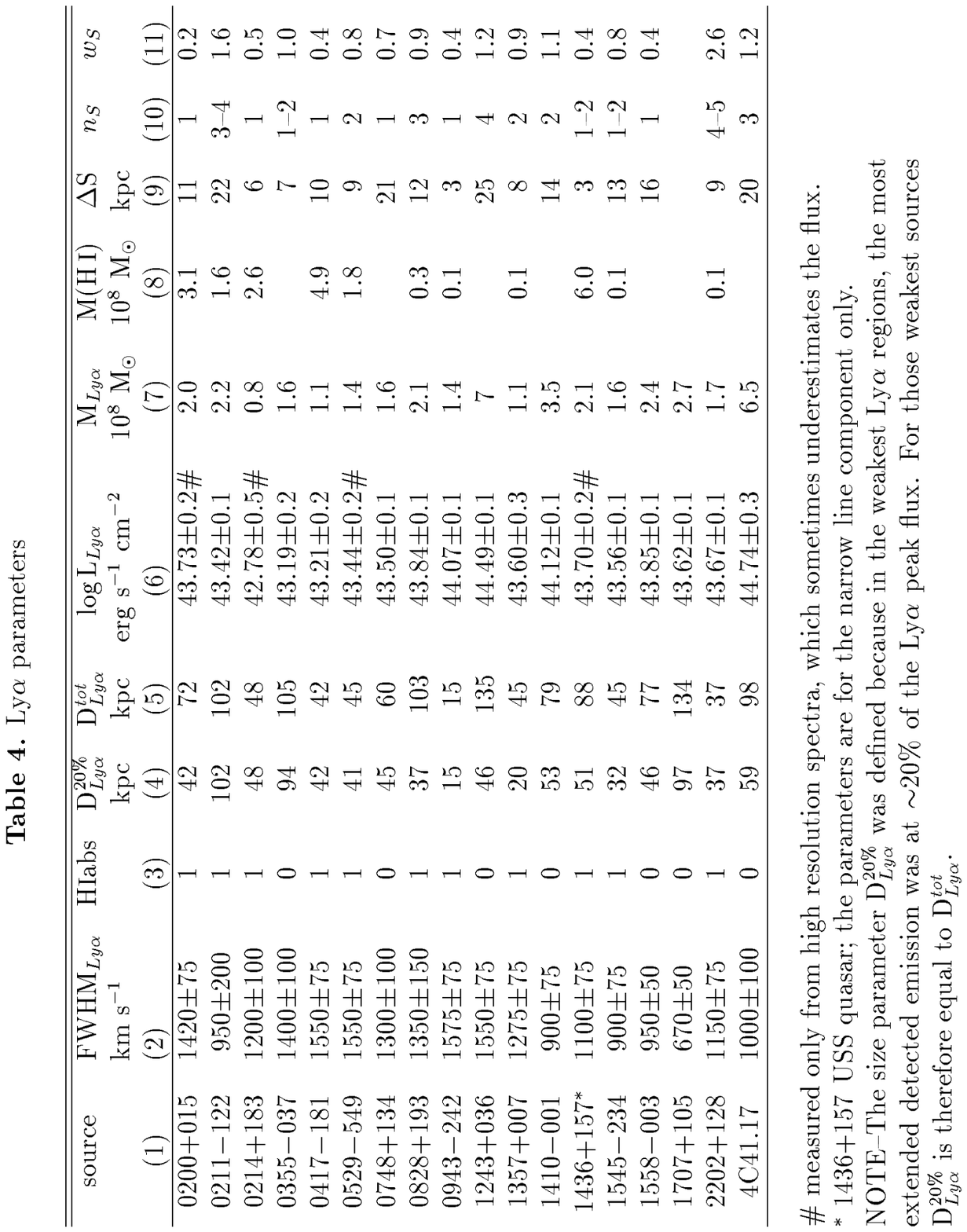,angle=180,height=27cm}
}
\end{table}

\begin{table}
\begin{center}
{\bf Table 5.} Radio parameters
\begin{tabular}{ccccccllc} \\ \hline\hline
source& D$_{Radio}$& D$^{trans}_{Radio}$&$\xi_{radio}$& SI& CF &Q & R &log\,P$_{1.5GHz}$  \\ 
      & (kpc)       & (kpc) & &  &\%  &   &  & (W Hz$^{-1}$) \\
  (1)       & (2) & (3) & (4)   & (5)     & (6)   & (7) &(8) &(9)  \\ \hline
0200+015    & ~41 & 8   & 1     &$-$1.17  &$<0.5$ &1.1**&1.3 &27.87 \\
0211$-$122  & 128 & 26  & 4     &$-$1.27  &3.8    &1.8  &2.0 &28.54 \\
0214+183    & ~48 &  9  & 1--2  &$-$1.05  &1.4    &1.4  &1.7 &27.92 \\
0355$-$037  & ~95 &$<8$ &       &$-$1.43  &       &     &1.8+&28.46 \\
0417$-$181  & ~28 & 11  & 1--2  &$-$1.26  &0.6    &1.7  &2.5 &28.32 \\
0529$-$549  & ~23 &$<8$ &       &$-$1.15  &       &     &    &28.54 \\
0748+134    & 103 & 12  & 2     &$-$1.46  &       &1.9**&1.0 &28.50 \\
0828+193    & ~92 & 11  & 3     &$-$1.23  &21     &1.3  &1.0 &27.65 \\
0943$-$242  & ~26 &$<5$ & 1     &$-$1.20  &$<0.5$ &1.0**&2.8 &28.30 \\
1243+036    & ~47 & 17  & 3     &$-$1.31  &1.0    &1.3  &2.0 &28.68 \\
1357+007    & ~35 &$<8$ &       &$-$1.21  &       &     &    &28.19 \\
1410$-$001  & 185 & 28  & 2     &$-$1.19  &6.7    &1.0  &2.7 &27.93 \\
1436+157    & ~42 & 11  & 2--3  &$-$1.18  &2.1    &1.3  &2.5 &27.90 \\
1545$-$234  & ~48 & 11  & 1     &$-$1.20  &11     &1.7  &1.7 & 28.19 \\
1558$-$003  & ~59 &$<8$ &       &$-$1.17  &       &2.8* &2.2+&28.27 \\
1707+105    & 171 &$<8$ &       &$-$1.20  &       &1.7* &8.8+&28.55 \\
2202+128    & ~26 & 11  & 4     &$-$1.25  &0.9    &1.8  &1.8 &28.06 \\
4C41.17     & ~92 & 20  & 3--4  &$-$1.33  &0.6    &1.5* &3.1 &28.58 \\ \hline
\end{tabular}
\end{center}
*  measured from $R$-band-ID + 1.5GHz radio map. \newline
** measured from $R$-band-ID on 4.7GHz radio map. \newline
+  measued from 1.5 GHz radio map. \newline
NOTE-- 0355$-$037: no position accurate enough could be measured of $R$-band ID. \newline
       1357+007 and 0529-549: not well enough resolved on 20cm radio map and 
                             no high frequency radio map available. \newline
\end{table}

\clearpage 
\newpage

\section*{Electronic Appendix}

\noindent {\bf 0200+015}: This galaxy is associated with a small ($5''$) 
double radio source and is a
typical example of luminous Ly$\alpha$ emission with strong H\,{\sc i} absorption.
Apart from one strong absorber 
(N(H\,{\sc i})$\sim10^{18}$ cm$^{-2}$)
it has two additional absorption systems of lower column density in the
Ly$\alpha$ emission. On the two-dimensional spectrum we see that the main
absorber does not cover the entire extent Ly$\alpha$ emission with the same
absorbing strength, but shows some
Ly$\alpha$ emission at the north-east extremity. This indicates 
that the column density or covering factor of the absorbing clouds changes
over the extent of the Ly$\alpha$ emission.

\noindent {\bf 0211$-$122}: This radio galaxy has a peculiar optical spectrum
in which the Ly$\alpha$ emission is anomalously weak compared to the higher 
ionization lines. This anomaly 
has been interpreted as being produced by dust mixed through the emission 
line gas which partly absorbs the Ly$\alpha$ emission \cite{oji94a}.
The two-dimensional high resolution 
spectrum of the Ly$\alpha$ shows a clearly different structure than the rest 
of the galaxies in our sample. 
There is one small region with strong Ly$\alpha$ 
emission that is relatively narrow (300 km s$^{-1}$ FWHM) and is responsible
for more than one third of the flux. Furthermore there are several weaker 
patches of Ly$\alpha$ emission distributed around the bright peak.
The bright narrow peak is spatially offset by $\sim 1''$ from the 
peak of the continuum emission.
This somewhat peculiar two-dimensional Ly$\alpha$ profile is consistent with the
interpretation of van Ojik et al. (1994)\nocite{oji94a} 
that dust is mixed through the
Ly$\alpha$ emitting gas of 0211$-$122.
The bright narrow peak may be a region where the dust
content is sufficiently low so that the Ly$\alpha$ photons can escape. 
In spite of the different appearance of the velocity field, the
integrated line profile does show a double peaked shape.
Although this double peaked appearance may be produced by dust within 
the Ly$\alpha$ region, 
it is also possible that an H\,{\sc i} absorption system plays a role.

If we model the observed
profile as being due to H\,{\sc i} absorption of an originally Gaussian profile, we 
obtain a fit as displayed in Fig. 2 with a very narrow 
($\sim$700
km s$^{-1}$ FWHM) Ly$\alpha$ having a strong 
H\,{\sc i} absorption of column density N(H\,{\sc i})=10$^{19.5}$ cm$^{-2}$. 
However, the centre of this absorption is not as opaque
as it should be with such a strong absorption (see also the two-dimensional 
spectrum in Fig. 1). There is more Ly$\alpha$ emission from
the bottom of the trough than in any of the other galaxies with strong HI
absorption. Thus it seems unlikely that this simple model
is correct. An alternative fit is given by a less luminous
original emission profile, thus
ignoring most of the emission from the bright narrow peak and giving an HI
column density of $\sim10^{18}$ cm$^{-2}$ (displayed in Fig. 4). 
However, this fit is also not well matched to the observed profile. 
It may be that the profile is purely due to dust 
mixed with the ionized gas, extinguishing the resonant scattering 
Ly$\alpha$ photons, although we cannot exclude that
H\,{\sc i} absorptions contribute to the shape of the Ly$\alpha$ profile
of 0211$-$122.

\noindent {\bf 0214+183}: The Ly$\alpha$ emission is double peaked over its 
entire spatial extent, although the spectrum does not have a 
high S/N 
due to the very blue wavelength of the redshifted Ly$\alpha$. 

\noindent {\bf 0355$-$037}: The Ly$\alpha$ profile is smooth and 
shows no signs of H\,{\sc i} absorptions. Apart from
the brightest Ly$\alpha$ emission, there is low surface brightness
Ly$\alpha$ emission extending over about $10''$ in the north-western direction, 
i.e. the direction of the brightest radio lobe.

\noindent {\bf 0417$-$181}: Although the spectrum has relatively low S/N, 
the Ly$\alpha$ emission has a deep and wide trough over its entire 
spatial extent of $\sim$42 kpc. In the red part of the Ly$\alpha$ emission on 
the two-dimensional spectrum, there
is a hint of another small absorption system.
On the spectrum, the faint continuum of a galaxy at $\sim5''$ north of 
0417$-$181 is also visible.

\noindent {\bf 0529$-$549}: The $\sim40$ kpc extended 
Ly$\alpha$ emission has a clear deep trough in its profile.

\noindent {\bf 0748+134}: This object shows no signs of H\,{\sc i} absorption, 
but there is a hint of a velocity shear in the Ly$\alpha$ emission. A more
sensitive spectrum is required to confirm this.

\noindent {\bf 0828+193}: 
The Ly$\alpha$ profile has a
spectacular shape. The emission is very strong but drops steeply
on the blue side of the peak, while slightly further bluewards some Ly$\alpha$
emission is visible again. Thus, it appears that almost the entire
blue wing of the Ly$\alpha$ emission profile has been absorbed. Also the most 
extended and fainter emission shows the same sudden drop at 4343 \AA.
This is a remarkable radio galaxy that 
has a close ($\sim3''$) companion along its radio axis \cite{rot95}.
The presence of a close companion, from which no Ly$\alpha$ emission is 
detected, suggests that a neutral gaseous halo of this galaxy might be
responsible for the absorption of the blue wing of the Ly$\alpha$ emission from
0828+193. However, it is not certain that the companion is at the same redshift,
because no emission lines are detected.
Also in the red wing of the Ly$\alpha$ profile, a broad shoulder is
observed that may be due to multiple H\,{\sc i} absorption systems or may be caused by 
intrinsic velocity structure of the Ly$\alpha$ emitting gas.

The steepness of the absorption trough next to the Ly$\alpha$
peak requires the absorber to have 
an H\,{\sc i} column density of $\sim10^{18.3}$ cm$^{-2}$. But the 
broadness of the absorption, extinguishing nearly all emission in the blue 
wing of Ly$\alpha$, requires the Doppler parameter for a single absorber to 
be $\sim162$ km s$^{-1}$. This fit is displayed in Fig. 2.
This absorption fit has removed too much of the original Ly$\alpha$
profile, as there is clearly emission observed just blueward of the sharp drop 
to zero at $\sim4340$ \AA. Thus, the absorption may be 
due to the combination of several absorption systems at slightly different 
velocities with respect to the Ly$\alpha$ peak.
The few small emission peaks that are left of the
Ly$\alpha$ blue wing, are significant and can be well modelled by assuming
three distinct absorbers (see Fig. 4).
The main absorber has a column density of $\sim10^{18.1}$ cm$^{-2}$
and the other two are $\sim10^{16.3}$ cm$^{-2}$ with more reasonable 
Doppler parameters of 16 to 80 km s$^{-1}$. Although the absorption may
be even more complex, this model is the simplest one that gives a satisfactory 
fit so we adopted these values. The broad shoulder in the red wing 
is also modelled by HI
absorption with a large gas velocity dispersion, but we cannot exclude that 
it is caused by true velocity structure in the Ly$\alpha$ emitting gas.

\noindent {\bf 1357+007}: This galaxy which is associated 
with a small ($3''$) radio source
has a deep trough in the Ly$\alpha$ profile.
The Ly$\alpha$ profile has a relatively low S/N.
The one-dimensional spectrum of this object 
is a good example for demonstrating the difference 
between modelling the troughs as being due to H\,{\sc i} absorption and as being due 
to true velocity structure.
The limitations of the models are illustrated by comparing the best fits
obtained by a two velocity component emission model (Fig. 3)
with those from an absorption model (Fig. 4).
The fit of the Ly$\alpha$ emission profile by a combination of two
Gaussian emission profiles at different velocities
is less satisfactory than the H\,{\sc i} absorption model. This is
because the trough between the two peaks on the profile is steeper than the
wings on the outside of the emission peaks. Although we cannot exclude the 
possibility of a non-Gaussian but symmetric double peaked emission profile,
this difference in steepness of
the trough and the outer wings of the emission profile is accounted for by a 
Voigt absorption profile which is also the best fit for the troughs in the 
profiles with the best S/N.
The actual situation may be more complicated than implied by the idealized 
assumptions of the models, but it seems that also in 1357+007
H\,{\sc i} absorption is the more plausible
interpretation in spite of the lower S/N ratio than in our best
spectra.

\noindent {\bf 1410$-$001}: This galaxy has a large (185 kpc) associated radio 
source and very extended Ly$\alpha$ 
emission ($\sim$80 kpc) with no signs of strong H\,{\sc i} absorption. However,
it shows an obvious velocity
shear in Ly$\alpha$, that may well be due to rotation.
The amplitude of the velocity shear is almost equal to 
the overall velocity width (FWHM) of the line. 

\noindent {\bf 1436+157}: This is a quasar--galaxy pair oriented along the 
radio axis (like the galaxy-galaxy pair 0828+193). 
There is no direct evidence that the galaxy is at the same 
redshift as the quasar; only
the quasar shows strong Ly$\alpha$ emission. 
Apart from very broad Ly$\alpha$
emission and strong continuum, as is common for quasars, the Ly$\alpha$ has
a spatially extended narrow component. This is why we include it in our sample.
A strong H\,{\sc i} absorption feature is present in the narrow line component 
over the entire Ly$\alpha$ extent. 
The extended Ly$\alpha$ emission is
larger than the radio source. Part of the emission
may be due to the companion galaxy or a tidal interaction if the companion 
galaxy is indeed associated with the quasar. 
 
The strong H\,{\sc i} absorption feature in the narrow line Ly$\alpha$
component requires a very large Doppler parameter of $\sim$200 km s$^{-1}$
when modelled by one single absorber. Although we cannot exclude the 
possibility of an absorber with such a large
intrinsic velocity dispersion, a better fit to the data is obtained 
by a model of two or more absorbers. 
In Fig. 3 the fit is shown that 
we obtain for two adjacent absorbers, 
each with an H\,{\sc i} column density of about 10$^{19.3}$ cm$^{-2}$ and 
Doppler parameters of 12--75 km s$^{-1}$.
We have also extracted a spectrum of the off-nuclear Ly$\alpha$ emission
by only summing the spectra beyond a distance of $2''$ from the peak of the 
continuum (shown in Fig. 4).
Although there is still some
contamination from the quasar continuum in this spectrum, the contribution
of broad Ly$\alpha$ emission is negligible.
To fit the off-nuclear profile, either a single absorber with large Doppler 
parameter ($\sim$250 km s$^{-2}$) is needed or more than one 
absorber. As a most plausible fit we have
adopted a two absorbers model, whose parameters are in Table 3.

\noindent {\bf 1545$-$234}: The Ly$\alpha$ profile of this radio galaxy
has an absorption in the blue wing. 
The absorption does not cover the entire Ly$\alpha$ extent.

\noindent {\bf 1558$-$003}: This object shows no signs of H\,{\sc i} absorption. The 
Ly$\alpha$ emission line appears 
to have a broad component, especially in the red wing. This galaxy also shows 
signs of a velocity shear in the Ly$\alpha$ kinematics.

\noindent {\bf 1707+105}: A galaxy with large associated radio sources and
a companion galaxy along the radio axis.
The Ly$\alpha$ emission is very extended ($\sim$130 kpc), 
where not only the radio galaxy
itself (A) but also the companion galaxy (B) shows strong Ly$\alpha$ 
emission. On the opposite side of the radio galaxy is a region of 
faint Ly$\alpha$
emission with a relative blueshift that seems to increase with distance from
the radio galaxy. The Ly$\alpha$ emitting gas has a relatively low velocity 
dispersion and shows no signs of H\,{\sc i} absorption.

\noindent {\bf 2202+128}: The radio and Ly$\alpha$ size of this galaxy are 
small ($\sim$35 kpc)
and the Ly$\alpha$ profile shows a strong H\,{\sc i} absorption feature just bluewards
of the peak and a somewhat weaker absorption feature in the red wing
at 4518 \AA. Further the profile shows a 
flattened ``shoulder'' in the red wing (see Fig. 4).

Taking the small peak next to the absorption at 4518 \AA\ to be real 
and assuming that the ``shoulder'' is also due to absorption,
implies a very broad absorption (Doppler parameter $\sim 450$ km s$^{-1}$).
An alternative model ignoring the small peak next to the absorption at 
4518 \AA\ gives similar values for the two sharp absorptions as in the first
model, but involves a much lower Doppler parameter ($\sim 230$ km s$^{-1}$
to account for the ``shoulder''. 
It is possible that the shoulder in the red wing of the emission profile is
indeed caused by an H\,{\sc i} absorption system with a large intrinsic velocity
dispersion or by a superposition of many small absorbers over this velocity
range. However the underlying Ly$\alpha$ emission profile may not be strictly
Gaussian and the ``shoulder'' may be due to true velocity structure of the
Ly$\alpha$ emitting gas.

\end{document}